\documentclass[reprint, superscriptaddress, amsmath, amssymb, prd, floatfix, onecolumn, notitlepage, 12pt]{revtex4-1}

\usepackage{fontawesome} 
\usepackage[breaklinks,colorlinks,urlcolor=blue,citecolor=blue,linkcolor=blue]{hyperref}
\usepackage[caption=false]{subfig}

\usepackage{graphicx}
\usepackage{dcolumn}
\usepackage{bm}
\usepackage[usenames,dvipsnames]{color}
\usepackage[normalem]{ulem}
\usepackage[breaklinks,colorlinks,urlcolor=blue,citecolor=blue,linkcolor=blue]{hyperref}
\usepackage{float}
\usepackage[T1]{fontenc}
\usepackage{times}
\usepackage{soul}

\usepackage{makecell, booktabs, multirow, siunitx} 

\newcommand{\dif}{\mathop{}\!{d}}

\begin{document}

\title[GMM-MI: interpret deep learning with mutual information]{A robust estimator of mutual information for deep learning interpretability}

\author{Davide Piras}
\email[]{d.piras@ucl.ac.uk}
\affiliation{\footnotesize{Department of Physics \& Astronomy, University College London, Gower Street, London WC1E 6BT, UK}}
\affiliation{\footnotesize{Département de Physique Théorique, Université de Genève, 24 quai Ernest Ansermet, 1211 Genève 4, Switzerland}}

\author{Hiranya V. Peiris}
\affiliation{\footnotesize{Department of Physics \& Astronomy, University College London, Gower Street, London WC1E 6BT, UK}}
\affiliation{\footnotesize{The Oskar Klein Centre for Cosmoparticle Physics, Department of Physics, Stockholm University, AlbaNova, Stockholm, SE-106 91, Sweden}}

\author{Andrew Pontzen}
\affiliation{\footnotesize{Department of Physics \& Astronomy, University College London, Gower Street, London WC1E 6BT, UK}}

\author{Luisa Lucie-Smith}
\affiliation{\footnotesize{Max-Planck-Institut für Astrophysik, Karl-Schwarzschild-Str. 1, 85748 Garching, Germany}}

\author{Ningyuan Guo}
\affiliation{\footnotesize{Department of Physics \& Astronomy, University College London, Gower Street, London WC1E 6BT, UK}}

\author{Brian Nord}
\affiliation{\footnotesize{Fermi National Accelerator Laboratory, P.O. Box 500, Batavia, IL 60510, USA}}
\affiliation{\footnotesize{Department of Astronomy \&  Astrophysics, University of Chicago, Chicago, IL 60637, USA}}
\affiliation{\footnotesize{Kavli Institute for Cosmological Physics, University of Chicago, Chicago, IL 60637, USA}}

\date{\today}

\begin{abstract}
We develop the use of mutual information (MI), a well-established metric in information theory, to interpret the inner workings of deep learning models. To accurately estimate MI from a finite number of samples, we present GMM-MI (pronounced ``Jimmie''), an algorithm based on Gaussian mixture models that can be applied to both discrete and continuous settings. GMM-MI is computationally efficient, robust to the choice of hyperparameters and provides the uncertainty on the MI estimate due to the finite sample size. We extensively validate GMM-MI on toy data for which the ground truth MI is known, comparing its performance against established mutual information estimators. We then demonstrate the use of our MI estimator in the context of representation learning, working with synthetic data and physical datasets describing highly non-linear processes. We train deep learning models to encode high-dimensional data within a meaningful compressed (latent) representation, and use GMM-MI to quantify both the level of disentanglement between the latent variables, and their association with relevant physical quantities, thus unlocking the interpretability of the latent representation. We make GMM-MI publicly available in this GitHub repository. \href{https://github.com/dpiras/GMM-MI}{\faicon{github}}
\end{abstract}

%
%
%
%
%
\maketitle

\section{Introduction}
\label{sec:introduction}
The flexibility and expressiveness of deep learning (DL) models are attractive features, which have led to their application to a variety of scientific problems (see e.g.\ \citet{Raghu20} for a recent review). Despite this recent progress, deep neural networks remain opaque models, and their power as universal approximators \cite{Cybenko89, Hornik89, Hornik91} comes at the expense of interpretability \cite{Molnar22}. Many techniques have been developed to gain insight into such black-box models \cite{Zeiler14, Simonyan14, Zhou16, Ribeiro16, Selvaraju17, Shrikumar17, Lundberg17, Chattopadhay18}. These solutions vary in their computational efficiency and in the range of tasks to which they can be applied; however, there is no consensus as to which method provides the most trustworthy interpretations, and a general framework to interpret deep neural networks is still an avenue of active investigation (see e.g.\ \citet{Li21, Linardatos21} for recent reviews).


DL models are also widely used in representation learning, where a high-dimensional dataset is compressed to a smaller set of variables; this latent representation should contain all the relevant information for downstream tasks such as reconstruction, classification or regression \cite{Schmidhuber92, Bengio13}. Disentanglement of these compressed variables is also often imposed, in order to associate each latent to a physical quantity of domain interest \cite{Bengio13, Louizos15, Chen16, Lample17, Higgins17, Jha18, Locatello19, Lezama19}. However, how best to access the information captured by these latent vectors and connect it to the relevant factors remain open questions.

In this work, we focus on representation learning and link the latent variables to relevant physical quantities by estimating their mutual information (MI), a well-established information-theoretic measure of the relationship between two random variables. MI allows us to interpret what the DL model has learned about the domain-specific parameters relevant to the problem: by interrogating the model through MI, we aim to discover what information is used by the model in making predictions, thus achieving the interpretation of its inner workings. We also use MI to quantify the level of disentanglement of the latent variables. 

MI has found applications in a variety of scientific fields, including astrophysics \cite{Pandey17, Sarkar20, Bhattacharjee20, Upham21, Malz21, Sarkar21, Jeffrey21, LucieSmith22, Sarkar22}, biophysics \cite{Fairhall12, Charzynska16, Tkacik16, Levchenko16, Wegner18, Holmes19, Uda20}, and dynamical complex systems \cite{Wicks07, Dunleavy12, Runge15, Myers19, Svenkeson19, Diego19, Jiang19, Jia20}, to name a few. However, estimating the mutual information $I (X, Y)$ between two random variables $X$ and $Y$, given samples from their joint distribution $p_{(X, Y)}$, remains a long-standing challenge, since it requires $p_{(X, Y)}$ to be known or estimated accurately \cite{Paninski03, Vergara15}. When $X$ and $Y$ are continuous variables with values over $\mathcal{X} \times \mathcal{Y}$, $I (X, Y)$ is defined as:
\begin{align}
    I (X, Y) \equiv \int_{\mathcal{X} \times \mathcal{Y}}  p_{(X, Y)}(x, y) \ln{\frac{p_{(X, Y)}(x, y)}{ p_X(x) p_Y(y)}} \dif x \dif y \ ,
    \label{eq:MI}
\end{align}
where $p_X$ and $p_Y$ are the marginal distributions of $X$ and $Y$, respectively, and $\ln$ refers to the natural logarithm, so that MI is measured in natural units (nat). $I (X, Y)$ represents the amount of information one gains about $Y$ by observing $X$ (or vice versa): it captures the full dependence between two variables going beyond the Pearson correlation coefficient, since $I (X, Y) = 0$ if and only if $X$ and $Y$ are statistically independent \cite{Cover06}. A comprehensive summary of MI and its properties can be found in \citet{Vergara15}.

The most straightforward estimator of $I (X, Y)$ given samples of $p_{(X, Y)}$ consists of binning the data and approximating Eq.~(\ref{eq:MI}) with a finite sum over the bins. This approach is heavily dependent on the binning scheme, and is prone to systematic errors \cite{Fraser86, Moon95, Darbellay99, Kwak02, Kraskov04, Suzuki08, Saxe19, Holmes19, Pichler22}. \citet{Kraskov04} proposed an estimator (hereafter referred to as KSG), based on $k$-nearest neighbors, which rewrites $I (X, Y)$ in terms of the Shannon entropy, and then applies the Kozachenko-Leonenko entropy estimator \cite{Kozachenko87} to calculate each term (see Sect.~\ref{sec:ksg_mine} for more details). However, the KSG estimator only returns a point estimate, is strongly dependent on the number of chosen neighbors, and does not scale well with sample size \cite{Gao14}. Bayesian approaches to obtain the full distribution of MI have also been discussed \cite{Hutter01, Hutter05, Archer13}, but they are not easily applicable to continuous data, and have been shown to be strongly dependent on the chosen prior \cite{Archer13}. 

More recently, MI estimators based on bounds approximated by neural networks have gained interest \cite{Tishby15, Chen16, Alemi16, Brakel17, Kolchinsky19, Belghazi18, Oord18, Moyer18, Poole19, Peng19, Hjelm19, Song20, Gokmen21}. In particular, \citet{Belghazi18} proposed a neural estimator of $I (X, Y)$ (hereafter referred to as MINE) rewriting it as a Kullback-Leibler (KL) divergence \cite{Kullback51}, and considering its Donsker-Varadhan representation \cite{Donsker83} (see Sect.~\ref{sec:ksg_mine} for more details). While yielding differentiable MI estimates (essential e.g.\ for backpropagation when training DL models), such neural-network-based estimators do not necessarily return an accurate estimate of Eq.~(\ref{eq:MI}), are heavily dependent on the training hyperparameters, and have been shown to suffer from a poor variance-bias tradeoff \cite{Song20}. The use of MI estimates for interpreting deep representation learning has recently been investigated as well \cite{Chen18, Chen16, LucieSmith22, Sedaghat21}; however, exploiting MI to interpret deep representation learning requires a robust density estimate of the joint probability distribution between latent variables and relevant physical parameters, and the uncertainties on the MI estimate to be quantified, ensuring that any trends in MI are statistically significant.

To address these requirements, we present GMM-MI (pronounced ``Jimmie''), an algorithm to estimate the full distribution of $I (X, Y)$ based on fitting samples drawn from the distribution with Gaussian mixture models (GMMs). While the use of GMMs to estimate MI is not new \cite{Kerroum10, Eirola14, Lan06, Leiva04, Nilsson02, Polo22}, these previous works only considered MI in the context of feature selection, and did not carry out uncertainty quantification on the relevant MI estimates, which is critical when using MI to interpret deep learning models. GMM-MI has been designed to be a robust and flexible tool that can be applied to multiple settings where MI estimation is required. Crucially, it also returns error estimates which we verified to be statistically correct on test datasets including bivariate distributions of various shapes and non-linear transformations of Gaussian samples. We first extensively validate GMM-MI on these toy data for which the ground truth MI is known, including comparisons to the KSG and MINE estimators in terms of both efficiency and accuracy, additionally showing that GMM-MI is unbiased and the MI uncertainty scales as expected with the sample size. We then train representation-learning models on high-dimensional datasets including simulations of dark matter halos formed through non-linear physical processes, real astrophysical spectra and synthetic shape images with known labels. We demonstrate the use of GMM-MI to achieve the interpretability of such models.

The paper is structured as follows. In Sect.~\ref{sec:proposed_procedure} we describe GMM-MI, and recall the essential details of the KSG and MINE estimators in Sect.~\ref{sec:ksg_mine}. In Sect.~\ref{sec:validation}, we present extensive experiments where we validate our MI estimator on toy data, and then in Sect.~\ref{sec:results} we use MI to interpret the latent space of DL models trained on synthetic and real data. We conclude in Sect.~\ref{sec:conclusion}, including an outlook over planned extensions of our algorithm.

\section{Method}
\label{sec:method}
\subsection{Estimation procedure (GMM-MI)}
\label{sec:proposed_procedure}
Our algorithm uses a GMM with $c$ components to obtain a fit of the joint distribution $p_{(X, Y)}$:
\begin{equation}
    p_{(X, Y)}(x, y | \theta) = \sum_{i=1}^c w_i \mathcal{N} (x, y| \mu_i, \Sigma_i) \ ,
    \label{eq:gmm}
\end{equation}
where 
$\theta$ is the set of weights $w_{1 : c}$, means $\mu_{1:c}$ and covariance matrices $\Sigma_{1:c}$. With this choice, the marginals $p(x)$ and $p(y)$ are also GMMs, with parameters determined by $\theta$. Our procedure for estimating MI and its associated uncertainty is as follows.

\begin{enumerate}
    \item For a given number of GMM components $c$, we randomly initialize $n_{\rm{init}}$ different GMM models. Each set of initial GMM parameters is obtained by first randomly assigning the responsibilities, namely the probabilities that each point belongs to a component $i$, sampling from a uniform distribution. The starting values of each $\mu_i$ and $\Sigma_i$ are calculated as the sample mean and covariance matrix of all points, weighted by the responsibilities, while each $w_i$ is initialized as the average responsibility across all points. Having multiple initializations is crucial to reduce the risk of stopping at local optima during the optimization procedure \cite{Ueda98, Bovy11, Emilie16, Baudry15, Melchior18}.
    \item We fit the data using $k$-fold cross-validation: this means that we train a GMM on $k-1$ subsets of the data (or ``folds''), and evaluate the trained model on the remaining validation fold. Each fit is performed with the expectation-maximization algorithm \cite{Dempster77}, and terminates when the change in log-likelihood on the training data is smaller than a chosen threshold. We also add a small regularization constant $\omega$ to the diagonal of each covariance matrix, as described e.g.\ in \citet{Melchior18}, to avoid singular covariance matrices.
    \item We select the model with the highest mean validation log-likelihood across folds $\hat{\ell}_c$, since it has the best generalization performance. Among the $k$ models corresponding to $\hat{\ell}_c$, we also store the final GMM parameters with the highest validation log-likelihood on a single fold: these will be used to initialize each bootstrap fit in step 5, thus reducing the risk of stopping at local optima and significantly accelerating convergence.
    \item We repeat steps 1--3 iteratively increasing the number of GMM components from $c=1$. We stop when $\hat{\ell}_{c} - \hat{\ell}_{c-1}$ is smaller than a user-specified positive threshold, and select the value of $c-1$ as the optimal number of GMM components to fit. In this way, we avoid overfitting the training data and adding too many components, which would considerably slow down the procedure while not significantly improving the density estimation.
    \item We bootstrap the data $n_{\rm{b}}$ times, and fit a GMM to each bootstrapped realization. Each fit is initialized with the set of parameters selected in step 3, and with the number of components found in step 4. We use bootstrap to capture not just a point estimate of MI, but its full distribution.
    \item For each fitted model, we calculate MI by solving the integral in Eq.~(\ref{eq:MI}) using Monte Carlo (MC) integration over $M$ samples.
    \item We return the sample mean and standard deviation of the distribution of MI values.
\end{enumerate}

\begin{figure}
\vskip -0.2in
        \centering
        \includegraphics[trim={0 0.7cm 0 0.4cm},clip, width=\columnwidth]{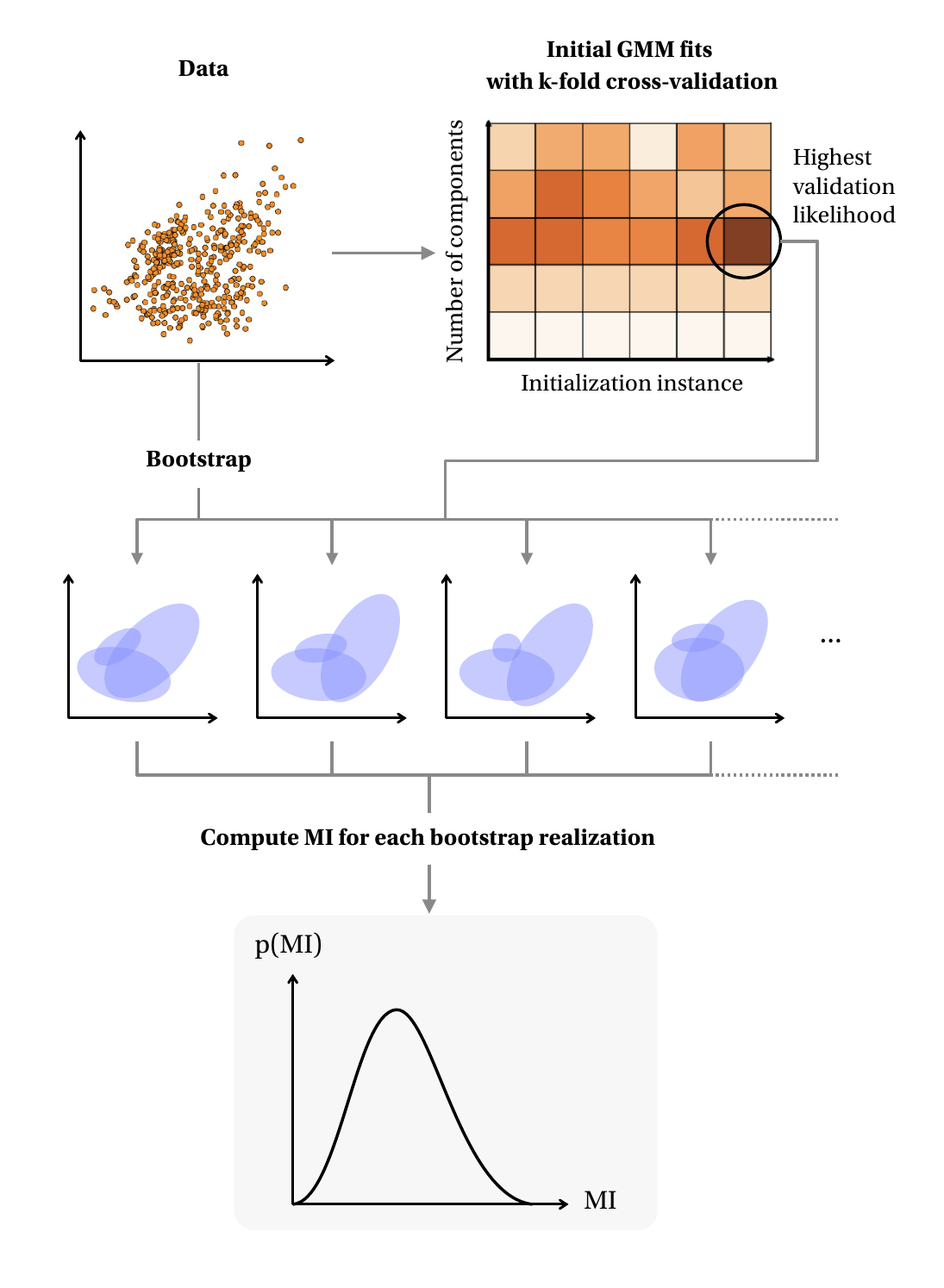}
        \caption{Flowchart describing GMM-MI, our algorithm for estimating the distribution of mutual information (MI) using Gaussian mixture models (GMMs). 
        All implementation details can be found in Sect.~\ref{sec:proposed_procedure}.}
        \label{fig:flowchart}
\end{figure}

A flowchart summarizing the GMM-MI procedure is shown in Fig.~\ref{fig:flowchart}. We choose the initialization procedure described in step 1 for its speed, but in our implementation of GMM-MI other initialization procedures are also available and could be alternatively used. For instance, it is possible that the random initialization we set as default returns overlapping components which inhibit the optimization procedure; in those cases, we recommend switching to an initialization based on $k$-means \cite{Lloyd82}. On the other hand, $k$-means itself is known to only guarantee convergence to local optima \cite{Arthur07}; for this reason, we also provide the possibility to perturb the means by a user-specified scale after an initial call to $k$-means. We call this approach ``randomized $k$-means'', and offer full flexibility to select the most appropriate initialization type based on the data being analyzed.

Our implementation also allows the user to set a higher patience, i.e.\ consider more than one additional component in step 4 after the validation loss has started to decrease; alternatively, it is possible to select the number of components yielding the lowest Akaike information criterion (AIC, \cite{Akaike74}) or Bayesian information criterion (BIC, \cite{Schwarz78}), with details in Appendix~\ref{app:AIC_BIC}. All three methods implemented are computationally efficient, and aim to prevent the model from overfitting the available samples; in Fig.~\ref{fig:ll_components} we further show that even in a case where the three metrics disagree on the number of GMM components to use, the final MI estimates agree with each other within the uncertainties, thus demonstrating that GMM-MI is robust to the metric being used. The number of folds ($k$) should be set based on the number of available samples, so that each fold is representative of the data. The number of initializations ($n_{\rm{init}}$), bootstrap realizations ($n_{\rm{b}}$), and MC samples ($M$) should be chosen based on the available computational budget. 

In many instances, the factors of variation that are used to generate the data are discrete variables \cite{Ross14}; in these cases, we will need to estimate MI between a continuous variable $X$ and a categorical variable $F$ which can take $v$ different values $f_{1:v}$. In this case, assuming the $v$ values have equal probability (as will be the case when considering the 3D shapes dataset in Sect.~\ref{sec:3D_shapes}), the mutual information $I(X, F)$ can be expressed as:
\begin{align}
     I(X, F) = &\frac{1}{v} \sum_{i=1}^{v}  \int_{\mathcal{X}} \dif x \  p_{(X|F)}(x|f_i) \left [ \ln p_{(X|F)}(x|f_i) - \ln \frac{1}{v} \sum_{j=1}^{v} p_{(X|F)}(x|f_j) \right ]  \ ,
    \label{eq:MI_cat}
\end{align}
where we use a GMM to fit each conditional probability $p_{(X|F)}(x|f_i)$. The full derivation of Eq.~(\ref{eq:MI_cat}) can be found in Appendix~\ref{app:MI_der}.

\subsection{Alternative estimators}
\label{sec:ksg_mine}
In order to validate our algorithm, we compare it with two established estimators of MI. The KSG estimator, first proposed in \citet{Kraskov04}, rewrites MI as:
\begin{align}
    I(X, Y) = H(X) + H(Y) - H(X, Y) \ ,
    \label{eq:MI_ksg}
\end{align}
where $H(\cdot)$ refers to the Shannon entropy, defined for a single variable as:
\begin{align}
\label{eq:entropy}
    H (X) \equiv - \int_{\mathcal{X}} p_X(x) \ln{p_X(x)} \ .
\end{align}
The Kozachenko-Leonenko estimator \cite{Kozachenko87} is then used  to evaluate the entropy in Eq.~(\ref{eq:entropy}):
\begin{align}
    \label{eq:kl_entropy}
    \widehat{H} (X) = - \psi(k) + \psi(N) + \ln{c_{\rm{d}}} + \frac{d}{N} \sum_{i=1}^{N} \ln {\epsilon^{(k)}} (i) \ ,
\end{align}
where $\psi(\cdot)$ is the digamma function, $k$ is the chosen number of nearest neighbors, $N$ is the number of available samples, $d$ is the dimensionality of $X$, $c_{\rm{d}}$ is the volume of the unit ball in $d$ dimensions, and $\epsilon^{(k)}$ is twice the distance between the $i^{\rm{th}}$ data point and its $k^{\rm{th}}$ neighbor. Applying Eq.~(\ref{eq:kl_entropy}) to each term in Eq.~(\ref{eq:MI_ksg}) would lead to biased estimates of MI \cite{Kraskov04, Holmes19}; for this reason, the KSG estimator actually considers a ball containing the $k$-nearest neighbors around each sample, and counts the number of points within it in both the $x$ and $y$ direction. The resulting estimator of MI then becomes \cite{Kraskov04, Holmes19}:
\begin{align}
    \widehat{I} (X, Y) =  \psi(k) + \psi(N) - \frac{1}{k} - \langle \psi(n_x^{(k)}) + \psi(n_y^{(k)}) \rangle  \ ,
\end{align}
where $n_x^{(k)}$ $(n_y^{(k)})$ represents the number of points in the $x$ $(y)$ direction, and $\langle \cdot \rangle$ indicates the mean over the available samples. In our experiments, we consider the implementation of the KSG estimator available from \textsc{sklearn} in this \href{https://scikit-learn.org/stable/modules/generated/sklearn.feature_selection.mutual_info_regression.html}{https link}.

We also compare our algorithm against the MINE estimator proposed in \citet{Belghazi18}. MI as defined in Eq.~(\ref{eq:MI}) can be interpreted as the KL divergence $D_{\rm{KL}}$ between the joint distribution and the product of the marginals:
\begin{align}
     I (X, Y) = D_{\rm{KL}}  \left[ p_{(X, Y)} || p_X p_Y \right] \ ,  
\end{align}
where the KL divergence between two generic probability distributions $p_X$ and $q_X$ defined over $\mathcal{X}$ is defined as:
\begin{align}
   D_{\rm{KL}} \left[ p_X || q_X \right] \equiv  \int_{\mathcal{X}} \dif x \ p_X(x) \ln{\frac{p_X(x)}{q_X(x)}}  \ .
\end{align}
The MINE estimator then considers the Donsker-Varadhan representation \cite{Donsker83} of the KL divergence:
\begin{align}
     D_{\rm{KL}} \left[ p_X || q_X \right]  = \sup_{T} \mathbb{E}_{p_X} \left [ T \right] - \ln{\mathbb{E}_{q_X} \left [ e^T \right]}  \ ,
\end{align}
where the supremum is taken over all the functions $T$ such that the expectations $\mathbb{E}\left[ \cdot \right ]$ are finite, and parameterizes $T$ with a neural network. In our experiments, we consider the implementation available in this \href{https://github.com/gtegner/mine-pytorch}{https link}, which includes the mitigation of the gradient bias through the use of an exponential moving average, as suggested in \citet{Belghazi18}.

\subsection{Representation learning}
\label{sec:bvae}
We apply our MI estimator GMM-MI to interpret the latent space of representation-learning models. Specifically, we consider $\beta$-variational autoencoders ($\beta$-VAEs, \cite{Kingma14, Higgins17}), where one neural network is trained to encode high-dimensional data $D$ into a distribution over disentangled latent variables $\mathbf{z}$, and a second network decodes samples of the latent distribution back into data points $\widetilde{D}$. The two networks are trained together to minimize the following loss function:
\begin{align}
    \mathcal{L} = \rm{MSE}(D, \widetilde{D}) + \beta D_{\rm{KL}} \left[p_{\phi}(\mathbf{z} | D) || p(\mathbf{z})\right] \ ,
    \label{eq:VAE}
\end{align}
where $\rm{MSE}$ indicates the mean squared error, $p_{\phi}(\mathbf{z} | D)$ represents the encoder parameterized by a set of weights $\phi$, $p(\mathbf{z})$ is the prior over the latent variables $\mathbf{z}$, and $\beta$ is a regularization constant which controls the level of disentanglement of $\mathbf{z}$. 

We will also reproduce the results of \citet{LucieSmith22} in Sect.~\ref{sec:dm_haloes}, for which the architecture is slightly different: the latent samples are combined with a given query (the radius $r$) and fed through the decoder to predict dark matter halo density profiles at each given $r$. This model is referred to as the interpretable variational encoder (IVE), with an analogous loss function to Eq.~(\ref{eq:VAE}).

\section{Validation}
\label{sec:validation}
In this section, we validate GMM-MI on toy data for which the MI can be computed analytically: we show that GMM-MI is in good agreement with the ground truth, as well as other MI estimators, while returning the full distribution of MI including its uncertainty. We run all the MI estimations on a single CPU node with 40 2.40GHz Intel Xeon Gold 6148 cores using no more than 300 MB of RAM, reporting the speed performance in each case.

\begin{figure}[ht!]
\subfloat[Bivariate Gaussian distribution.]{%
  \includegraphics[trim={0 0.1cm 0 0.1cm},clip, width=0.5\columnwidth]{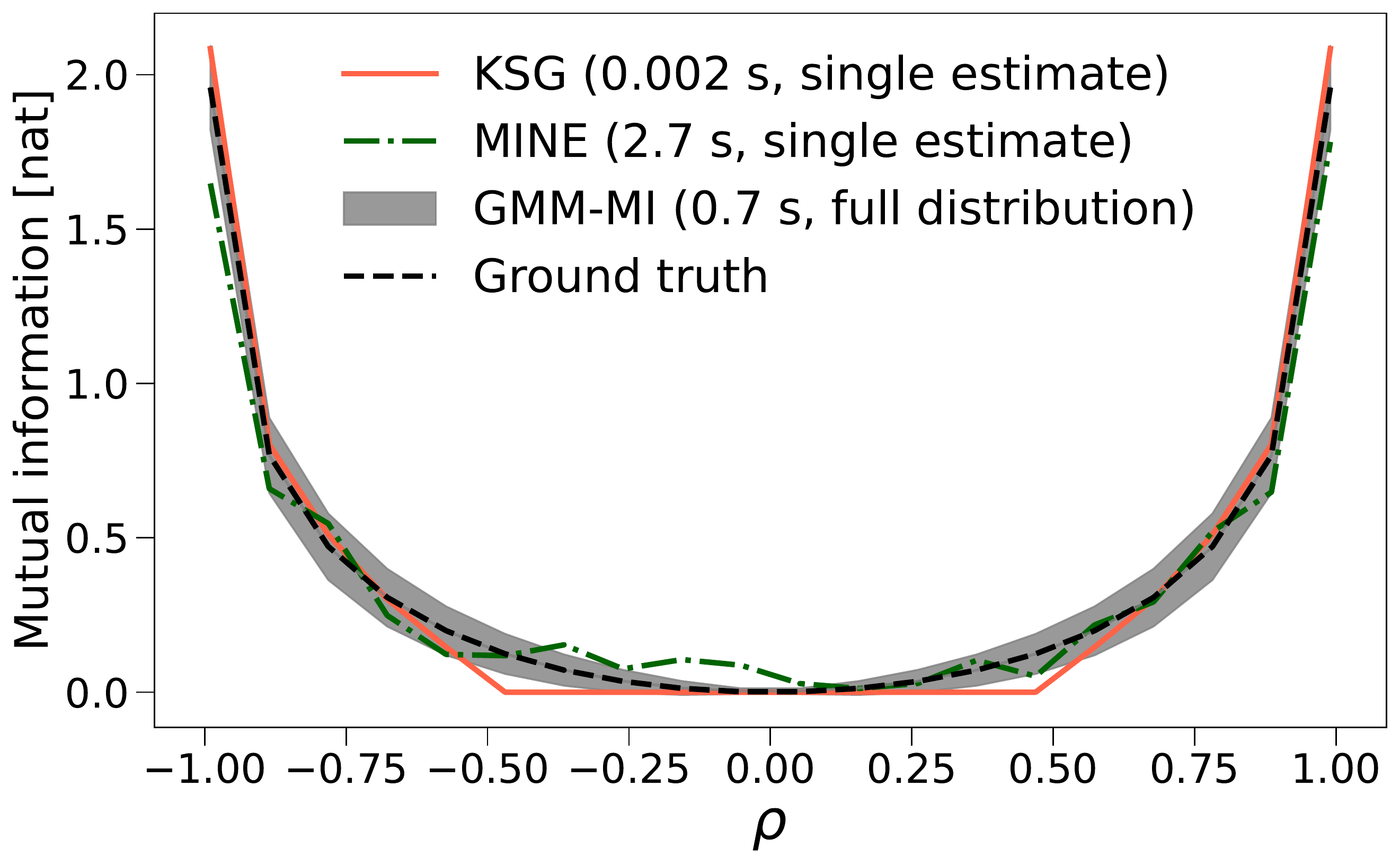}%
}
\subfloat[Gamma-exponential distribution.]{%
  \includegraphics[trim={0 0.1cm 0 0.1cm},clip, width=0.5\columnwidth]{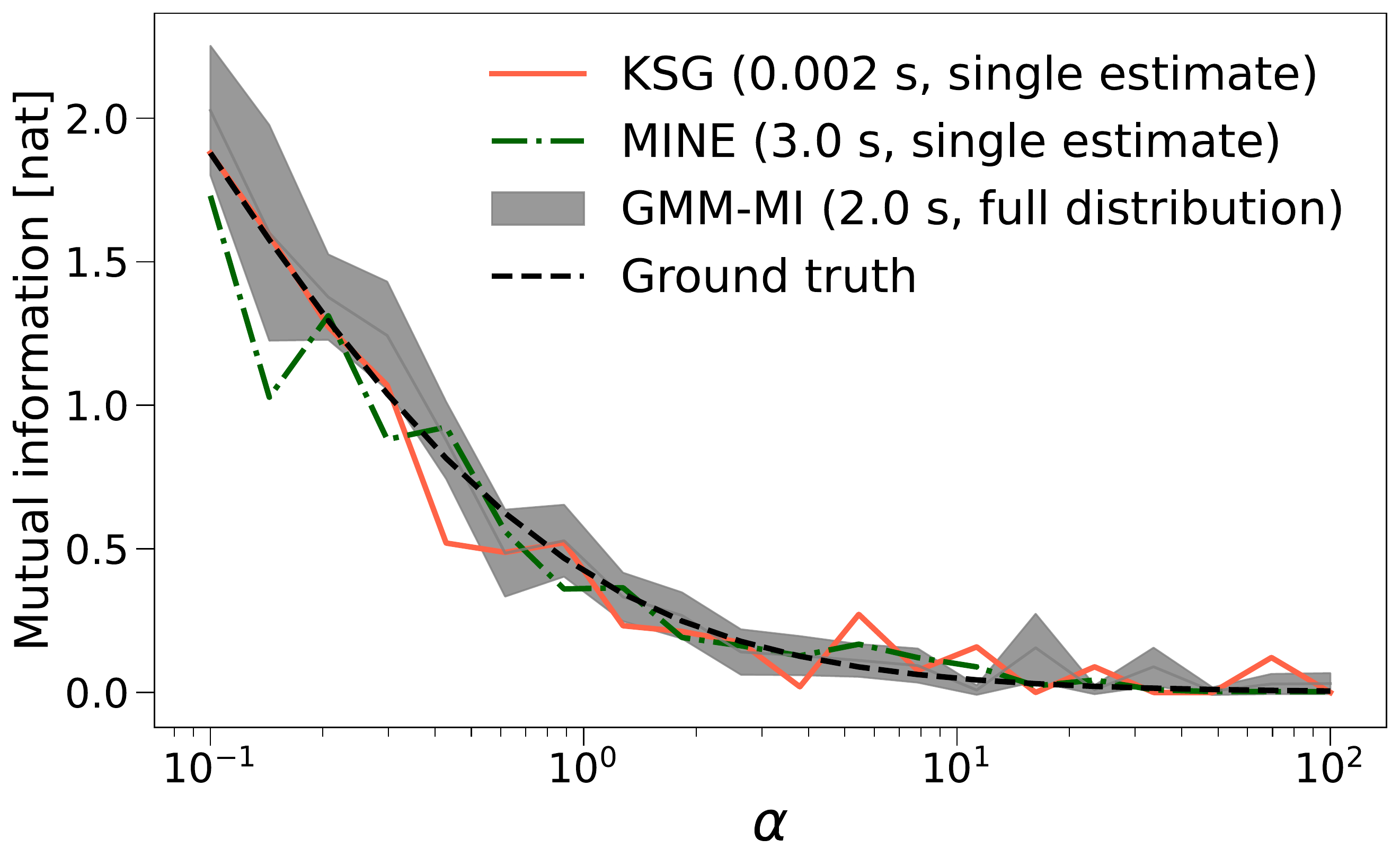}%
}

\subfloat[Ordered Weinman exponential distribution.]{%
  \includegraphics[trim={0 0.1cm 0 0.1cm},clip, width=0.5\columnwidth]{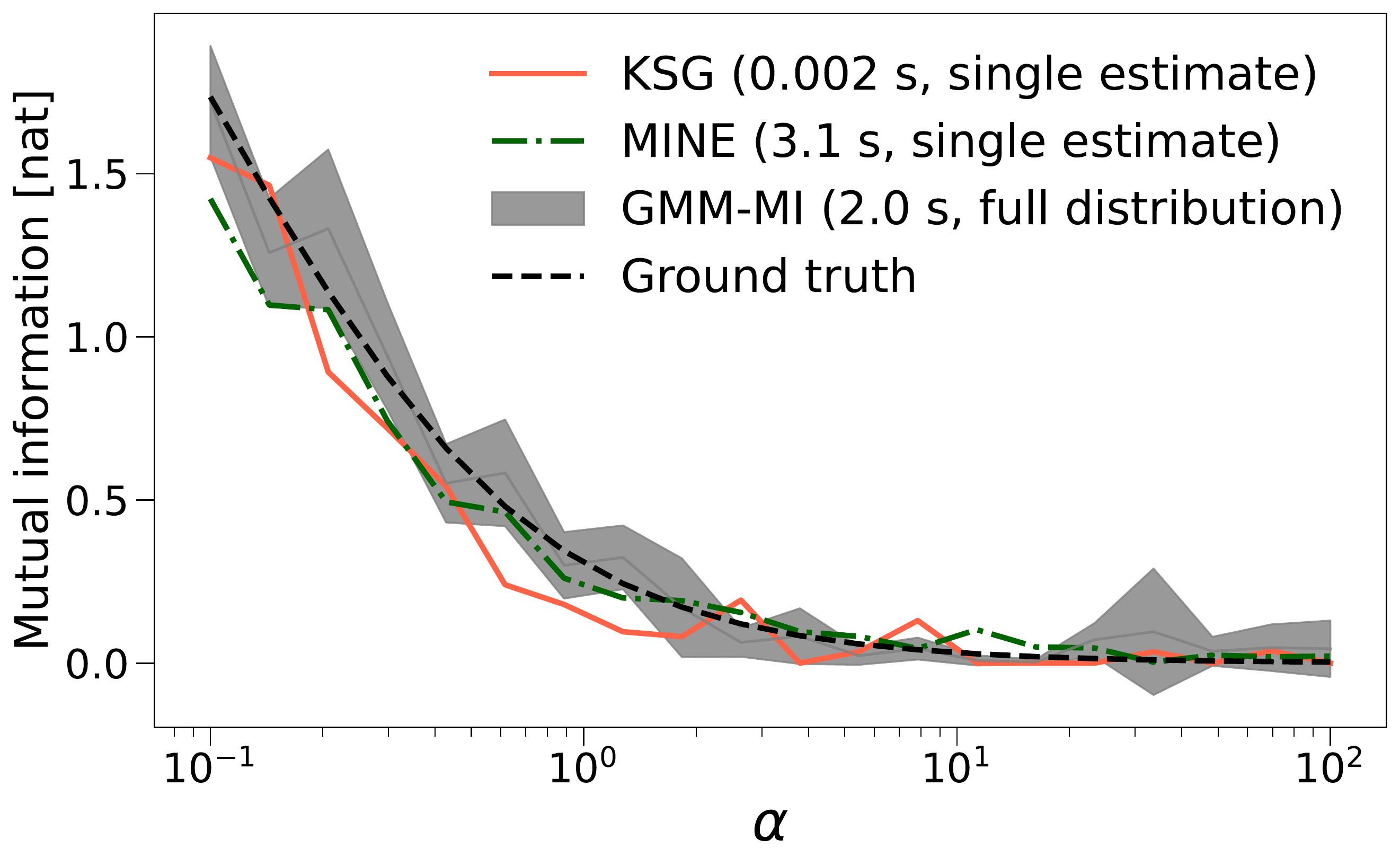}%
}
\caption{Estimates of mutual information (MI) for: (a) a bivariate Gaussian distribution with varying correlation coefficient $\rho$; (b) a gamma-exponential distribution with varying $\alpha$, as in Eq.~(\ref{eq:gammaexp}); (c) an ordered Weinman exponential distribution with varying $\alpha$, as in Eq.~(\ref{eq:weinman}). The dashed black line indicates the ground truth MI. We compare the KSG estimator (\cite{Kraskov04}, solid red line), the MINE estimator (\cite{Belghazi18}, dotted-dash green line), and our estimator GMM-MI, indicated with the gray shaded area (mean $\pm$ two standard deviations). The numbers in parentheses indicate the time to obtain a single estimate with KSG and MINE, and the full distribution of MI in the case of GMM-MI, for each $\rho$ or $\alpha$. These estimates are obtained from $N=200$ samples, and with hyperparameters reported in Sect.~\ref{sec:validation}.}\label{fig:D1_compare}
\end{figure}


We first consider a bivariate Gaussian distribution with unit variance of each marginal and varying level of correlation $\rho \in \left[ -1, 1 \right]$, following \citet{Belghazi18}. In this case, the true value of I(X, Y) can be obtained analytically by solving the integral in Eq.~(\ref{eq:MI}), yielding:
\begin{align}
    I(X, Y)_{\rm{true}} = -\frac{1}{2} \ln {\left(1-\rho^2 \right)} \ .
    \label{eq:mi_true}
\end{align}
We consider two additional bivariate distributions, the gamma-exponential distribution \cite{Darbellay99, Darbellay00, Kraskov04, Haeri14}, with density ($\alpha>0$ is a free parameter):
\begin{align}
\label{eq:gammaexp}
    p_{(X, Y)} (x, y | \alpha) = \begin{cases} \frac{1}{\Gamma(\alpha)} x^{\alpha} e^{-x-xy} & x>0, y>0 \\
    0 & \mbox{otherwise}
    \end{cases}\ ,
\end{align}
where $\Gamma$ is the gamma function, and the ordered Weinman exponential distribution \cite{Darbellay99, Darbellay00, Kraskov04, Haeri14}, with density:
\begin{align}
\label{eq:weinman}
    p_{(X, Y)} (x, y | \alpha) = \begin{cases} \frac{2}{\alpha} e^{-2x-\frac{y-x}{\alpha}} & y>x>0 \\
    0 & \mbox{otherwise}
    \end{cases}\ .
\end{align}
The true value of $I(X, Y)$ for these distributions can be obtained analytically, and is reported in Appendix~\ref{app:true_mi}. Since $I(X, Y)$ is invariant under invertible transformations of each random variable \cite{Kraskov04}, we consider $\ln(X)$ and $\ln{(Y)}$ when estimating MI in the case of the last two distributions \cite{Kraskov04}. To demonstrate the power of our estimator, we restrict ourselves to the case with only $N=200$ samples. To estimate MI, we consider the KSG estimator with 1 neighbor (to minimize the bias, and following \citet{Kraskov04}), the MINE estimator trained for 50 epochs with a learning rate of $10^{-3}$ and a batch size of 32, and our estimator GMM-MI with $k=2$ folds, $n_{\rm{init}} =3$ different initializations, a log-likelihood threshold on each individual fit of $10^{-5}$, a threshold on the mean validation log-likelihood to select the number of GMM components of $10^{-5}$, $n_{\rm{b}} = 100$ bootstrap realizations, $M=10^4$ MC samples, and a regularization scale of $\omega=10^{-12}$.

The results are reported in Fig.~\ref{fig:D1_compare}. The KSG estimator is the fastest, and yields MI values closely matching the ground truth, but returns biased estimates around e.g.\ $|\rho| = 0.4$ in the bivariate Gaussian case, and $\alpha\simeq 1$ in the ordered Weinman case. The MINE estimator is more computationally expensive and shows a relatively high variance, which is expected since MINE has been shown to be prone to variance overestimation due to the use of batches \cite{Poole19}. 
GMM-MI, on the other hand, returns a distribution of MI in good agreement with the ground truth in $\mathcal{O}(1)$ s, and provides an uncertainty estimate due to the finite sample size. We also found the results of GMM-MI to be robust to the choice of hyperparameters: changing the values of the likelihood threshold, MC samples, bootstrap realizations or regularization scale by one order of magnitude, or doubling the number of folds and initializations, did not significantly change the results obtained with GMM-MI.

We further validate GMM-MI by testing that it is  unbiased, and that the estimated MI variance scales as $N^{-1}$, when the number of available samples $N$ increases. We additionally show that GMM-MI satisfies the MI property of invariance under invertible non-linear transformations \cite{Kraskov04}. We consider a bivariate Gaussian distribution with $\rho=0.6$, and three different functions applied to one marginal variable $Y$: $f(y) = y$ (identity), $f(y) = y+0.5y^3$ (cubic) and $f(y) = \ln{(y+5.5)}$ (logarithmic). To deal with these datasets, we change the GMM-MI hyperparameters to $k=3$, $n_{\rm{init}} =5$, and $M=10^5$; however, we find no significant variations in the results even with different sets of hyperparameters. We repeat the estimation procedure of MI 500 times, drawing $N$ samples with a different seed every time, and considering $N=200$, $N=2\,000$ and $N=20\,000$. For each estimate, we calculate the bias, i.e.\ the difference between the estimated value of MI and the ground truth.

\begin{figure*}[t!]
\subfloat[$f(y) = y$.]{%
  \includegraphics[width=0.34\columnwidth]{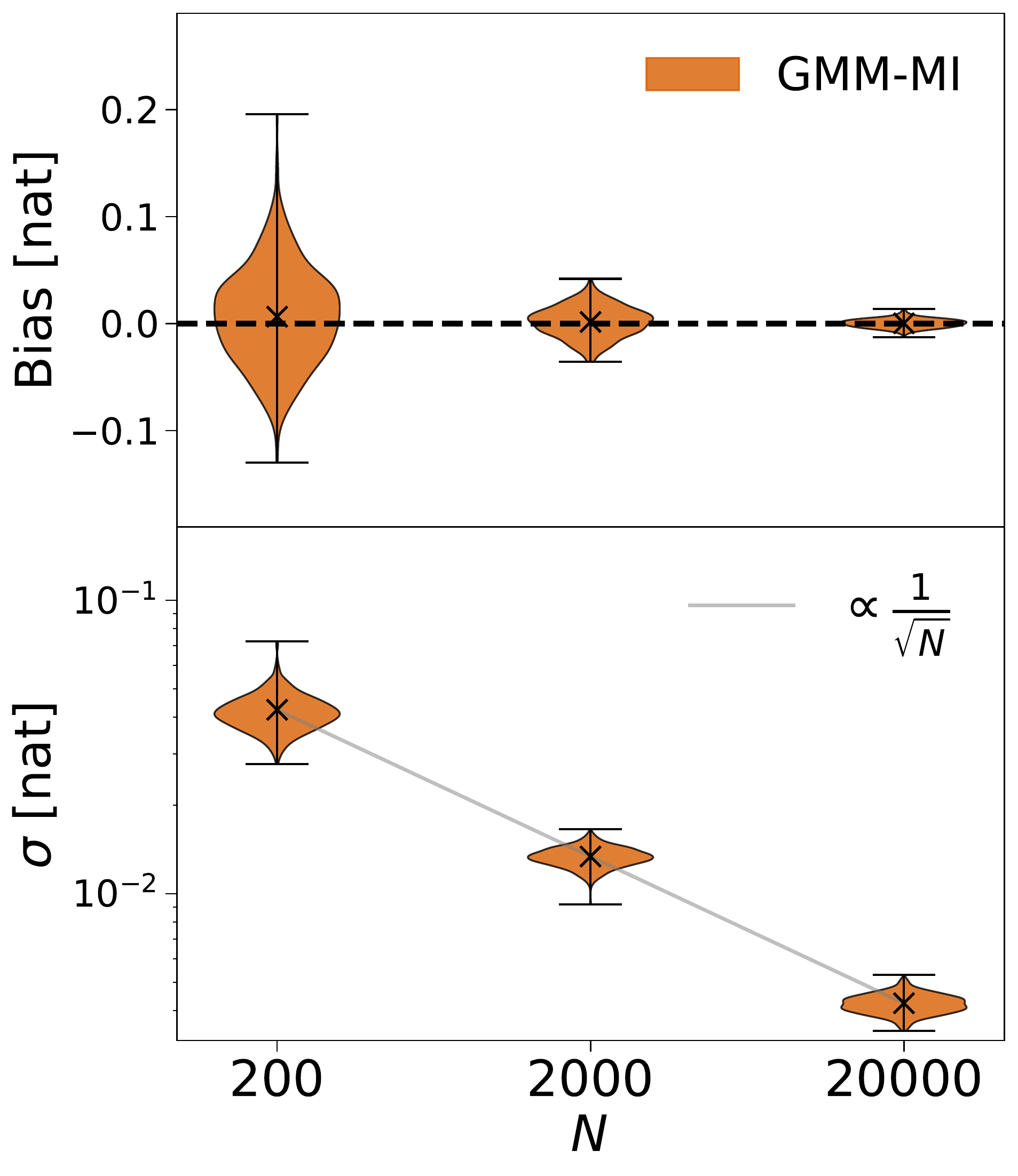}%
}
\subfloat[$f(y) = y + 0.5y^3$.]{%
  \includegraphics[width=0.34\columnwidth]{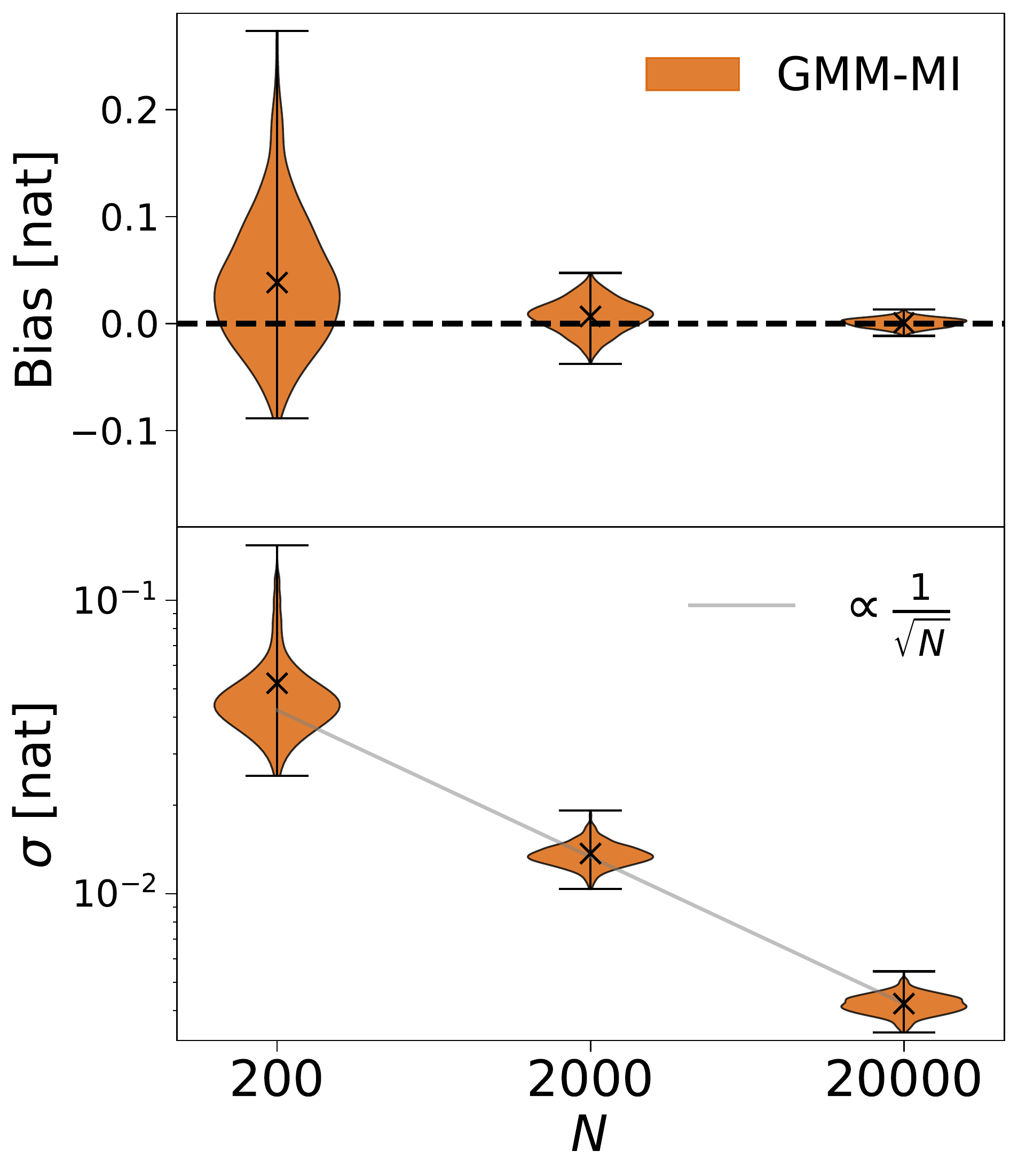}%
}
\subfloat[$f(y) = \ln{(y + 5.5)}$.]{%
  \includegraphics[width=0.34\columnwidth]{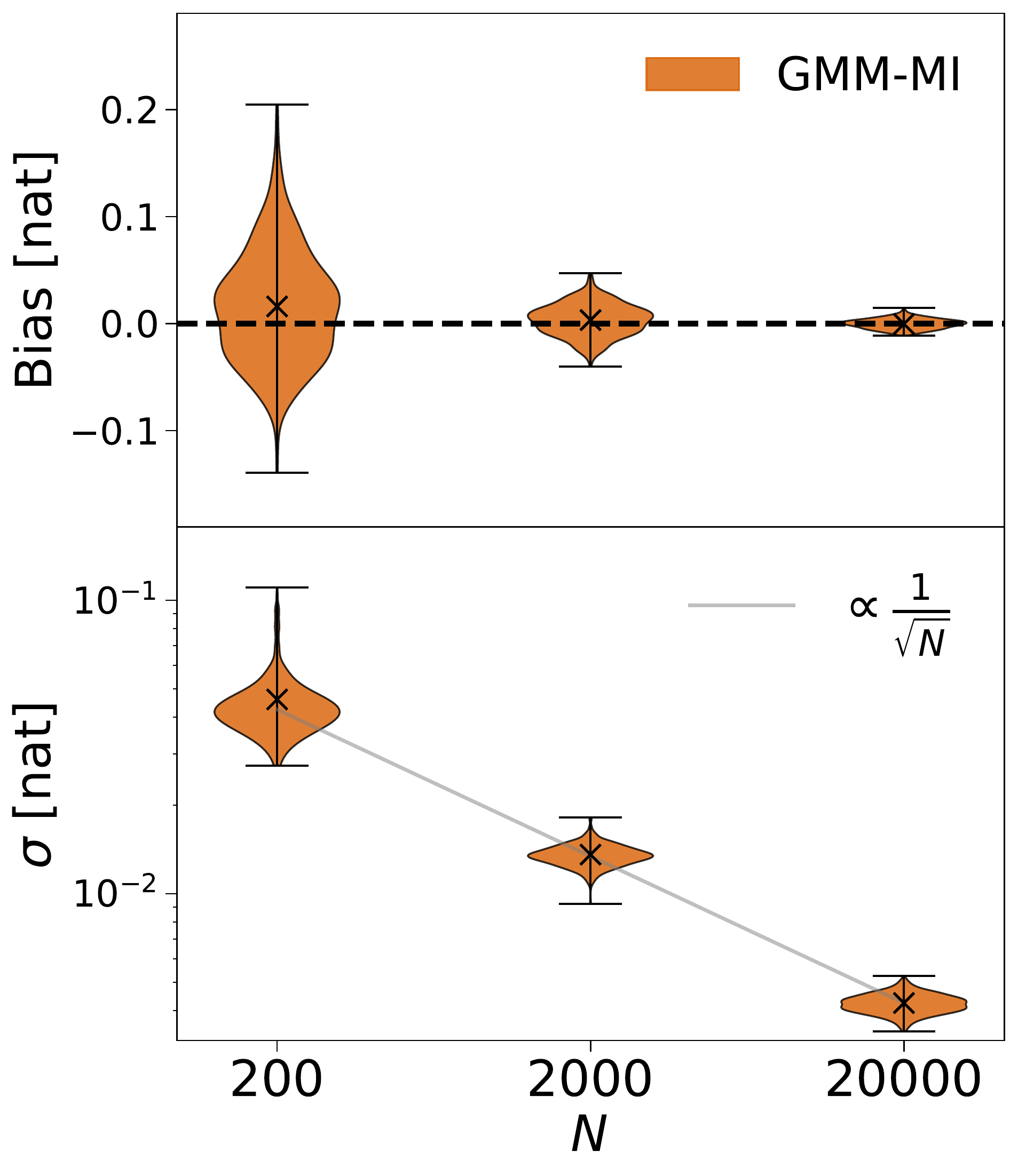}%
}
\caption{\textit{Left panel}: violin plots of the bias and standard deviation returned by GMM-MI, when applied to samples from a bivariate Gaussian distribution with $\rho=0.6$, with varying number of samples $N$. The bias is defined as the difference between the MI estimate returned by GMM-MI and the ground truth MI. The black crosses represent the mean of each distribution, while the gray line in the bottom plots indicates the expected power law $\propto \frac{1}{\sqrt{N}}$ passing through the estimated mean MI value at $N=20\,000$. \textit{Middle panel}: same as in the left panel, but all marginal $y$ samples are mapped through a cubic function, to demonstrate the invariance of MI under invertible transformations. \textit{Right panel}: same as in the middle panel, but with a logarithmic function.}
\label{fig:invariance_compare}
\end{figure*}

We report violin plots of the bias and of the MI standard deviation as returned by GMM-MI across the 500 trials in Fig.~\ref{fig:invariance_compare}. The mean bias, indicated as a black cross, converges to 0 as $N$ grows, and it is always well below the typical value of the standard deviation, thus demonstrating that GMM-MI is unbiased. This is true even when considering the cubic and the logarithmic transformations, further confirming that GMM-MI correctly captures the invariance property of MI. Moreover, in all cases the standard deviations returned by GMM-MI follow a power law $\propto \frac{1}{\sqrt{N}}$ as expected, represented as a gray line in the bottom plots. Remarkably, we found that even with very low numbers of samples ($N=50$), GMM-MI returns MI values consistent with the ground truth, even when applying the non-linear transformations considered in this section.

\begin{figure}[ht!]
        \centering
        \includegraphics[trim={0 0.1cm 0 0.1cm},clip, width=\columnwidth]{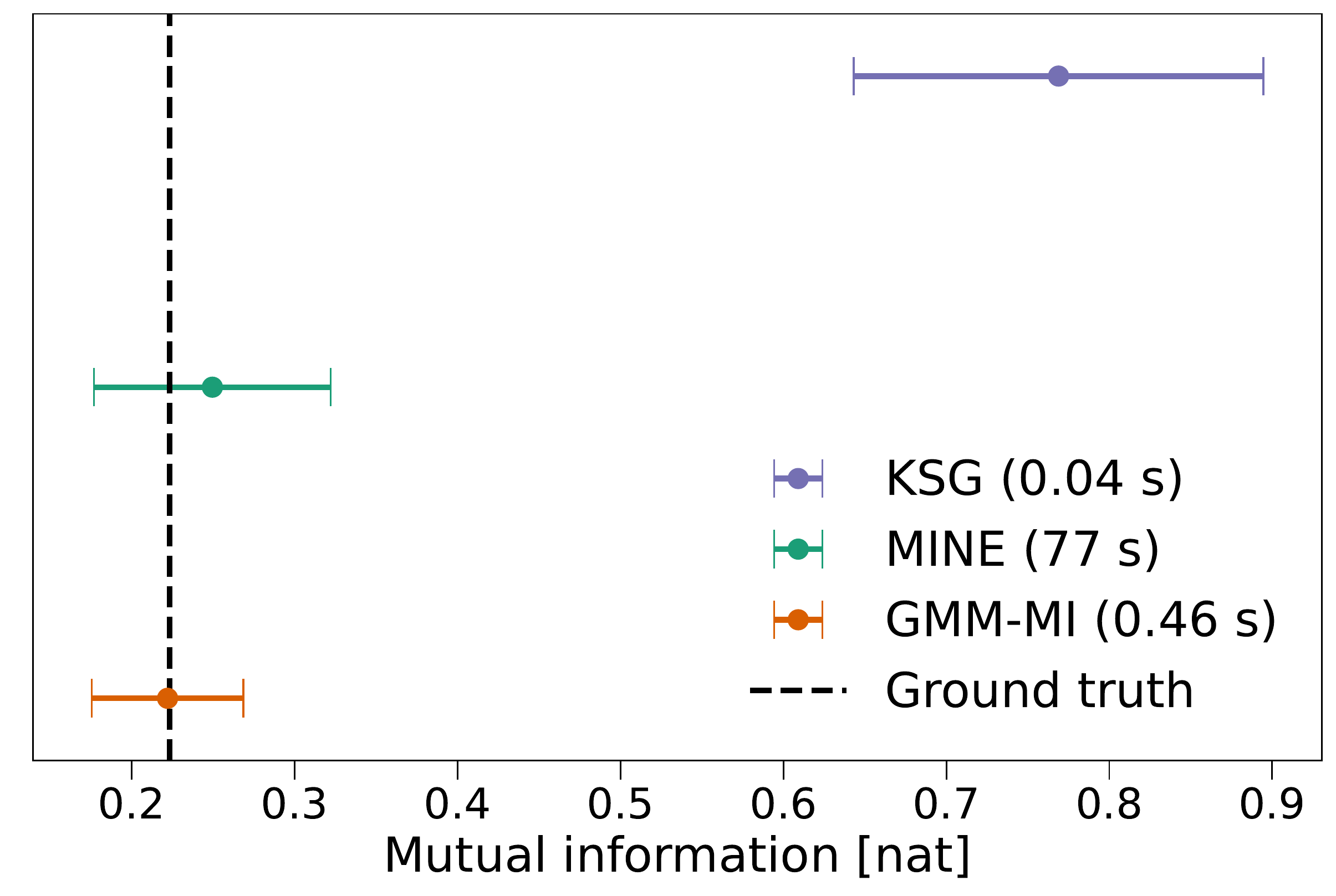}
        \caption{Mutual information distributions (means $\pm$ one standard deviation) when bootstrapping $N=200$ samples from a bivariate Gaussian distribution using three different estimators. As reported in \citet{Holmes19}, the KSG estimator (purple line) returns a biased estimate of MI. On the other hand, the MINE estimator (green line) and our algorithm based on Gaussian mixture models (red line) all agree with the ground truth (dashed black line). However, MINE takes two orders of magnitude more time than our estimator, and returns a higher-variance estimate since it includes the variability due to the neural network initialization.}
        \label{fig:bs_compare}
\end{figure}

\subsection{A note on bootstrap}
As reported in \citet{Holmes19}, using bootstrap to associate an error bar to MI estimates can lead to catastrophic failures: duplicate points can be interpreted as fine-scale features, introducing spurious extra MI. In this section, we address this concern and empirically show that, despite including a bootstrap step, our procedure does not lead to biased estimates of MI. 

We consider the same experiment described in \citet{Holmes19}, where a single data set of $N=200$ bivariate Gaussian samples with $\rho=0.6$ is bootstrapped 20 times. We apply the KSG (with 3 neighbors, following \citet{Holmes19}) and MINE estimators to each bootstrapped realization, and compare it against our estimator with $n_{\mathrm{b}} = 20$. The results are reported in Fig.~\ref{fig:bs_compare}. The KSG estimator returns a mean MI biased by a factor of 4, while both MINE and our procedure return an accurate estimate. However, MINE is two orders of magnitude more computationally demanding, and returns an error bar which is larger than with our procedure, since it tends to overestimate the variance, as discussed in Sect.~\ref{sec:validation}. 

\section{Results}
\label{sec:results}
In this section, we apply our estimator to interpret the latent space of representation-learning models trained on three different datasets, ranging from synthetic images to cosmological simulations. We use our MI estimator to quantify the level of disentanglement of latent variables, and link them to relevant physical parameters. In the following experiments, we consider $k=3$ folds, $n_{\rm{init}} =5$ different initializations, a log-likelihood threshold on each individual fit of $10^{-5}$, $n_{\rm{b}} = 100$ bootstrap realizations, $M=10^5$ MC samples, and a regularization scale of $\omega=10^{-15}$; as in the experiments described in Sect.~\ref{sec:validation}, we found GMM-MI to be robust to the hyperparameter choices. Obtaining the full distribution of MI with our algorithm typically takes $\mathcal{O}(10)$ s on the datasets we analyze, using the same hardware described in Sect~\ref{sec:validation}.

\subsection{3D Shapes}
\label{sec:3D_shapes}

We consider the 3D Shapes dataset \cite{Kim18, Burgess18}, which consists of images of various shapes that were generated by the following factors: shape (4 values), scale (8 values), orientation (15 values), floor color (10 values), wall color (10 values), and object color (10 values). Each combination of factors is included in the dataset exactly once, for a total of 480\,000 images. We train a $\beta$-VAE, as described in Sect.~\ref{sec:bvae}, on this dataset, using a 6-dimensional latent space and setting the value of $\beta$ using cross-validation.

\begin{figure}[t!]
        \centering
        \includegraphics[width=\columnwidth]{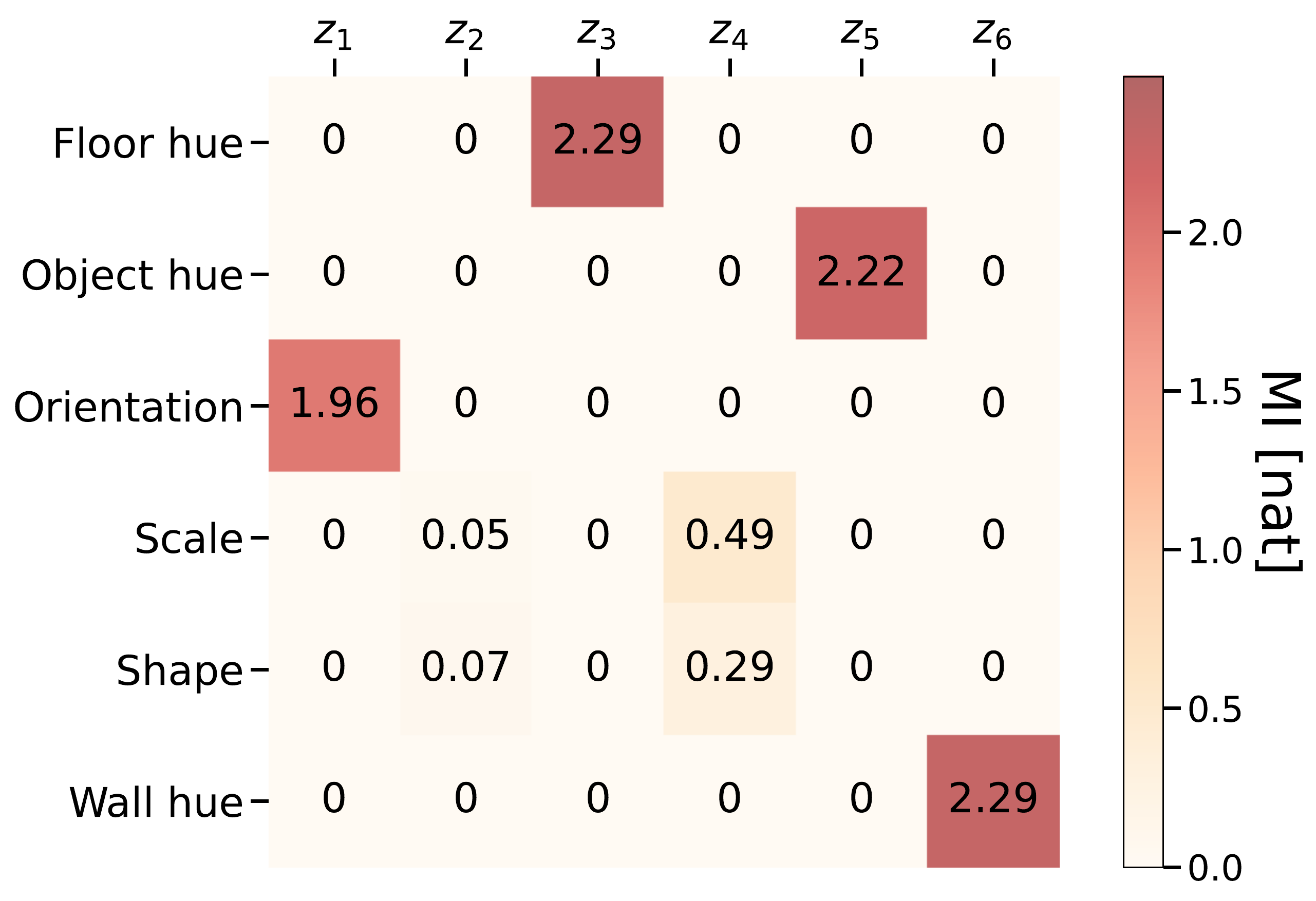}
        \caption{Mutual information (MI) between each ground truth factor and each latent variable of the $\beta$-VAE model trained on the 3D Shapes dataset. The values are obtained using GMM-MI, and the associated error estimates (not shown) are typically $\mathcal{O}(10^{-3})$--$\mathcal{O}(10^{-4})$ nat. All zeros indicate values of MI below 0.01 nat. Each latent is dependent upon only a single factor, except for $z_2$ and $z_4$, which appear entangled with scale and shape, as also found in \citet{Kim18}.}
        \label{fig:MI_s}
\end{figure}

After training, we encode 10\% of the data, which were not used for training or validation, and sample one point from each latent distribution. To interpret what each latent variable $z_i$ has learned about each generative factor of variation $f_j$, we measure the mutual information $I(z_i, f_j)$ using Eq.~(\ref{eq:MI_cat}). In Fig.~\ref{fig:MI_s} we report the MI values for all latents and factors using GMM-MI: except for scale and shape, each latent variable carries information about a single factor of variation. The difficulty in disentangling scale and shape was also reported in \citet{Kim18}. To assess the level of dependence between latent variables, we calculate $I(z_i, z_j)$: these values are below $10^{-4}$ nat for all pairs, except for the one carrying information about both scale and shape, i.e.\ $I(z_2, z_4) = 0.04 \pm 0.01$ nat.

\subsection{Dark matter halo density profiles}
\label{sec:dm_haloes}

\begin{figure}
        \centering
        \includegraphics[width=\columnwidth]{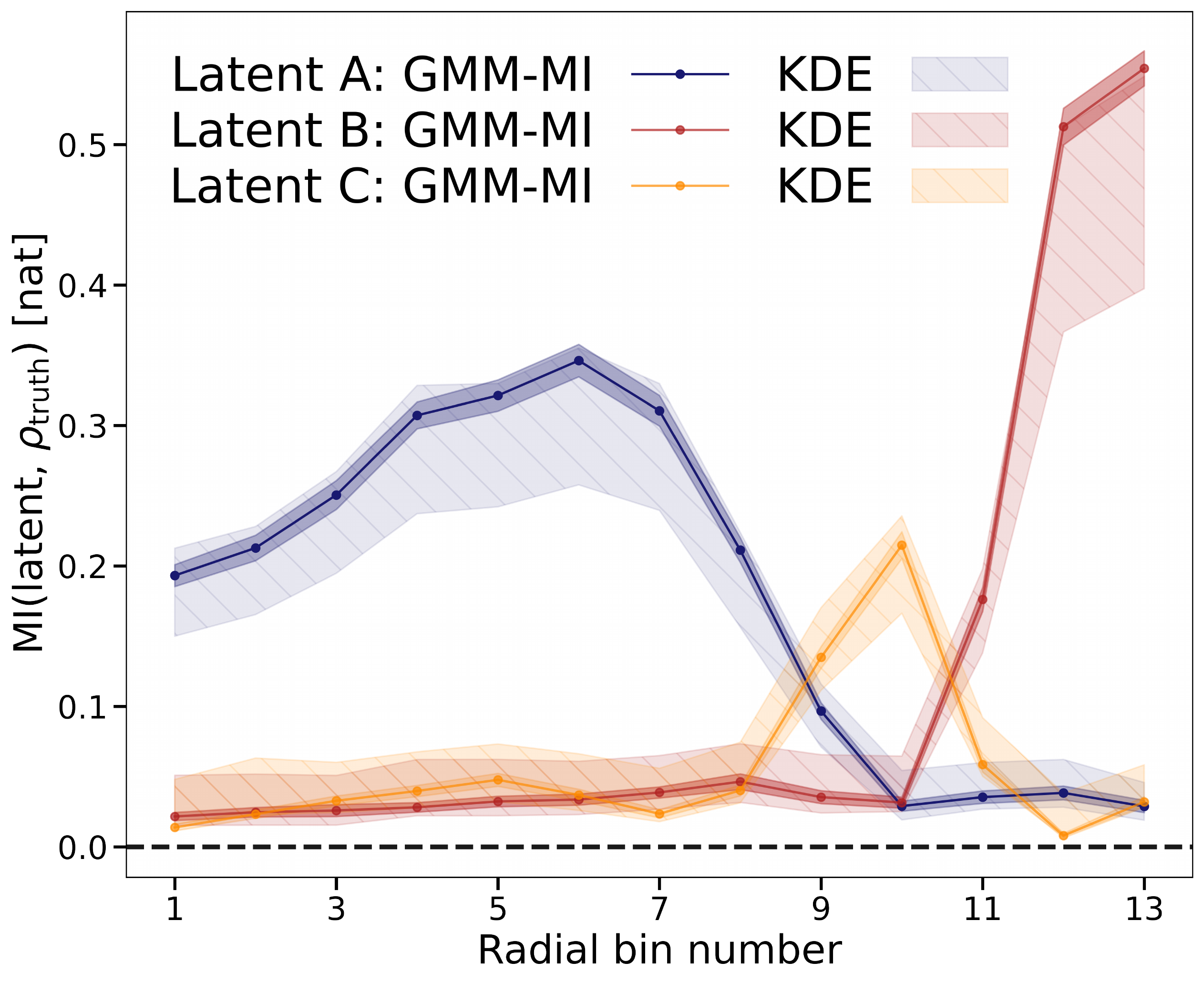}   
        \caption{Mutual information (MI) between each latent variable and dark matter halo density $\rho_{\rm{truth}}$ in each radial bin for the $\mbox{IVE}_{\rm{infall}}$ model \cite{LucieSmith22}. The points with darker error bars correspond to the sample mean and standard deviation obtained with our algorithm (GMM-MI). The striped areas indicate the kernel density estimate (KDE) values shown in \citet{LucieSmith22}, with bandwidths of 0.3 (lower limit) and 0.1 (upper limit). There is good agreement between the two approaches; in particular, the GMM-MI estimates overlap with the KDE estimates at lower (higher) bandwidth when the MI estimates are higher (lower), due to the different KDE bandwidths sometimes underfitting and sometimes overfitting the data. GMM-MI is designed to avoid such cases, as we explain in Sect.~\ref{sec:dm_haloes}.
        }
        \label{fig:MI_trend}
\end{figure}

In the standard model of cosmology, only 5\% of our Universe consists of baryonic matter, while the remainder consists of dark matter (25\%) and dark energy (70\%) \cite{Dodelson03}. In particular, dark matter only interacts via the gravitational force, and gathers into stable large-scale structures, called `halos', where galaxy formation typically occurs. Given the highly non-linear physical processes taking place during the formation of such structures, a common tool to analyze dark matter halos are cosmological $N$-body simulations, where particles representing the dark matter are evolved in a box under the influence of gravity \cite{Navarro96, Tormen97, Jenkins98}.

Dark matter halos forming within such simulations exhibit a universal spherically-averaged density profile as a function of their radius \cite{Navarro97, Huss99, Wang09}; this universality encompasses a huge range of halo masses and persists within different cosmological models. While the universality of the density profile is still not fully understood, \citet[][LS22 hereafter]{LucieSmith22} showed that it is possible to train a deep representation learning model to compress raw dark matter halo data into a compact disentangled representation that contains all the information needed to predict dark matter density profiles. Following LS22, we consider 4332 dark matter halos from a single $N$-body simulation, and encode them using their $\mbox{IVE}_{\rm{infall}}$ model with 3 latent variables. The latent representation is used to predict the dark matter halo density profile in 13 different radial bins.

We calculate the MI between the ground-truth halo density in each radial bin and each latent variable, aiming to reproduce the middle panel of fig. 4 in LS22, where further details can be found. We show the trend of MI for all radial bins and latent variables in Fig.~\ref{fig:MI_trend}.
We compare the estimates from GMM-MI with those obtained using kernel density estimation (KDE) with different bandwidths, as done in LS22. A major difference between the two approaches is that our bands indicate the error coming from the limited sample size, while their bands represent the sensitivity of the KDE to different bandwidths. The results are in good agreement: in particular, GMM-MI returns estimates closer to the KDE approach with smaller bandwidth when MI is high; in this case, the higher KDE bandwidth value underfits the data. On the other hand, for lower values of MI, GMM-MI yields estimates consistent with the KDE ones at higher bandwidth, since the lower bandwidth overfits the data. This confirms that GMM-MI avoids underfitting and overfitting of the data by design. We also checked that the latent variables of the $\mbox{IVE}_{\rm{infall}}$ model are independent: as in LS22, the MI between each pair of latents is $\mathcal{O}(10^{-2})$ nat.

\subsection{Stellar spectra}
\label{sec:stellar_spectra}

We consider the model presented in \citet[][S21 hereafter]{Sedaghat21}, where a $\beta$-VAE is trained on about $7\,000$ real unique spectra with a 128-dimensional latent space. These spectra were collected by the High-Accuracy Radial-velocity Planet Searcher instrument (HARPS, \cite{Pepe02, Mayor03}) in the spectral range 378--691 nm, and include mainly stellar spectra, even though Jupiter and asteroid contaminants are present in the dataset. All details about the data, the preprocessing steps and the training procedure can be found in S21. 

To select the most informative latent variables, the median absolute deviation (MAD) is calculated for each of them; the rest of the analysis is carried out on the six most informative latents only. We calculate MI between each of these six variables and six known physical factors, all treated as continuous variables. These are the star radial velocity, its effective temperature $T_{\rm{eff}}$, its mass, its metallicity $\left[\mbox{M}/ \mbox{H}\right]$, the atmospheric air mass and the signal-to-noise ratio (SNR). 

\begin{figure}

\subfloat[Our results, using GMM-MI.]{%
  \includegraphics[width=0.5\columnwidth]{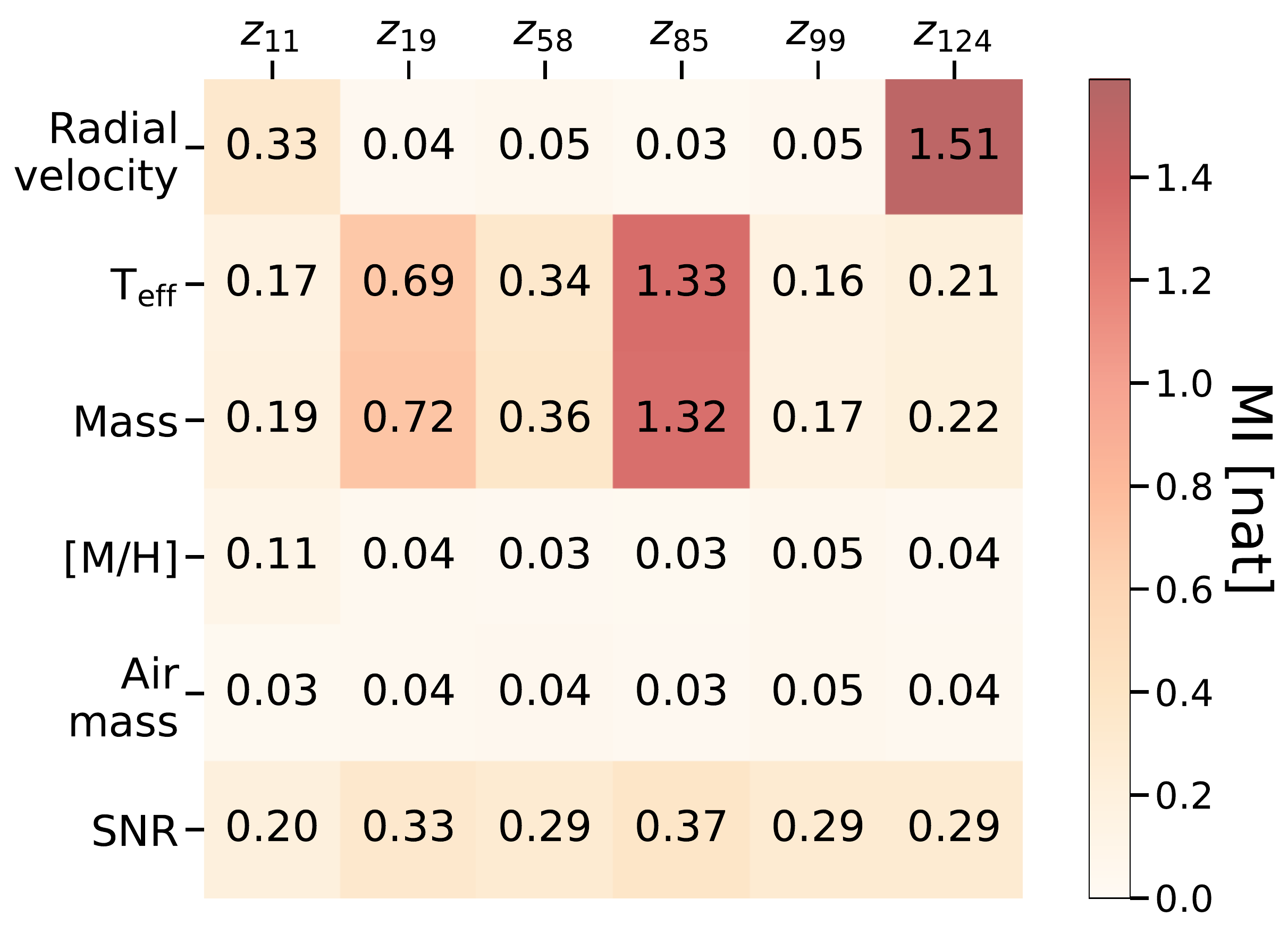}%
}
\subfloat[Results from \citet{Sedaghat21}, using histograms.]{%
  \includegraphics[width=0.5\columnwidth]{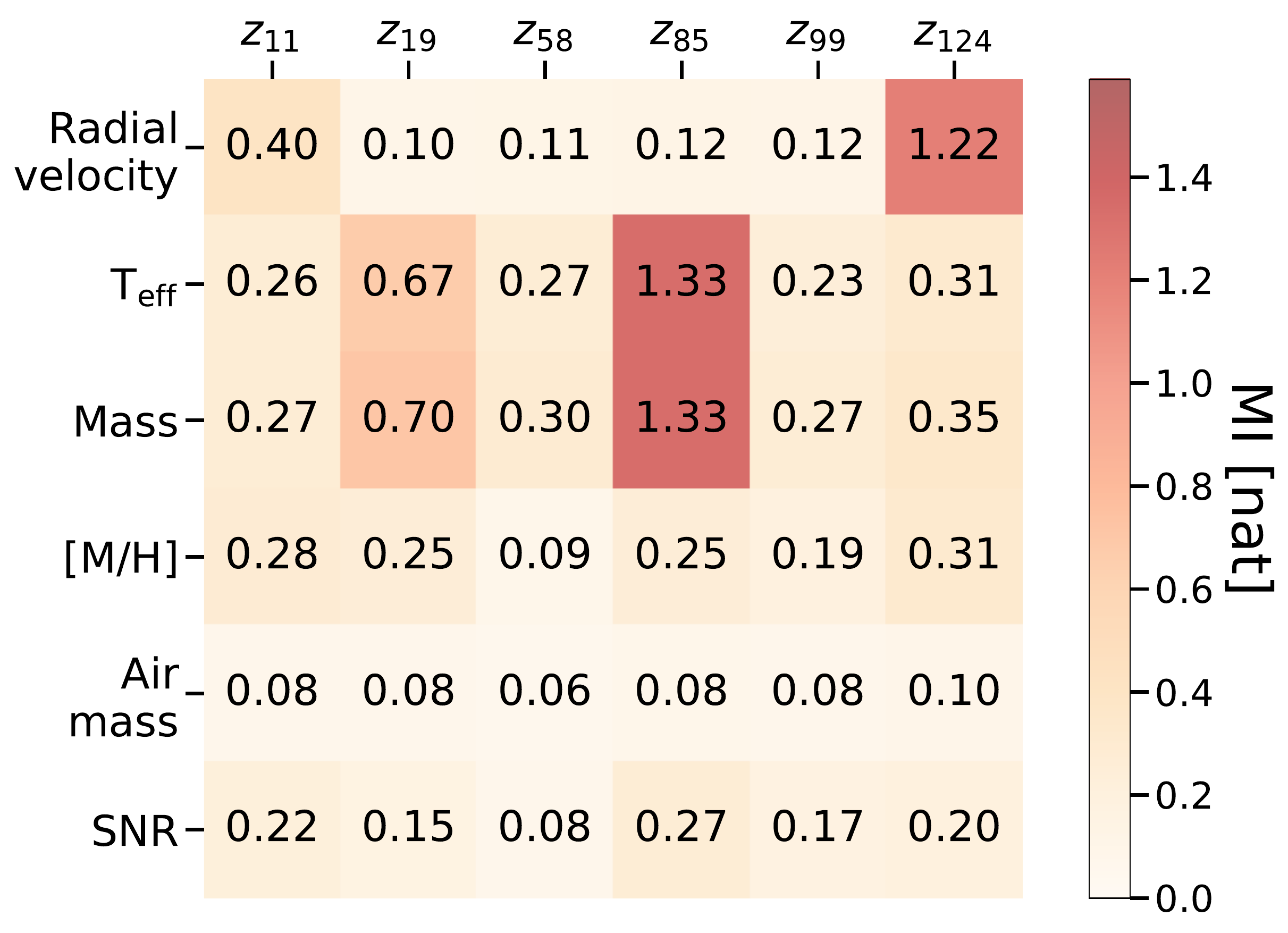}%
}

\caption{\textit{Left panel}: Mutual information (MI) between the six most-informative latent variables (based on the median absolute deviation) and six astrophysical parameters related to the stellar spectra described in Sect.~\ref{sec:stellar_spectra}, calculated using our algorithm GMM-MI; the corresponding error bars (not shown) are typically $\mathcal{O}(10^{-2})$ nat. \textit{Right panel}: Same as the top panel, using the MI estimator of \citet{Sedaghat21} based on histograms. The results agree overall: the 124$^{\rm{th}}$ variable is highly correlated with the radial velocity, while the effective temperature $T_{\rm{eff}}$ and mass seem mostly correlated with the 85$^{\rm{th}}$ latent. The metallicity $\left[\mbox{M}/ \mbox{H}\right]$, the atmospheric air mass and the signal-to-noise ratio (SNR) do not display strong dependence on any of the selected latent variables.}\label{fig:MI_sed}
\end{figure}

The MI estimates obtained with GMM-MI are shown in the top panel of Fig.~\ref{fig:MI_sed}: the 124$^{\rm{th}}$ latent variable shows high dependence on the radial velocity, while the 85$^{\rm{th}}$ latent appears entangled with both the effective temperature and the mass. The other physical parameters do not show a dependence on a particular latent amongst the ones with the highest MAD, even though in S21 a more complete analysis exploring latent traversals and investigating subsets of data is presented. The bottom panel of the same figure shows the results obtained with the procedure outlined in S21, which uses histograms with a certain number of bins (40 in this case) as density estimators. The trend agrees with our results, even though the particularly high number of bins chosen might overfit the data and overestimate MI (compare e.g.\ the $\left[\mbox{M}/ \mbox{H}\right]$ MI estimates), analogously to the KDE results in Fig.~\ref{fig:MI_trend}. On the other hand, our algorithm provides a robust way to select the hyperparameters, thus avoiding underfitting or overfitting the samples.

\section{Conclusions}
\label{sec:conclusion}
We presented GMM-MI (pronounced ``Jimmie''), an efficient and robust algorithm to estimate the mutual information (MI) between two random variables given samples of their joint distribution. Our algorithm uses Gaussian mixture models (GMMs) to fit the available samples, and returns the full distribution of MI through bootstrapping, thus including the uncertainty on MI due to the finite sample size. GMM-MI is demonstrably accurate, and benefits from the flexibility and computational efficiency of GMMs. Moreover, it can be applied to both discrete and continuous settings, and is robust to the choice of hyperparameters.

We extensively validated GMM-MI on toy datasets for which the ground truth MI is known, showing equal or better performance with respect to established estimators like KSG \cite{Kraskov04} and MINE \cite{Belghazi18}; we also tested that GMM-MI respects MI invariance under invertible transformations, is unbiased and returns MI errors that scale as expected with sample size. We demonstrated the application of our estimator to interpret the latent space of three different deep representation-learning models trained on synthetic shape images, large-scale structure in cosmological simulations and real spectra of stars. We calculated both the MI between latent variables and physical factors, and the MI between the latent variables themselves, to investigate their degree of disentanglement, reproducing MI estimates obtained with various techniques, including histograms and kernel density estimators. These results further validate the accuracy of GMM-MI and confirm the power of MI for gaining interpretability of deep learning models. 

We plan to extend our work by improving the density estimation with more flexible tools such as normalizing flows (NFs, \cite{Dinh14, Rezende15}), which can be seamlessly integrated into neural network-based settings and can benefit from graphics processing unit (GPU) acceleration. Moreover, combining NFs with a differentiable numerical integrator would make our estimator amenable to backpropagation, thus allowing its use in the context of MI optimization. We will explore this avenue in future work.

\section*{Data availability statement}
GMM-MI is publicly available in this GitHub repository (https://github.com/dpiras/GMM-MI, also accessible by clicking the icon \href{https://github.com/dpiras/GMM-MI}{\faicon{github}}), together with all data and results from the paper.

\begin{acknowledgments}
We thank Nima Sedaghat, Martino Romaniello and Vojtech Cvrcek for sharing the stellar spectra model and data. We are also grateful to Justin Alsing for useful discussions about initialization procedures for GMM fitting. DP was supported by the UCL Provost’s Strategic Development Fund, and by a Swiss National Science Foundation (SNSF) Professorship grant (No.~202671). The work of HVP was supported by the G\"{o}ran Gustafsson Foundation for Research in Natural Sciences and Medicine and the European Research Council (ERC) under the European Union’s Horizon 2020 research and innovation programme (grant agreement no.~101018897 CosmicExplorer). HVP and LLS acknowledge the hospitality of the Aspen Center for Physics, which is supported by National Science Foundation grant PHY-1607611. The participation of HVP and LLS at the Aspen Center for Physics was supported by the Simons Foundation. This study was supported by the European Union’s Horizon 2020 research and innovation programme under grant agreement No. 818085 GMGalaxies. AP is additionally supported by the Royal Society. NG was funded by the UCL Graduate Research Scholarship (GRS) and UCL Overseas Research Scholarship (ORS). This manuscript has been authored by Fermi Research Alliance, LLC under Contract No. DE-AC02-07CH11359 with the U.S. Department of Energy, Office of Science, Office of High Energy Physics. This work used computing equipment funded by the Research Capital Investment Fund (RCIF) provided by UKRI, and partially funded by the UCL Cosmoparticle Initiative. This work used facilities provided by the UCL Cosmoparticle Initiative.
\end{acknowledgments}

\section*{Author contributions}
\textbf{D.P.}: formal analysis; investigation; validation; software; writing -- original draft preparation, review \& editing; visualization.
\textbf{H.V.P.}: conceptualization; methodology; validation; writing -- review \& editing; funding acquisition.
\textbf{A.P.}: conceptualization; methodology; validation; writing -- review \& editing; funding acquisition.
\textbf{L.L.-S}: methodology; validation; resources; writing -- review \& editing.
\textbf{N.G.}: software; validation; writing -- review \& editing.
\textbf{B.N.}: writing -- review \& editing.

\bibliography{bibliography}

\providecommand{\noopsort}[1]{}\providecommand{\singleletter}[1]{#1}%
\begin{thebibliography}{115}%
\makeatletter
\providecommand \@ifxundefined [1]{%
 \@ifx{#1\undefined}
}%
\providecommand \@ifnum [1]{%
 \ifnum #1\expandafter \@firstoftwo
 \else \expandafter \@secondoftwo
 \fi
}%
\providecommand \@ifx [1]{%
 \ifx #1\expandafter \@firstoftwo
 \else \expandafter \@secondoftwo
 \fi
}%
\providecommand \natexlab [1]{#1}%
\providecommand \enquote  [1]{``#1''}%
\providecommand \bibnamefont  [1]{#1}%
\providecommand \bibfnamefont [1]{#1}%
\providecommand \citenamefont [1]{#1}%
\providecommand \href@noop [0]{\@secondoftwo}%
\providecommand \href [0]{\begingroup \@sanitize@url \@href}%
\providecommand \@href[1]{\@@startlink{#1}\@@href}%
\providecommand \@@href[1]{\endgroup#1\@@endlink}%
\providecommand \@sanitize@url [0]{\catcode `\\12\catcode `\$12\catcode
  `\&12\catcode `\#12\catcode `\^12\catcode `\_12\catcode `\%12\relax}%
\providecommand \@@startlink[1]{}%
\providecommand \@@endlink[0]{}%
\providecommand \url  [0]{\begingroup\@sanitize@url \@url }%
\providecommand \@url [1]{\endgroup\@href {#1}{\urlprefix }}%
\providecommand \urlprefix  [0]{URL }%
\providecommand \Eprint [0]{\href }%
\providecommand \doibase [0]{http://dx.doi.org/}%
\providecommand \selectlanguage [0]{\@gobble}%
\providecommand \bibinfo  [0]{\@secondoftwo}%
\providecommand \bibfield  [0]{\@secondoftwo}%
\providecommand \translation [1]{[#1]}%
\providecommand \BibitemOpen [0]{}%
\providecommand \bibitemStop [0]{}%
\providecommand \bibitemNoStop [0]{.\EOS\space}%
\providecommand \EOS [0]{\spacefactor3000\relax}%
\providecommand \BibitemShut  [1]{\csname bibitem#1\endcsname}%
\let\auto@bib@innerbib\@empty
\bibitem [{\citenamefont {{Raghu}}\ and\ \citenamefont
  {{Schmidt}}(2020)}]{Raghu20}%
  \BibitemOpen
  \bibfield  {author} {\bibinfo {author} {\bibfnamefont {M.}~\bibnamefont
  {{Raghu}}}\ and\ \bibinfo {author} {\bibfnamefont {E.}~\bibnamefont
  {{Schmidt}}},\ }\href@noop {} {\bibfield  {journal} {\bibinfo  {journal}
  {arXiv e-prints}\ ,\ \bibinfo {eid} {arXiv:2003.11755}} (\bibinfo {year}
  {2020})},\ \Eprint {http://arxiv.org/abs/2003.11755} {arXiv:2003.11755
  [cs.LG]} \BibitemShut {NoStop}%
\bibitem [{\citenamefont {Cybenko}(1989)}]{Cybenko89}%
  \BibitemOpen
  \bibfield  {author} {\bibinfo {author} {\bibfnamefont {G.}~\bibnamefont
  {Cybenko}},\ }\href {\doibase 10.1007/BF02551274} {\bibfield  {journal}
  {\bibinfo  {journal} {Mathematics of Control, Signals, and Systems (MCSS)}\
  }\textbf {\bibinfo {volume} {2}},\ \bibinfo {pages} {303} (\bibinfo {year}
  {1989})}\BibitemShut {NoStop}%
\bibitem [{\citenamefont {Hornik}\ \emph {et~al.}(1989)\citenamefont {Hornik},
  \citenamefont {Stinchcombe},\ and\ \citenamefont {White}}]{Hornik89}%
  \BibitemOpen
  \bibfield  {author} {\bibinfo {author} {\bibfnamefont {K.}~\bibnamefont
  {Hornik}}, \bibinfo {author} {\bibfnamefont {M.}~\bibnamefont {Stinchcombe}},
  \ and\ \bibinfo {author} {\bibfnamefont {H.}~\bibnamefont {White}},\ }\href
  {\doibase https://doi.org/10.1016/0893-6080(89)90020-8} {\bibfield  {journal}
  {\bibinfo  {journal} {Neural Networks}\ }\textbf {\bibinfo {volume} {2}},\
  \bibinfo {pages} {359} (\bibinfo {year} {1989})}\BibitemShut {NoStop}%
\bibitem [{\citenamefont {Hornik}(1991)}]{Hornik91}%
  \BibitemOpen
  \bibfield  {author} {\bibinfo {author} {\bibfnamefont {K.}~\bibnamefont
  {Hornik}},\ }\href {\doibase https://doi.org/10.1016/0893-6080(91)90009-T}
  {\bibfield  {journal} {\bibinfo  {journal} {Neural Networks}\ }\textbf
  {\bibinfo {volume} {4}},\ \bibinfo {pages} {251} (\bibinfo {year}
  {1991})}\BibitemShut {NoStop}%
\bibitem [{\citenamefont {Molnar}(2022)}]{Molnar22}%
  \BibitemOpen
  \bibfield  {author} {\bibinfo {author} {\bibfnamefont {C.}~\bibnamefont
  {Molnar}},\ }\href {https://christophm.github.io/interpretable-ml-book}
  {\emph {\bibinfo {title} {Interpretable Machine Learning}}},\ \bibinfo
  {edition} {2nd}\ ed.\ (\bibinfo  {publisher} {Leanpub},\ \bibinfo {year}
  {2022})\BibitemShut {NoStop}%
\bibitem [{\citenamefont {Zeiler}\ and\ \citenamefont
  {Fergus}(2014)}]{Zeiler14}%
  \BibitemOpen
  \bibfield  {author} {\bibinfo {author} {\bibfnamefont {M.~D.}\ \bibnamefont
  {Zeiler}}\ and\ \bibinfo {author} {\bibfnamefont {R.}~\bibnamefont
  {Fergus}},\ }in\ \href@noop {} {\emph {\bibinfo {booktitle} {Computer Vision
  -- ECCV 2014}}},\ \bibinfo {editor} {edited by\ \bibinfo {editor}
  {\bibfnamefont {D.}~\bibnamefont {Fleet}}, \bibinfo {editor} {\bibfnamefont
  {T.}~\bibnamefont {Pajdla}}, \bibinfo {editor} {\bibfnamefont
  {B.}~\bibnamefont {Schiele}}, \ and\ \bibinfo {editor} {\bibfnamefont
  {T.}~\bibnamefont {Tuytelaars}}}\ (\bibinfo  {publisher} {Springer
  International Publishing},\ \bibinfo {address} {Cham},\ \bibinfo {year}
  {2014})\ pp.\ \bibinfo {pages} {818--833}\BibitemShut {NoStop}%
\bibitem [{\citenamefont {Simonyan}\ \emph {et~al.}(2014)\citenamefont
  {Simonyan}, \citenamefont {Vedaldi},\ and\ \citenamefont
  {Zisserman}}]{Simonyan14}%
  \BibitemOpen
  \bibfield  {author} {\bibinfo {author} {\bibfnamefont {K.}~\bibnamefont
  {Simonyan}}, \bibinfo {author} {\bibfnamefont {A.}~\bibnamefont {Vedaldi}}, \
  and\ \bibinfo {author} {\bibfnamefont {A.}~\bibnamefont {Zisserman}},\ }in\
  \href@noop {} {\emph {\bibinfo {booktitle} {In Workshop at International
  Conference on Learning Representations}}}\ (\bibinfo {year}
  {2014})\BibitemShut {NoStop}%
\bibitem [{\citenamefont {Zhou}\ \emph {et~al.}(2016)\citenamefont {Zhou},
  \citenamefont {Khosla}, \citenamefont {Lapedriza}, \citenamefont {Oliva},\
  and\ \citenamefont {Torralba}}]{Zhou16}%
  \BibitemOpen
  \bibfield  {author} {\bibinfo {author} {\bibfnamefont {B.}~\bibnamefont
  {Zhou}}, \bibinfo {author} {\bibfnamefont {A.}~\bibnamefont {Khosla}},
  \bibinfo {author} {\bibfnamefont {A.}~\bibnamefont {Lapedriza}}, \bibinfo
  {author} {\bibfnamefont {A.}~\bibnamefont {Oliva}}, \ and\ \bibinfo {author}
  {\bibfnamefont {A.}~\bibnamefont {Torralba}},\ }in\ \href {\doibase
  10.1109/CVPR.2016.319} {\emph {\bibinfo {booktitle} {2016 IEEE Conference on
  Computer Vision and Pattern Recognition (CVPR)}}}\ (\bibinfo  {publisher}
  {IEEE Computer Society},\ \bibinfo {address} {Los Alamitos, CA, USA},\
  \bibinfo {year} {2016})\ pp.\ \bibinfo {pages} {2921--2929}\BibitemShut
  {NoStop}%
\bibitem [{\citenamefont {Ribeiro}\ \emph {et~al.}(2016)\citenamefont
  {Ribeiro}, \citenamefont {Singh},\ and\ \citenamefont
  {Guestrin}}]{Ribeiro16}%
  \BibitemOpen
  \bibfield  {author} {\bibinfo {author} {\bibfnamefont {M.~T.}\ \bibnamefont
  {Ribeiro}}, \bibinfo {author} {\bibfnamefont {S.}~\bibnamefont {Singh}}, \
  and\ \bibinfo {author} {\bibfnamefont {C.}~\bibnamefont {Guestrin}},\ }in\
  \href {\doibase 10.1145/2939672.2939778} {\emph {\bibinfo {booktitle}
  {Proceedings of the 22nd ACM SIGKDD International Conference on Knowledge
  Discovery and Data Mining}}},\ \bibinfo {series and number} {KDD '16}\
  (\bibinfo  {publisher} {Association for Computing Machinery},\ \bibinfo
  {address} {New York, NY, USA},\ \bibinfo {year} {2016})\ p.\ \bibinfo {pages}
  {1135–1144}\BibitemShut {NoStop}%
\bibitem [{\citenamefont {Selvaraju}\ \emph {et~al.}(2017)\citenamefont
  {Selvaraju}, \citenamefont {Cogswell}, \citenamefont {Das}, \citenamefont
  {Vedantam}, \citenamefont {Parikh},\ and\ \citenamefont
  {Batra}}]{Selvaraju17}%
  \BibitemOpen
  \bibfield  {author} {\bibinfo {author} {\bibfnamefont {R.~R.}\ \bibnamefont
  {Selvaraju}}, \bibinfo {author} {\bibfnamefont {M.}~\bibnamefont {Cogswell}},
  \bibinfo {author} {\bibfnamefont {A.}~\bibnamefont {Das}}, \bibinfo {author}
  {\bibfnamefont {R.}~\bibnamefont {Vedantam}}, \bibinfo {author}
  {\bibfnamefont {D.}~\bibnamefont {Parikh}}, \ and\ \bibinfo {author}
  {\bibfnamefont {D.}~\bibnamefont {Batra}},\ }in\ \href {\doibase
  10.1109/ICCV.2017.74} {\emph {\bibinfo {booktitle} {2017 IEEE International
  Conference on Computer Vision (ICCV)}}}\ (\bibinfo {year} {2017})\ pp.\
  \bibinfo {pages} {618--626}\BibitemShut {NoStop}%
\bibitem [{\citenamefont {Shrikumar}\ \emph {et~al.}(2017)\citenamefont
  {Shrikumar}, \citenamefont {Greenside},\ and\ \citenamefont
  {Kundaje}}]{Shrikumar17}%
  \BibitemOpen
  \bibfield  {author} {\bibinfo {author} {\bibfnamefont {A.}~\bibnamefont
  {Shrikumar}}, \bibinfo {author} {\bibfnamefont {P.}~\bibnamefont
  {Greenside}}, \ and\ \bibinfo {author} {\bibfnamefont {A.}~\bibnamefont
  {Kundaje}},\ }in\ \href@noop {} {\emph {\bibinfo {booktitle} {Proceedings of
  the 34th International Conference on Machine Learning - Volume 70}}},\
  \bibinfo {series and number} {ICML'17}\ (\bibinfo  {publisher} {JMLR.org},\
  \bibinfo {year} {2017})\ p.\ \bibinfo {pages} {3145–3153}\BibitemShut
  {NoStop}%
\bibitem [{\citenamefont {Lundberg}\ and\ \citenamefont
  {Lee}(2017)}]{Lundberg17}%
  \BibitemOpen
  \bibfield  {author} {\bibinfo {author} {\bibfnamefont {S.~M.}\ \bibnamefont
  {Lundberg}}\ and\ \bibinfo {author} {\bibfnamefont {S.-I.}\ \bibnamefont
  {Lee}},\ }in\ \href@noop {} {\emph {\bibinfo {booktitle} {Proceedings of the
  31st International Conference on Neural Information Processing Systems}}},\
  \bibinfo {series and number} {NIPS'17}\ (\bibinfo  {publisher} {Curran
  Associates Inc.},\ \bibinfo {address} {Red Hook, NY, USA},\ \bibinfo {year}
  {2017})\ p.\ \bibinfo {pages} {4768–4777}\BibitemShut {NoStop}%
\bibitem [{\citenamefont {Chattopadhay}\ \emph {et~al.}(2018)\citenamefont
  {Chattopadhay}, \citenamefont {Sarkar}, \citenamefont {Howlader},\ and\
  \citenamefont {Balasubramanian}}]{Chattopadhay18}%
  \BibitemOpen
  \bibfield  {author} {\bibinfo {author} {\bibfnamefont {A.}~\bibnamefont
  {Chattopadhay}}, \bibinfo {author} {\bibfnamefont {A.}~\bibnamefont
  {Sarkar}}, \bibinfo {author} {\bibfnamefont {P.}~\bibnamefont {Howlader}}, \
  and\ \bibinfo {author} {\bibfnamefont {V.~N.}\ \bibnamefont
  {Balasubramanian}},\ }in\ \href {\doibase 10.1109/WACV.2018.00097} {\emph
  {\bibinfo {booktitle} {2018 IEEE Winter Conference on Applications of
  Computer Vision (WACV)}}}\ (\bibinfo {year} {2018})\ pp.\ \bibinfo {pages}
  {839--847}\BibitemShut {NoStop}%
\bibitem [{\citenamefont {{Li}}\ \emph {et~al.}(2021)\citenamefont {{Li}},
  \citenamefont {{Xiong}}, \citenamefont {{Li}}, \citenamefont {{Wu}},
  \citenamefont {{Zhang}}, \citenamefont {{Liu}}, \citenamefont {{Bian}},\ and\
  \citenamefont {{Dou}}}]{Li21}%
  \BibitemOpen
  \bibfield  {author} {\bibinfo {author} {\bibfnamefont {X.}~\bibnamefont
  {{Li}}}, \bibinfo {author} {\bibfnamefont {H.}~\bibnamefont {{Xiong}}},
  \bibinfo {author} {\bibfnamefont {X.}~\bibnamefont {{Li}}}, \bibinfo {author}
  {\bibfnamefont {X.}~\bibnamefont {{Wu}}}, \bibinfo {author} {\bibfnamefont
  {X.}~\bibnamefont {{Zhang}}}, \bibinfo {author} {\bibfnamefont
  {J.}~\bibnamefont {{Liu}}}, \bibinfo {author} {\bibfnamefont
  {J.}~\bibnamefont {{Bian}}}, \ and\ \bibinfo {author} {\bibfnamefont
  {D.}~\bibnamefont {{Dou}}},\ }\href@noop {} {\bibfield  {journal} {\bibinfo
  {journal} {arXiv e-prints}\ ,\ \bibinfo {eid} {arXiv:2103.10689}} (\bibinfo
  {year} {2021})},\ \Eprint {http://arxiv.org/abs/2103.10689} {arXiv:2103.10689
  [cs.LG]} \BibitemShut {NoStop}%
\bibitem [{\citenamefont {Linardatos}\ \emph {et~al.}(2021)\citenamefont
  {Linardatos}, \citenamefont {Papastefanopoulos},\ and\ \citenamefont
  {Kotsiantis}}]{Linardatos21}%
  \BibitemOpen
  \bibfield  {author} {\bibinfo {author} {\bibfnamefont {P.}~\bibnamefont
  {Linardatos}}, \bibinfo {author} {\bibfnamefont {V.}~\bibnamefont
  {Papastefanopoulos}}, \ and\ \bibinfo {author} {\bibfnamefont
  {S.}~\bibnamefont {Kotsiantis}},\ }\href {\doibase 10.3390/e23010018}
  {\bibfield  {journal} {\bibinfo  {journal} {Entropy}\ }\textbf {\bibinfo
  {volume} {23}} (\bibinfo {year} {2021}),\ 10.3390/e23010018}\BibitemShut
  {NoStop}%
\bibitem [{\citenamefont {Schmidhuber}(1992)}]{Schmidhuber92}%
  \BibitemOpen
  \bibfield  {author} {\bibinfo {author} {\bibfnamefont {J.}~\bibnamefont
  {Schmidhuber}},\ }\href {\doibase 10.1162/neco.1992.4.6.863} {\bibfield
  {journal} {\bibinfo  {journal} {Neural Computation}\ }\textbf {\bibinfo
  {volume} {4}},\ \bibinfo {pages} {863} (\bibinfo {year} {1992})},\ \Eprint
  {http://arxiv.org/abs/https://direct.mit.edu/neco/article-pdf/4/6/863/812421/neco.1992.4.6.863.pdf}
  {https://direct.mit.edu/neco/article-pdf/4/6/863/812421/neco.1992.4.6.863.pdf}
  \BibitemShut {NoStop}%
\bibitem [{\citenamefont {Bengio}\ \emph {et~al.}(2013)\citenamefont {Bengio},
  \citenamefont {Courville},\ and\ \citenamefont {Vincent}}]{Bengio13}%
  \BibitemOpen
  \bibfield  {author} {\bibinfo {author} {\bibfnamefont {Y.}~\bibnamefont
  {Bengio}}, \bibinfo {author} {\bibfnamefont {A.}~\bibnamefont {Courville}}, \
  and\ \bibinfo {author} {\bibfnamefont {P.}~\bibnamefont {Vincent}},\ }\href
  {\doibase 10.1109/TPAMI.2013.50} {\bibfield  {journal} {\bibinfo  {journal}
  {IEEE transactions on pattern analysis and machine intelligence}\ }\textbf
  {\bibinfo {volume} {35}},\ \bibinfo {pages} {1798} (\bibinfo {year}
  {2013})}\BibitemShut {NoStop}%
\bibitem [{\citenamefont {Louizos}\ \emph {et~al.}(2016)\citenamefont
  {Louizos}, \citenamefont {Swersky}, \citenamefont {Li}, \citenamefont
  {Welling},\ and\ \citenamefont {Zemel}}]{Louizos15}%
  \BibitemOpen
  \bibfield  {author} {\bibinfo {author} {\bibfnamefont {C.}~\bibnamefont
  {Louizos}}, \bibinfo {author} {\bibfnamefont {K.}~\bibnamefont {Swersky}},
  \bibinfo {author} {\bibfnamefont {Y.}~\bibnamefont {Li}}, \bibinfo {author}
  {\bibfnamefont {M.}~\bibnamefont {Welling}}, \ and\ \bibinfo {author}
  {\bibfnamefont {R.~S.}\ \bibnamefont {Zemel}},\ }in\ \href
  {http://arxiv.org/abs/1511.00830} {\emph {\bibinfo {booktitle} {4th
  International Conference on Learning Representations, {ICLR} 2016, San Juan,
  Puerto Rico, May 2-4, 2016, Conference Track Proceedings}}},\ \bibinfo
  {editor} {edited by\ \bibinfo {editor} {\bibfnamefont {Y.}~\bibnamefont
  {Bengio}}\ and\ \bibinfo {editor} {\bibfnamefont {Y.}~\bibnamefont {LeCun}}}\
  (\bibinfo {year} {2016})\BibitemShut {NoStop}%
\bibitem [{\citenamefont {Chen}\ \emph {et~al.}(2016)\citenamefont {Chen},
  \citenamefont {Duan}, \citenamefont {Houthooft}, \citenamefont {Schulman},
  \citenamefont {Sutskever},\ and\ \citenamefont {Abbeel}}]{Chen16}%
  \BibitemOpen
  \bibfield  {author} {\bibinfo {author} {\bibfnamefont {X.}~\bibnamefont
  {Chen}}, \bibinfo {author} {\bibfnamefont {Y.}~\bibnamefont {Duan}}, \bibinfo
  {author} {\bibfnamefont {R.}~\bibnamefont {Houthooft}}, \bibinfo {author}
  {\bibfnamefont {J.}~\bibnamefont {Schulman}}, \bibinfo {author}
  {\bibfnamefont {I.}~\bibnamefont {Sutskever}}, \ and\ \bibinfo {author}
  {\bibfnamefont {P.}~\bibnamefont {Abbeel}},\ }\href@noop {} {\bibfield
  {journal} {\bibinfo  {journal} {Advances in Neural Information Processing
  Systems}\ }\textbf {\bibinfo {volume} {29}} (\bibinfo {year}
  {2016})}\BibitemShut {NoStop}%
\bibitem [{\citenamefont {Lample}\ \emph {et~al.}(2017)\citenamefont {Lample},
  \citenamefont {Zeghidour}, \citenamefont {Usunier}, \citenamefont {Bordes},
  \citenamefont {Denoyer},\ and\ \citenamefont {Ranzato}}]{Lample17}%
  \BibitemOpen
  \bibfield  {author} {\bibinfo {author} {\bibfnamefont {G.}~\bibnamefont
  {Lample}}, \bibinfo {author} {\bibfnamefont {N.}~\bibnamefont {Zeghidour}},
  \bibinfo {author} {\bibfnamefont {N.}~\bibnamefont {Usunier}}, \bibinfo
  {author} {\bibfnamefont {A.}~\bibnamefont {Bordes}}, \bibinfo {author}
  {\bibfnamefont {L.}~\bibnamefont {Denoyer}}, \ and\ \bibinfo {author}
  {\bibfnamefont {M.}~\bibnamefont {Ranzato}},\ }in\ \href@noop {} {\emph
  {\bibinfo {booktitle} {Proceedings of the 31st International Conference on
  Neural Information Processing Systems}}},\ \bibinfo {series and number}
  {NIPS'17}\ (\bibinfo  {publisher} {Curran Associates Inc.},\ \bibinfo
  {address} {Red Hook, NY, USA},\ \bibinfo {year} {2017})\ p.\ \bibinfo {pages}
  {5969–5978}\BibitemShut {NoStop}%
\bibitem [{\citenamefont {Higgins}\ \emph {et~al.}(2017)\citenamefont
  {Higgins}, \citenamefont {Matthey}, \citenamefont {Pal}, \citenamefont
  {Burgess}, \citenamefont {Glorot}, \citenamefont {Botvinick}, \citenamefont
  {Mohamed},\ and\ \citenamefont {Lerchner}}]{Higgins17}%
  \BibitemOpen
  \bibfield  {author} {\bibinfo {author} {\bibfnamefont {I.}~\bibnamefont
  {Higgins}}, \bibinfo {author} {\bibfnamefont {L.}~\bibnamefont {Matthey}},
  \bibinfo {author} {\bibfnamefont {A.}~\bibnamefont {Pal}}, \bibinfo {author}
  {\bibfnamefont {C.~P.}\ \bibnamefont {Burgess}}, \bibinfo {author}
  {\bibfnamefont {X.}~\bibnamefont {Glorot}}, \bibinfo {author} {\bibfnamefont
  {M.~M.}\ \bibnamefont {Botvinick}}, \bibinfo {author} {\bibfnamefont
  {S.}~\bibnamefont {Mohamed}}, \ and\ \bibinfo {author} {\bibfnamefont
  {A.}~\bibnamefont {Lerchner}},\ }in\ \href@noop {} {\emph {\bibinfo
  {booktitle} {ICLR}}}\ (\bibinfo {year} {2017})\BibitemShut {NoStop}%
\bibitem [{\citenamefont {Jha}\ \emph {et~al.}(2018)\citenamefont {Jha},
  \citenamefont {Anand}, \citenamefont {Singh},\ and\ \citenamefont
  {Veeravasarapu}}]{Jha18}%
  \BibitemOpen
  \bibfield  {author} {\bibinfo {author} {\bibfnamefont {A.~H.}\ \bibnamefont
  {Jha}}, \bibinfo {author} {\bibfnamefont {S.}~\bibnamefont {Anand}}, \bibinfo
  {author} {\bibfnamefont {M.}~\bibnamefont {Singh}}, \ and\ \bibinfo {author}
  {\bibfnamefont {V.}~\bibnamefont {Veeravasarapu}},\ }in\ \href@noop {} {\emph
  {\bibinfo {booktitle} {Computer Vision -- ECCV 2018}}},\ \bibinfo {editor}
  {edited by\ \bibinfo {editor} {\bibfnamefont {V.}~\bibnamefont {Ferrari}},
  \bibinfo {editor} {\bibfnamefont {M.}~\bibnamefont {Hebert}}, \bibinfo
  {editor} {\bibfnamefont {C.}~\bibnamefont {Sminchisescu}}, \ and\ \bibinfo
  {editor} {\bibfnamefont {Y.}~\bibnamefont {Weiss}}}\ (\bibinfo  {publisher}
  {Springer International Publishing},\ \bibinfo {address} {Cham},\ \bibinfo
  {year} {2018})\ pp.\ \bibinfo {pages} {829--845}\BibitemShut {NoStop}%
\bibitem [{\citenamefont {Locatello}\ \emph {et~al.}(2019)\citenamefont
  {Locatello}, \citenamefont {Bauer}, \citenamefont {Lucic}, \citenamefont
  {Raetsch}, \citenamefont {Gelly}, \citenamefont {Sch{\"o}lkopf},\ and\
  \citenamefont {Bachem}}]{Locatello19}%
  \BibitemOpen
  \bibfield  {author} {\bibinfo {author} {\bibfnamefont {F.}~\bibnamefont
  {Locatello}}, \bibinfo {author} {\bibfnamefont {S.}~\bibnamefont {Bauer}},
  \bibinfo {author} {\bibfnamefont {M.}~\bibnamefont {Lucic}}, \bibinfo
  {author} {\bibfnamefont {G.}~\bibnamefont {Raetsch}}, \bibinfo {author}
  {\bibfnamefont {S.}~\bibnamefont {Gelly}}, \bibinfo {author} {\bibfnamefont
  {B.}~\bibnamefont {Sch{\"o}lkopf}}, \ and\ \bibinfo {author} {\bibfnamefont
  {O.}~\bibnamefont {Bachem}},\ }in\ \href
  {https://proceedings.mlr.press/v97/locatello19a.html} {\emph {\bibinfo
  {booktitle} {Proceedings of the 36th International Conference on Machine
  Learning}}},\ \bibinfo {series} {Proceedings of Machine Learning Research},
  Vol.~\bibinfo {volume} {97},\ \bibinfo {editor} {edited by\ \bibinfo {editor}
  {\bibfnamefont {K.}~\bibnamefont {Chaudhuri}}\ and\ \bibinfo {editor}
  {\bibfnamefont {R.}~\bibnamefont {Salakhutdinov}}}\ (\bibinfo  {publisher}
  {PMLR},\ \bibinfo {year} {2019})\ pp.\ \bibinfo {pages}
  {4114--4124}\BibitemShut {NoStop}%
\bibitem [{\citenamefont {Lezama}(2019)}]{Lezama19}%
  \BibitemOpen
  \bibfield  {author} {\bibinfo {author} {\bibfnamefont {J.}~\bibnamefont
  {Lezama}},\ }in\ \href {https://openreview.net/forum?id=Hkg4W2AcFm} {\emph
  {\bibinfo {booktitle} {7th International Conference on Learning
  Representations, {ICLR} 2019, New Orleans, LA, USA, May 6-9, 2019}}}\
  (\bibinfo  {publisher} {OpenReview.net},\ \bibinfo {year} {2019})\BibitemShut
  {NoStop}%
\bibitem [{\citenamefont {{Pandey}}\ and\ \citenamefont
  {{Sarkar}}(2017)}]{Pandey17}%
  \BibitemOpen
  \bibfield  {author} {\bibinfo {author} {\bibfnamefont {B.}~\bibnamefont
  {{Pandey}}}\ and\ \bibinfo {author} {\bibfnamefont {S.}~\bibnamefont
  {{Sarkar}}},\ }\href {\doibase 10.1093/mnrasl/slw250} {\bibfield  {journal}
  {\bibinfo  {journal} {Monthly Notices of the Royal Astronomical Society}\
  }\textbf {\bibinfo {volume} {467}},\ \bibinfo {pages} {L6} (\bibinfo {year}
  {2017})},\ \Eprint {http://arxiv.org/abs/1611.00283} {arXiv:1611.00283
  [astro-ph.CO]} \BibitemShut {NoStop}%
\bibitem [{\citenamefont {{Sarkar}}\ and\ \citenamefont
  {{Pandey}}(2020)}]{Sarkar20}%
  \BibitemOpen
  \bibfield  {author} {\bibinfo {author} {\bibfnamefont {S.}~\bibnamefont
  {{Sarkar}}}\ and\ \bibinfo {author} {\bibfnamefont {B.}~\bibnamefont
  {{Pandey}}},\ }\href {\doibase 10.1093/mnras/staa2236} {\bibfield  {journal}
  {\bibinfo  {journal} {Monthly Notices of the Royal Astronomical Society}\
  }\textbf {\bibinfo {volume} {497}},\ \bibinfo {pages} {4077} (\bibinfo {year}
  {2020})},\ \Eprint {http://arxiv.org/abs/2003.13974} {arXiv:2003.13974
  [astro-ph.GA]} \BibitemShut {NoStop}%
\bibitem [{\citenamefont {{Bhattacharjee}}\ \emph {et~al.}(2020)\citenamefont
  {{Bhattacharjee}}, \citenamefont {{Pandey}},\ and\ \citenamefont
  {{Sarkar}}}]{Bhattacharjee20}%
  \BibitemOpen
  \bibfield  {author} {\bibinfo {author} {\bibfnamefont {S.}~\bibnamefont
  {{Bhattacharjee}}}, \bibinfo {author} {\bibfnamefont {B.}~\bibnamefont
  {{Pandey}}}, \ and\ \bibinfo {author} {\bibfnamefont {S.}~\bibnamefont
  {{Sarkar}}},\ }\href {\doibase 10.1088/1475-7516/2020/09/039} {\bibfield
  {journal} {\bibinfo  {journal} {Journal of Cosmology and Astroparticle
  Physics}\ }\textbf {\bibinfo {volume} {2020}},\ \bibinfo {eid} {039}
  (\bibinfo {year} {2020})},\ \Eprint {http://arxiv.org/abs/2004.05016}
  {arXiv:2004.05016 [astro-ph.GA]} \BibitemShut {NoStop}%
\bibitem [{\citenamefont {{Upham}}\ \emph {et~al.}(2021)\citenamefont
  {{Upham}}, \citenamefont {{Brown}},\ and\ \citenamefont
  {{Whittaker}}}]{Upham21}%
  \BibitemOpen
  \bibfield  {author} {\bibinfo {author} {\bibfnamefont {R.~E.}\ \bibnamefont
  {{Upham}}}, \bibinfo {author} {\bibfnamefont {M.~L.}\ \bibnamefont
  {{Brown}}}, \ and\ \bibinfo {author} {\bibfnamefont {L.}~\bibnamefont
  {{Whittaker}}},\ }\href {\doibase 10.1093/mnras/stab522} {\bibfield
  {journal} {\bibinfo  {journal} {Monthly Notices of the Royal Astronomical
  Society}\ }\textbf {\bibinfo {volume} {503}},\ \bibinfo {pages} {1999}
  (\bibinfo {year} {2021})},\ \Eprint {http://arxiv.org/abs/2012.06267}
  {arXiv:2012.06267 [astro-ph.CO]} \BibitemShut {NoStop}%
\bibitem [{\citenamefont {{Malz}}\ \emph {et~al.}(2021)\citenamefont {{Malz}},
  \citenamefont {{Lanusse}}, \citenamefont {{Crenshaw}},\ and\ \citenamefont
  {{Graham}}}]{Malz21}%
  \BibitemOpen
  \bibfield  {author} {\bibinfo {author} {\bibfnamefont {A.~I.}\ \bibnamefont
  {{Malz}}}, \bibinfo {author} {\bibfnamefont {F.}~\bibnamefont {{Lanusse}}},
  \bibinfo {author} {\bibfnamefont {J.~F.}\ \bibnamefont {{Crenshaw}}}, \ and\
  \bibinfo {author} {\bibfnamefont {M.~L.}\ \bibnamefont {{Graham}}},\
  }\href@noop {} {\bibfield  {journal} {\bibinfo  {journal} {arXiv e-prints}\
  ,\ \bibinfo {eid} {arXiv:2104.08229}} (\bibinfo {year} {2021})},\ \Eprint
  {http://arxiv.org/abs/2104.08229} {arXiv:2104.08229 [astro-ph.IM]}
  \BibitemShut {NoStop}%
\bibitem [{\citenamefont {{Sarkar}}\ \emph {et~al.}(2021)\citenamefont
  {{Sarkar}}, \citenamefont {{Pandey}},\ and\ \citenamefont
  {{Bhattacharjee}}}]{Sarkar21}%
  \BibitemOpen
  \bibfield  {author} {\bibinfo {author} {\bibfnamefont {S.}~\bibnamefont
  {{Sarkar}}}, \bibinfo {author} {\bibfnamefont {B.}~\bibnamefont {{Pandey}}},
  \ and\ \bibinfo {author} {\bibfnamefont {S.}~\bibnamefont
  {{Bhattacharjee}}},\ }\href {\doibase 10.1093/mnras/staa3665} {\bibfield
  {journal} {\bibinfo  {journal} {Monthly Notices of the Royal Astronomical
  Society}\ }\textbf {\bibinfo {volume} {501}},\ \bibinfo {pages} {994}
  (\bibinfo {year} {2021})},\ \Eprint {http://arxiv.org/abs/2009.12797}
  {arXiv:2009.12797 [astro-ph.GA]} \BibitemShut {NoStop}%
\bibitem [{\citenamefont {{Jeffrey}}\ \emph {et~al.}(2021)\citenamefont
  {{Jeffrey}}, \citenamefont {{Alsing}},\ and\ \citenamefont
  {{Lanusse}}}]{Jeffrey21}%
  \BibitemOpen
  \bibfield  {author} {\bibinfo {author} {\bibfnamefont {N.}~\bibnamefont
  {{Jeffrey}}}, \bibinfo {author} {\bibfnamefont {J.}~\bibnamefont {{Alsing}}},
  \ and\ \bibinfo {author} {\bibfnamefont {F.}~\bibnamefont {{Lanusse}}},\
  }\href {\doibase 10.1093/mnras/staa3594} {\bibfield  {journal} {\bibinfo
  {journal} {Monthly Notices of the Royal Astronomical Society}\ }\textbf
  {\bibinfo {volume} {501}},\ \bibinfo {pages} {954} (\bibinfo {year}
  {2021})},\ \Eprint {http://arxiv.org/abs/2009.08459} {arXiv:2009.08459
  [astro-ph.CO]} \BibitemShut {NoStop}%
\bibitem [{\citenamefont {Lucie-Smith}\ \emph {et~al.}(2022)\citenamefont
  {Lucie-Smith}, \citenamefont {Peiris}, \citenamefont {Pontzen}, \citenamefont
  {Nord}, \citenamefont {Thiyagalingam},\ and\ \citenamefont
  {Piras}}]{LucieSmith22}%
  \BibitemOpen
  \bibfield  {author} {\bibinfo {author} {\bibfnamefont {L.}~\bibnamefont
  {Lucie-Smith}}, \bibinfo {author} {\bibfnamefont {H.~V.}\ \bibnamefont
  {Peiris}}, \bibinfo {author} {\bibfnamefont {A.}~\bibnamefont {Pontzen}},
  \bibinfo {author} {\bibfnamefont {B.}~\bibnamefont {Nord}}, \bibinfo {author}
  {\bibfnamefont {J.}~\bibnamefont {Thiyagalingam}}, \ and\ \bibinfo {author}
  {\bibfnamefont {D.}~\bibnamefont {Piras}},\ }\href {\doibase
  10.1103/PhysRevD.105.103533} {\bibfield  {journal} {\bibinfo  {journal}
  {Phys. Rev. D}\ }\textbf {\bibinfo {volume} {105}},\ \bibinfo {pages}
  {103533} (\bibinfo {year} {2022})}\BibitemShut {NoStop}%
\bibitem [{\citenamefont {{Sarkar}}\ \emph {et~al.}(2022)\citenamefont
  {{Sarkar}}, \citenamefont {{Pandey}},\ and\ \citenamefont
  {{Das}}}]{Sarkar22}%
  \BibitemOpen
  \bibfield  {author} {\bibinfo {author} {\bibfnamefont {S.}~\bibnamefont
  {{Sarkar}}}, \bibinfo {author} {\bibfnamefont {B.}~\bibnamefont {{Pandey}}},
  \ and\ \bibinfo {author} {\bibfnamefont {A.}~\bibnamefont {{Das}}},\ }\href
  {\doibase 10.1088/1475-7516/2022/03/024} {\bibfield  {journal} {\bibinfo
  {journal} {Journal of Cosmology and Astroparticle Physics}\ }\textbf
  {\bibinfo {volume} {2022}},\ \bibinfo {eid} {024} (\bibinfo {year} {2022})},\
  \Eprint {http://arxiv.org/abs/2111.11252} {arXiv:2111.11252 [astro-ph.GA]}
  \BibitemShut {NoStop}%
\bibitem [{\citenamefont {Fairhall}\ \emph {et~al.}(2012)\citenamefont
  {Fairhall}, \citenamefont {Shea-Brown},\ and\ \citenamefont
  {Barreiro}}]{Fairhall12}%
  \BibitemOpen
  \bibfield  {author} {\bibinfo {author} {\bibfnamefont {A.}~\bibnamefont
  {Fairhall}}, \bibinfo {author} {\bibfnamefont {E.}~\bibnamefont
  {Shea-Brown}}, \ and\ \bibinfo {author} {\bibfnamefont {A.}~\bibnamefont
  {Barreiro}},\ }\href {\doibase https://doi.org/10.1016/j.conb.2012.06.005}
  {\bibfield  {journal} {\bibinfo  {journal} {Current Opinion in Neurobiology}\
  }\textbf {\bibinfo {volume} {22}},\ \bibinfo {pages} {653} (\bibinfo {year}
  {2012})},\ \bibinfo {note} {microcircuits}\BibitemShut {NoStop}%
\bibitem [{\citenamefont {Charzyńska}\ and\ \citenamefont
  {Gambin}(2016)}]{Charzynska16}%
  \BibitemOpen
  \bibfield  {author} {\bibinfo {author} {\bibfnamefont {A.}~\bibnamefont
  {Charzyńska}}\ and\ \bibinfo {author} {\bibfnamefont {A.}~\bibnamefont
  {Gambin}},\ }\href {\doibase 10.3390/e18010013} {\bibfield  {journal}
  {\bibinfo  {journal} {Entropy}\ }\textbf {\bibinfo {volume} {18}} (\bibinfo
  {year} {2016}),\ 10.3390/e18010013}\BibitemShut {NoStop}%
\bibitem [{\citenamefont {Tka\v{c}ik}\ and\ \citenamefont
  {Bialek}(2016)}]{Tkacik16}%
  \BibitemOpen
  \bibfield  {author} {\bibinfo {author} {\bibfnamefont {G.}~\bibnamefont
  {Tka\v{c}ik}}\ and\ \bibinfo {author} {\bibfnamefont {W.}~\bibnamefont
  {Bialek}},\ }\href {\doibase 10.1146/annurev-conmatphys-031214-014803}
  {\bibfield  {journal} {\bibinfo  {journal} {Annual Review of Condensed Matter
  Physics}\ }\textbf {\bibinfo {volume} {7}},\ \bibinfo {pages} {89} (\bibinfo
  {year} {2016})},\ \Eprint
  {http://arxiv.org/abs/https://doi.org/10.1146/annurev-conmatphys-031214-014803}
  {https://doi.org/10.1146/annurev-conmatphys-031214-014803} \BibitemShut
  {NoStop}%
\bibitem [{\citenamefont {Levchenko}\ and\ \citenamefont
  {Nemenman}(2014)}]{Levchenko16}%
  \BibitemOpen
  \bibfield  {author} {\bibinfo {author} {\bibfnamefont {A.}~\bibnamefont
  {Levchenko}}\ and\ \bibinfo {author} {\bibfnamefont {I.}~\bibnamefont
  {Nemenman}},\ }\href {\doibase https://doi.org/10.1016/j.copbio.2014.05.002}
  {\bibfield  {journal} {\bibinfo  {journal} {Current Opinion in
  Biotechnology}\ }\textbf {\bibinfo {volume} {28}},\ \bibinfo {pages} {156}
  (\bibinfo {year} {2014})},\ \bibinfo {note} {nanobiotechnology • Systems
  biology}\BibitemShut {NoStop}%
\bibitem [{\citenamefont {von Wegner}\ \emph {et~al.}(2018)\citenamefont {von
  Wegner}, \citenamefont {Laufs},\ and\ \citenamefont
  {Tagliazucchi}}]{Wegner18}%
  \BibitemOpen
  \bibfield  {author} {\bibinfo {author} {\bibfnamefont {F.}~\bibnamefont {von
  Wegner}}, \bibinfo {author} {\bibfnamefont {H.}~\bibnamefont {Laufs}}, \ and\
  \bibinfo {author} {\bibfnamefont {E.}~\bibnamefont {Tagliazucchi}},\ }\href
  {\doibase 10.1103/PhysRevE.97.022415} {\bibfield  {journal} {\bibinfo
  {journal} {Phys. Rev. E}\ }\textbf {\bibinfo {volume} {97}},\ \bibinfo
  {pages} {022415} (\bibinfo {year} {2018})}\BibitemShut {NoStop}%
\bibitem [{\citenamefont {Holmes}\ and\ \citenamefont
  {Nemenman}(2019)}]{Holmes19}%
  \BibitemOpen
  \bibfield  {author} {\bibinfo {author} {\bibfnamefont {C.~M.}\ \bibnamefont
  {Holmes}}\ and\ \bibinfo {author} {\bibfnamefont {I.}~\bibnamefont
  {Nemenman}},\ }\href {\doibase 10.1103/physreve.100.022404} {\bibfield
  {journal} {\bibinfo  {journal} {Physical Review E}\ }\textbf {\bibinfo
  {volume} {100}} (\bibinfo {year} {2019}),\
  10.1103/physreve.100.022404}\BibitemShut {NoStop}%
\bibitem [{\citenamefont {Uda}(2020)}]{Uda20}%
  \BibitemOpen
  \bibfield  {author} {\bibinfo {author} {\bibfnamefont {S.}~\bibnamefont
  {Uda}},\ }\href {\doibase 10.1007/s12551-020-00665-w} {\bibfield  {journal}
  {\bibinfo  {journal} {Biophysical Reviews}\ }\textbf {\bibinfo {volume}
  {12}},\ \bibinfo {pages} {377} (\bibinfo {year} {2020})}\BibitemShut
  {NoStop}%
\bibitem [{\citenamefont {Wicks}\ \emph {et~al.}(2007)\citenamefont {Wicks},
  \citenamefont {Chapman},\ and\ \citenamefont {Dendy}}]{Wicks07}%
  \BibitemOpen
  \bibfield  {author} {\bibinfo {author} {\bibfnamefont {R.~T.}\ \bibnamefont
  {Wicks}}, \bibinfo {author} {\bibfnamefont {S.~C.}\ \bibnamefont {Chapman}},
  \ and\ \bibinfo {author} {\bibfnamefont {R.~O.}\ \bibnamefont {Dendy}},\
  }\href {\doibase 10.1103/PhysRevE.75.051125} {\bibfield  {journal} {\bibinfo
  {journal} {Phys. Rev. E}\ }\textbf {\bibinfo {volume} {75}},\ \bibinfo
  {pages} {051125} (\bibinfo {year} {2007})}\BibitemShut {NoStop}%
\bibitem [{\citenamefont {Dunleavy}\ \emph {et~al.}(2012)\citenamefont
  {Dunleavy}, \citenamefont {Wiesner},\ and\ \citenamefont
  {Royall}}]{Dunleavy12}%
  \BibitemOpen
  \bibfield  {author} {\bibinfo {author} {\bibfnamefont {A.~J.}\ \bibnamefont
  {Dunleavy}}, \bibinfo {author} {\bibfnamefont {K.}~\bibnamefont {Wiesner}}, \
  and\ \bibinfo {author} {\bibfnamefont {C.~P.}\ \bibnamefont {Royall}},\
  }\href {\doibase 10.1103/PhysRevE.86.041505} {\bibfield  {journal} {\bibinfo
  {journal} {Phys. Rev. E}\ }\textbf {\bibinfo {volume} {86}},\ \bibinfo
  {pages} {041505} (\bibinfo {year} {2012})}\BibitemShut {NoStop}%
\bibitem [{\citenamefont {Runge}(2015)}]{Runge15}%
  \BibitemOpen
  \bibfield  {author} {\bibinfo {author} {\bibfnamefont {J.}~\bibnamefont
  {Runge}},\ }\href {\doibase 10.1103/PhysRevE.92.062829} {\bibfield  {journal}
  {\bibinfo  {journal} {Phys. Rev. E}\ }\textbf {\bibinfo {volume} {92}},\
  \bibinfo {pages} {062829} (\bibinfo {year} {2015})}\BibitemShut {NoStop}%
\bibitem [{\citenamefont {Myers}\ \emph {et~al.}(2019)\citenamefont {Myers},
  \citenamefont {Munch},\ and\ \citenamefont {Khasawneh}}]{Myers19}%
  \BibitemOpen
  \bibfield  {author} {\bibinfo {author} {\bibfnamefont {A.}~\bibnamefont
  {Myers}}, \bibinfo {author} {\bibfnamefont {E.}~\bibnamefont {Munch}}, \ and\
  \bibinfo {author} {\bibfnamefont {F.~A.}\ \bibnamefont {Khasawneh}},\ }\href
  {\doibase 10.1103/PhysRevE.100.022314} {\bibfield  {journal} {\bibinfo
  {journal} {Phys. Rev. E}\ }\textbf {\bibinfo {volume} {100}},\ \bibinfo
  {pages} {022314} (\bibinfo {year} {2019})}\BibitemShut {NoStop}%
\bibitem [{\citenamefont {Svenkeson}\ and\ \citenamefont
  {West}(2019)}]{Svenkeson19}%
  \BibitemOpen
  \bibfield  {author} {\bibinfo {author} {\bibfnamefont {A.}~\bibnamefont
  {Svenkeson}}\ and\ \bibinfo {author} {\bibfnamefont {B.~J.}\ \bibnamefont
  {West}},\ }\href {\doibase 10.1103/PhysRevE.100.022119} {\bibfield  {journal}
  {\bibinfo  {journal} {Phys. Rev. E}\ }\textbf {\bibinfo {volume} {100}},\
  \bibinfo {pages} {022119} (\bibinfo {year} {2019})}\BibitemShut {NoStop}%
\bibitem [{\citenamefont {Diego}\ \emph {et~al.}(2019)\citenamefont {Diego},
  \citenamefont {Haaga},\ and\ \citenamefont {Hannisdal}}]{Diego19}%
  \BibitemOpen
  \bibfield  {author} {\bibinfo {author} {\bibfnamefont {D.}~\bibnamefont
  {Diego}}, \bibinfo {author} {\bibfnamefont {K.~A.}\ \bibnamefont {Haaga}}, \
  and\ \bibinfo {author} {\bibfnamefont {B.}~\bibnamefont {Hannisdal}},\ }\href
  {\doibase 10.1103/PhysRevE.99.042212} {\bibfield  {journal} {\bibinfo
  {journal} {Phys. Rev. E}\ }\textbf {\bibinfo {volume} {99}},\ \bibinfo
  {pages} {042212} (\bibinfo {year} {2019})}\BibitemShut {NoStop}%
\bibitem [{\citenamefont {Jiang}\ and\ \citenamefont {Kumar}(2019)}]{Jiang19}%
  \BibitemOpen
  \bibfield  {author} {\bibinfo {author} {\bibfnamefont {P.}~\bibnamefont
  {Jiang}}\ and\ \bibinfo {author} {\bibfnamefont {P.}~\bibnamefont {Kumar}},\
  }\href {\doibase 10.1103/PhysRevE.99.012306} {\bibfield  {journal} {\bibinfo
  {journal} {Phys. Rev. E}\ }\textbf {\bibinfo {volume} {99}},\ \bibinfo
  {pages} {012306} (\bibinfo {year} {2019})}\BibitemShut {NoStop}%
\bibitem [{\citenamefont {Jia}\ \emph {et~al.}(2020)\citenamefont {Jia},
  \citenamefont {Lin}, \citenamefont {Liu}, \citenamefont {Jiao},\ and\
  \citenamefont {Wang}}]{Jia20}%
  \BibitemOpen
  \bibfield  {author} {\bibinfo {author} {\bibfnamefont {Z.}~\bibnamefont
  {Jia}}, \bibinfo {author} {\bibfnamefont {Y.}~\bibnamefont {Lin}}, \bibinfo
  {author} {\bibfnamefont {Y.}~\bibnamefont {Liu}}, \bibinfo {author}
  {\bibfnamefont {Z.}~\bibnamefont {Jiao}}, \ and\ \bibinfo {author}
  {\bibfnamefont {J.}~\bibnamefont {Wang}},\ }\href {\doibase
  10.1103/PhysRevE.101.062113} {\bibfield  {journal} {\bibinfo  {journal}
  {Phys. Rev. E}\ }\textbf {\bibinfo {volume} {101}},\ \bibinfo {pages}
  {062113} (\bibinfo {year} {2020})}\BibitemShut {NoStop}%
\bibitem [{\citenamefont {Paninski}(2003)}]{Paninski03}%
  \BibitemOpen
  \bibfield  {author} {\bibinfo {author} {\bibfnamefont {L.}~\bibnamefont
  {Paninski}},\ }\href {\doibase 10.1162/089976603321780272} {\bibfield
  {journal} {\bibinfo  {journal} {Neural Computation}\ }\textbf {\bibinfo
  {volume} {15}},\ \bibinfo {pages} {1191} (\bibinfo {year} {2003})},\ \Eprint
  {http://arxiv.org/abs/https://direct.mit.edu/neco/article-pdf/15/6/1191/815550/089976603321780272.pdf}
  {https://direct.mit.edu/neco/article-pdf/15/6/1191/815550/089976603321780272.pdf}
  \BibitemShut {NoStop}%
\bibitem [{\citenamefont {{Vergara}}\ and\ \citenamefont
  {{Est{\'e}vez}}(2015)}]{Vergara15}%
  \BibitemOpen
  \bibfield  {author} {\bibinfo {author} {\bibfnamefont {J.~R.}\ \bibnamefont
  {{Vergara}}}\ and\ \bibinfo {author} {\bibfnamefont {P.~A.}\ \bibnamefont
  {{Est{\'e}vez}}},\ }\href@noop {} {\bibfield  {journal} {\bibinfo  {journal}
  {arXiv e-prints}\ ,\ \bibinfo {eid} {arXiv:1509.07577}} (\bibinfo {year}
  {2015})},\ \Eprint {http://arxiv.org/abs/1509.07577} {arXiv:1509.07577
  [cs.LG]} \BibitemShut {NoStop}%
\bibitem [{\citenamefont {Cover}\ and\ \citenamefont {Thomas}(2006)}]{Cover06}%
  \BibitemOpen
  \bibfield  {author} {\bibinfo {author} {\bibfnamefont {T.~M.}\ \bibnamefont
  {Cover}}\ and\ \bibinfo {author} {\bibfnamefont {J.~A.}\ \bibnamefont
  {Thomas}},\ }\href@noop {} {\emph {\bibinfo {title} {{Elements of Information
  Theory (Wiley Series in Telecommunications and Signal Processing)}}}}\
  (\bibinfo  {publisher} {Wiley-Interscience},\ \bibinfo {address} {USA},\
  \bibinfo {year} {2006})\BibitemShut {NoStop}%
\bibitem [{\citenamefont {Fraser}\ and\ \citenamefont
  {Swinney}(1986)}]{Fraser86}%
  \BibitemOpen
  \bibfield  {author} {\bibinfo {author} {\bibfnamefont {A.~M.}\ \bibnamefont
  {Fraser}}\ and\ \bibinfo {author} {\bibfnamefont {H.~L.}\ \bibnamefont
  {Swinney}},\ }\href {\doibase 10.1103/PhysRevA.33.1134} {\bibfield  {journal}
  {\bibinfo  {journal} {Phys. Rev. A}\ }\textbf {\bibinfo {volume} {33}},\
  \bibinfo {pages} {1134} (\bibinfo {year} {1986})}\BibitemShut {NoStop}%
\bibitem [{\citenamefont {Moon}\ \emph {et~al.}(1995)\citenamefont {Moon},
  \citenamefont {Rajagopalan},\ and\ \citenamefont {Lall}}]{Moon95}%
  \BibitemOpen
  \bibfield  {author} {\bibinfo {author} {\bibfnamefont {Y.-I.}\ \bibnamefont
  {Moon}}, \bibinfo {author} {\bibfnamefont {B.}~\bibnamefont {Rajagopalan}}, \
  and\ \bibinfo {author} {\bibfnamefont {U.}~\bibnamefont {Lall}},\ }\href
  {\doibase 10.1103/PhysRevE.52.2318} {\bibfield  {journal} {\bibinfo
  {journal} {Phys. Rev. E}\ }\textbf {\bibinfo {volume} {52}},\ \bibinfo
  {pages} {2318} (\bibinfo {year} {1995})}\BibitemShut {NoStop}%
\bibitem [{\citenamefont {Darbellay}\ and\ \citenamefont
  {Vajda}(1999)}]{Darbellay99}%
  \BibitemOpen
  \bibfield  {author} {\bibinfo {author} {\bibfnamefont {G.}~\bibnamefont
  {Darbellay}}\ and\ \bibinfo {author} {\bibfnamefont {I.}~\bibnamefont
  {Vajda}},\ }\href {\doibase 10.1109/18.761290} {\bibfield  {journal}
  {\bibinfo  {journal} {IEEE Transactions on Information Theory}\ }\textbf
  {\bibinfo {volume} {45}},\ \bibinfo {pages} {1315} (\bibinfo {year}
  {1999})}\BibitemShut {NoStop}%
\bibitem [{\citenamefont {Kwak}\ and\ \citenamefont {Choi}(2002)}]{Kwak02}%
  \BibitemOpen
  \bibfield  {author} {\bibinfo {author} {\bibfnamefont {N.}~\bibnamefont
  {Kwak}}\ and\ \bibinfo {author} {\bibfnamefont {C.-H.}\ \bibnamefont
  {Choi}},\ }\href {\doibase 10.1109/TPAMI.2002.1114861} {\bibfield  {journal}
  {\bibinfo  {journal} {IEEE Transactions on Pattern Analysis and Machine
  Intelligence}\ }\textbf {\bibinfo {volume} {24}},\ \bibinfo {pages} {1667}
  (\bibinfo {year} {2002})}\BibitemShut {NoStop}%
\bibitem [{\citenamefont {Kraskov}\ \emph {et~al.}(2004)\citenamefont
  {Kraskov}, \citenamefont {St\"ogbauer},\ and\ \citenamefont
  {Grassberger}}]{Kraskov04}%
  \BibitemOpen
  \bibfield  {author} {\bibinfo {author} {\bibfnamefont {A.}~\bibnamefont
  {Kraskov}}, \bibinfo {author} {\bibfnamefont {H.}~\bibnamefont
  {St\"ogbauer}}, \ and\ \bibinfo {author} {\bibfnamefont {P.}~\bibnamefont
  {Grassberger}},\ }\href {\doibase 10.1103/PhysRevE.69.066138} {\bibfield
  {journal} {\bibinfo  {journal} {Phys. Rev. E}\ }\textbf {\bibinfo {volume}
  {69}},\ \bibinfo {pages} {066138} (\bibinfo {year} {2004})}\BibitemShut
  {NoStop}%
\bibitem [{\citenamefont {Suzuki}\ \emph {et~al.}(2008)\citenamefont {Suzuki},
  \citenamefont {Sugiyama}, \citenamefont {Sese},\ and\ \citenamefont
  {Kanamori}}]{Suzuki08}%
  \BibitemOpen
  \bibfield  {author} {\bibinfo {author} {\bibfnamefont {T.}~\bibnamefont
  {Suzuki}}, \bibinfo {author} {\bibfnamefont {M.}~\bibnamefont {Sugiyama}},
  \bibinfo {author} {\bibfnamefont {J.}~\bibnamefont {Sese}}, \ and\ \bibinfo
  {author} {\bibfnamefont {T.}~\bibnamefont {Kanamori}},\ }in\ \href
  {https://proceedings.mlr.press/v4/suzuki08a.html} {\emph {\bibinfo
  {booktitle} {Proceedings of the Workshop on New Challenges for Feature
  Selection in Data Mining and Knowledge Discovery at ECML/PKDD 2008}}},\
  \bibinfo {series} {Proceedings of Machine Learning Research}, Vol.~\bibinfo
  {volume} {4},\ \bibinfo {editor} {edited by\ \bibinfo {editor} {\bibfnamefont
  {Y.}~\bibnamefont {Saeys}}, \bibinfo {editor} {\bibfnamefont
  {H.}~\bibnamefont {Liu}}, \bibinfo {editor} {\bibfnamefont {I.}~\bibnamefont
  {Inza}}, \bibinfo {editor} {\bibfnamefont {L.}~\bibnamefont {Wehenkel}}, \
  and\ \bibinfo {editor} {\bibfnamefont {Y.~V.~d.}\ \bibnamefont {Pee}}}\
  (\bibinfo  {publisher} {PMLR},\ \bibinfo {address} {Antwerp, Belgium},\
  \bibinfo {year} {2008})\ pp.\ \bibinfo {pages} {5--20}\BibitemShut {NoStop}%
\bibitem [{\citenamefont {Saxe}\ \emph {et~al.}(2019)\citenamefont {Saxe},
  \citenamefont {Bansal}, \citenamefont {Dapello}, \citenamefont {Advani},
  \citenamefont {Kolchinsky}, \citenamefont {Tracey},\ and\ \citenamefont
  {Cox}}]{Saxe19}%
  \BibitemOpen
  \bibfield  {author} {\bibinfo {author} {\bibfnamefont {A.~M.}\ \bibnamefont
  {Saxe}}, \bibinfo {author} {\bibfnamefont {Y.}~\bibnamefont {Bansal}},
  \bibinfo {author} {\bibfnamefont {J.}~\bibnamefont {Dapello}}, \bibinfo
  {author} {\bibfnamefont {M.}~\bibnamefont {Advani}}, \bibinfo {author}
  {\bibfnamefont {A.}~\bibnamefont {Kolchinsky}}, \bibinfo {author}
  {\bibfnamefont {B.~D.}\ \bibnamefont {Tracey}}, \ and\ \bibinfo {author}
  {\bibfnamefont {D.~D.}\ \bibnamefont {Cox}},\ }\href {\doibase
  10.1088/1742-5468/ab3985} {\bibfield  {journal} {\bibinfo  {journal} {Journal
  of Statistical Mechanics: Theory and Experiment}\ }\textbf {\bibinfo {volume}
  {2019}},\ \bibinfo {pages} {124020} (\bibinfo {year} {2019})}\BibitemShut
  {NoStop}%
\bibitem [{\citenamefont {{Pichler}}\ \emph {et~al.}(2022)\citenamefont
  {{Pichler}}, \citenamefont {{Colombo}}, \citenamefont {{Boudiaf}},
  \citenamefont {{Koliander}},\ and\ \citenamefont {{Piantanida}}}]{Pichler22}%
  \BibitemOpen
  \bibfield  {author} {\bibinfo {author} {\bibfnamefont {G.}~\bibnamefont
  {{Pichler}}}, \bibinfo {author} {\bibfnamefont {P.}~\bibnamefont
  {{Colombo}}}, \bibinfo {author} {\bibfnamefont {M.}~\bibnamefont
  {{Boudiaf}}}, \bibinfo {author} {\bibfnamefont {G.}~\bibnamefont
  {{Koliander}}}, \ and\ \bibinfo {author} {\bibfnamefont {P.}~\bibnamefont
  {{Piantanida}}},\ }\href@noop {} {\bibfield  {journal} {\bibinfo  {journal}
  {arXiv e-prints}\ ,\ \bibinfo {eid} {arXiv:2202.06618}} (\bibinfo {year}
  {2022})},\ \Eprint {http://arxiv.org/abs/2202.06618} {arXiv:2202.06618
  [cs.LG]} \BibitemShut {NoStop}%
\bibitem [{\citenamefont {{Kozachenko}}\ and\ \citenamefont
  {{Leonenko}}(1987)}]{Kozachenko87}%
  \BibitemOpen
  \bibfield  {author} {\bibinfo {author} {\bibfnamefont {L.~F.}\ \bibnamefont
  {{Kozachenko}}}\ and\ \bibinfo {author} {\bibfnamefont {N.~N.}\ \bibnamefont
  {{Leonenko}}},\ }\href@noop {} {\bibfield  {journal} {\bibinfo  {journal}
  {{Probl. Inf. Transm.}}\ }\textbf {\bibinfo {volume} {23}},\ \bibinfo {pages}
  {95} (\bibinfo {year} {1987})}\BibitemShut {NoStop}%
\bibitem [{\citenamefont {{Gao}}\ \emph {et~al.}(2014)\citenamefont {{Gao}},
  \citenamefont {{Ver Steeg}},\ and\ \citenamefont {{Galstyan}}}]{Gao14}%
  \BibitemOpen
  \bibfield  {author} {\bibinfo {author} {\bibfnamefont {S.}~\bibnamefont
  {{Gao}}}, \bibinfo {author} {\bibfnamefont {G.}~\bibnamefont {{Ver Steeg}}},
  \ and\ \bibinfo {author} {\bibfnamefont {A.}~\bibnamefont {{Galstyan}}},\
  }\href@noop {} {\bibfield  {journal} {\bibinfo  {journal} {arXiv e-prints}\
  ,\ \bibinfo {eid} {arXiv:1411.2003}} (\bibinfo {year} {2014})},\ \Eprint
  {http://arxiv.org/abs/1411.2003} {arXiv:1411.2003 [cs.IT]} \BibitemShut
  {NoStop}%
\bibitem [{\citenamefont {Hutter}(2002)}]{Hutter01}%
  \BibitemOpen
  \bibfield  {author} {\bibinfo {author} {\bibfnamefont {M.}~\bibnamefont
  {Hutter}},\ }in\ \href {http://www.hutter1.net/ai/xentropy.htm} {\emph
  {\bibinfo {booktitle} {Advances in Neural Information Processing Systems
  14}}},\ \bibinfo {editor} {edited by\ \bibinfo {editor} {\bibfnamefont
  {T.~G.}\ \bibnamefont {Dietterich}}, \bibinfo {editor} {\bibfnamefont
  {S.}~\bibnamefont {Becker}}, \ and\ \bibinfo {editor} {\bibfnamefont
  {Z.}~\bibnamefont {Ghahramani}}}\ (\bibinfo  {publisher} {MIT Press},\
  \bibinfo {address} {Cambridge, MA},\ \bibinfo {year} {2002})\ pp.\ \bibinfo
  {pages} {399--406}\BibitemShut {NoStop}%
\bibitem [{\citenamefont {Hutter}\ and\ \citenamefont
  {Zaffalon}(2005)}]{Hutter05}%
  \BibitemOpen
  \bibfield  {author} {\bibinfo {author} {\bibfnamefont {M.}~\bibnamefont
  {Hutter}}\ and\ \bibinfo {author} {\bibfnamefont {M.}~\bibnamefont
  {Zaffalon}},\ }\href@noop {} {\bibfield  {journal} {\bibinfo  {journal}
  {Computational Statistics \& Data Analysis}\ }\textbf {\bibinfo {volume}
  {48}},\ \bibinfo {pages} {633} (\bibinfo {year} {2005})}\BibitemShut
  {NoStop}%
\bibitem [{\citenamefont {Archer}\ \emph {et~al.}(2013)\citenamefont {Archer},
  \citenamefont {Park},\ and\ \citenamefont {Pillow}}]{Archer13}%
  \BibitemOpen
  \bibfield  {author} {\bibinfo {author} {\bibfnamefont {E.}~\bibnamefont
  {Archer}}, \bibinfo {author} {\bibfnamefont {I.~M.}\ \bibnamefont {Park}}, \
  and\ \bibinfo {author} {\bibfnamefont {J.~W.}\ \bibnamefont {Pillow}},\
  }\href {\doibase 10.3390/e15051738} {\bibfield  {journal} {\bibinfo
  {journal} {Entropy}\ }\textbf {\bibinfo {volume} {15}},\ \bibinfo {pages}
  {1738} (\bibinfo {year} {2013})}\BibitemShut {NoStop}%
\bibitem [{\citenamefont {Tishby}\ and\ \citenamefont
  {Zaslavsky}(2015)}]{Tishby15}%
  \BibitemOpen
  \bibfield  {author} {\bibinfo {author} {\bibfnamefont {N.}~\bibnamefont
  {Tishby}}\ and\ \bibinfo {author} {\bibfnamefont {N.}~\bibnamefont
  {Zaslavsky}},\ }in\ \href@noop {} {\emph {\bibinfo {booktitle} {2015 IEEE
  Information Theory Workshop (ITW)}}}\ (\bibinfo {organization} {IEEE},\
  \bibinfo {year} {2015})\ pp.\ \bibinfo {pages} {1--5}\BibitemShut {NoStop}%
\bibitem [{\citenamefont {{Alemi}}\ \emph {et~al.}(2016)\citenamefont
  {{Alemi}}, \citenamefont {{Fischer}}, \citenamefont {{Dillon}},\ and\
  \citenamefont {{Murphy}}}]{Alemi16}%
  \BibitemOpen
  \bibfield  {author} {\bibinfo {author} {\bibfnamefont {A.~A.}\ \bibnamefont
  {{Alemi}}}, \bibinfo {author} {\bibfnamefont {I.}~\bibnamefont {{Fischer}}},
  \bibinfo {author} {\bibfnamefont {J.~V.}\ \bibnamefont {{Dillon}}}, \ and\
  \bibinfo {author} {\bibfnamefont {K.}~\bibnamefont {{Murphy}}},\ }\href@noop
  {} {\bibfield  {journal} {\bibinfo  {journal} {arXiv e-prints}\ ,\ \bibinfo
  {eid} {arXiv:1612.00410}} (\bibinfo {year} {2016})},\ \Eprint
  {http://arxiv.org/abs/1612.00410} {arXiv:1612.00410 [cs.LG]} \BibitemShut
  {NoStop}%
\bibitem [{\citenamefont {{Brakel}}\ and\ \citenamefont
  {{Bengio}}(2017)}]{Brakel17}%
  \BibitemOpen
  \bibfield  {author} {\bibinfo {author} {\bibfnamefont {P.}~\bibnamefont
  {{Brakel}}}\ and\ \bibinfo {author} {\bibfnamefont {Y.}~\bibnamefont
  {{Bengio}}},\ }\href@noop {} {\bibfield  {journal} {\bibinfo  {journal}
  {arXiv e-prints}\ ,\ \bibinfo {eid} {arXiv:1710.05050}} (\bibinfo {year}
  {2017})},\ \Eprint {http://arxiv.org/abs/1710.05050} {arXiv:1710.05050
  [stat.ML]} \BibitemShut {NoStop}%
\bibitem [{\citenamefont {Kolchinsky}\ \emph {et~al.}(2019)\citenamefont
  {Kolchinsky}, \citenamefont {Tracey},\ and\ \citenamefont
  {Wolpert}}]{Kolchinsky19}%
  \BibitemOpen
  \bibfield  {author} {\bibinfo {author} {\bibfnamefont {A.}~\bibnamefont
  {Kolchinsky}}, \bibinfo {author} {\bibfnamefont {B.~D.}\ \bibnamefont
  {Tracey}}, \ and\ \bibinfo {author} {\bibfnamefont {D.~H.}\ \bibnamefont
  {Wolpert}},\ }\href {\doibase 10.3390/e21121181} {\bibfield  {journal}
  {\bibinfo  {journal} {Entropy}\ }\textbf {\bibinfo {volume} {21}} (\bibinfo
  {year} {2019}),\ 10.3390/e21121181}\BibitemShut {NoStop}%
\bibitem [{\citenamefont {Belghazi}\ \emph {et~al.}(2018)\citenamefont
  {Belghazi}, \citenamefont {Baratin}, \citenamefont {Rajeshwar}, \citenamefont
  {Ozair}, \citenamefont {Bengio}, \citenamefont {Courville},\ and\
  \citenamefont {Hjelm}}]{Belghazi18}%
  \BibitemOpen
  \bibfield  {author} {\bibinfo {author} {\bibfnamefont {M.~I.}\ \bibnamefont
  {Belghazi}}, \bibinfo {author} {\bibfnamefont {A.}~\bibnamefont {Baratin}},
  \bibinfo {author} {\bibfnamefont {S.}~\bibnamefont {Rajeshwar}}, \bibinfo
  {author} {\bibfnamefont {S.}~\bibnamefont {Ozair}}, \bibinfo {author}
  {\bibfnamefont {Y.}~\bibnamefont {Bengio}}, \bibinfo {author} {\bibfnamefont
  {A.}~\bibnamefont {Courville}}, \ and\ \bibinfo {author} {\bibfnamefont
  {D.}~\bibnamefont {Hjelm}},\ }in\ \href
  {https://proceedings.mlr.press/v80/belghazi18a.html} {\emph {\bibinfo
  {booktitle} {Proceedings of the 35th International Conference on Machine
  Learning}}},\ \bibinfo {series} {Proceedings of Machine Learning Research},
  Vol.~\bibinfo {volume} {80},\ \bibinfo {editor} {edited by\ \bibinfo {editor}
  {\bibfnamefont {J.}~\bibnamefont {Dy}}\ and\ \bibinfo {editor} {\bibfnamefont
  {A.}~\bibnamefont {Krause}}}\ (\bibinfo  {publisher} {PMLR},\ \bibinfo {year}
  {2018})\ pp.\ \bibinfo {pages} {531--540}\BibitemShut {NoStop}%
\bibitem [{\citenamefont {van~den Oord}\ \emph {et~al.}(2018)\citenamefont
  {van~den Oord}, \citenamefont {Li},\ and\ \citenamefont {Vinyals}}]{Oord18}%
  \BibitemOpen
  \bibfield  {author} {\bibinfo {author} {\bibfnamefont {A.}~\bibnamefont
  {van~den Oord}}, \bibinfo {author} {\bibfnamefont {Y.}~\bibnamefont {Li}}, \
  and\ \bibinfo {author} {\bibfnamefont {O.}~\bibnamefont {Vinyals}},\ }\href
  {http://arxiv.org/abs/1807.03748} {\bibfield  {journal} {\bibinfo  {journal}
  {CoRR}\ }\textbf {\bibinfo {volume} {abs/1807.03748}} (\bibinfo {year}
  {2018})},\ \Eprint {http://arxiv.org/abs/1807.03748} {1807.03748}
  \BibitemShut {NoStop}%
\bibitem [{\citenamefont {Moyer}\ \emph {et~al.}(2018)\citenamefont {Moyer},
  \citenamefont {Gao}, \citenamefont {Brekelmans}, \citenamefont {Galstyan},\
  and\ \citenamefont {Ver~Steeg}}]{Moyer18}%
  \BibitemOpen
  \bibfield  {author} {\bibinfo {author} {\bibfnamefont {D.}~\bibnamefont
  {Moyer}}, \bibinfo {author} {\bibfnamefont {S.}~\bibnamefont {Gao}}, \bibinfo
  {author} {\bibfnamefont {R.}~\bibnamefont {Brekelmans}}, \bibinfo {author}
  {\bibfnamefont {A.}~\bibnamefont {Galstyan}}, \ and\ \bibinfo {author}
  {\bibfnamefont {G.}~\bibnamefont {Ver~Steeg}},\ }in\ \href
  {https://proceedings.neurips.cc/paper/2018/file/415185ea244ea2b2bedeb0449b926802-Paper.pdf}
  {\emph {\bibinfo {booktitle} {Advances in Neural Information Processing
  Systems}}},\ Vol.~\bibinfo {volume} {31},\ \bibinfo {editor} {edited by\
  \bibinfo {editor} {\bibfnamefont {S.}~\bibnamefont {Bengio}}, \bibinfo
  {editor} {\bibfnamefont {H.}~\bibnamefont {Wallach}}, \bibinfo {editor}
  {\bibfnamefont {H.}~\bibnamefont {Larochelle}}, \bibinfo {editor}
  {\bibfnamefont {K.}~\bibnamefont {Grauman}}, \bibinfo {editor} {\bibfnamefont
  {N.}~\bibnamefont {Cesa-Bianchi}}, \ and\ \bibinfo {editor} {\bibfnamefont
  {R.}~\bibnamefont {Garnett}}}\ (\bibinfo  {publisher} {Curran Associates,
  Inc.},\ \bibinfo {year} {2018})\BibitemShut {NoStop}%
\bibitem [{\citenamefont {{Poole}}\ \emph {et~al.}(2019)\citenamefont
  {{Poole}}, \citenamefont {{Ozair}}, \citenamefont {{van den Oord}},
  \citenamefont {{Alemi}},\ and\ \citenamefont {{Tucker}}}]{Poole19}%
  \BibitemOpen
  \bibfield  {author} {\bibinfo {author} {\bibfnamefont {B.}~\bibnamefont
  {{Poole}}}, \bibinfo {author} {\bibfnamefont {S.}~\bibnamefont {{Ozair}}},
  \bibinfo {author} {\bibfnamefont {A.}~\bibnamefont {{van den Oord}}},
  \bibinfo {author} {\bibfnamefont {A.~A.}\ \bibnamefont {{Alemi}}}, \ and\
  \bibinfo {author} {\bibfnamefont {G.}~\bibnamefont {{Tucker}}},\ }\href@noop
  {} {\bibfield  {journal} {\bibinfo  {journal} {arXiv e-prints}\ ,\ \bibinfo
  {eid} {arXiv:1905.06922}} (\bibinfo {year} {2019})},\ \Eprint
  {http://arxiv.org/abs/1905.06922} {arXiv:1905.06922 [cs.LG]} \BibitemShut
  {NoStop}%
\bibitem [{\citenamefont {Peng}\ \emph {et~al.}(2019)\citenamefont {Peng},
  \citenamefont {Kanazawa}, \citenamefont {Toyer}, \citenamefont {Abbeel},\
  and\ \citenamefont {Levine}}]{Peng19}%
  \BibitemOpen
  \bibfield  {author} {\bibinfo {author} {\bibfnamefont {X.~B.}\ \bibnamefont
  {Peng}}, \bibinfo {author} {\bibfnamefont {A.}~\bibnamefont {Kanazawa}},
  \bibinfo {author} {\bibfnamefont {S.}~\bibnamefont {Toyer}}, \bibinfo
  {author} {\bibfnamefont {P.}~\bibnamefont {Abbeel}}, \ and\ \bibinfo {author}
  {\bibfnamefont {S.}~\bibnamefont {Levine}},\ }in\ \href
  {https://openreview.net/forum?id=HyxPx3R9tm} {\emph {\bibinfo {booktitle}
  {7th International Conference on Learning Representations, {ICLR} 2019, New
  Orleans, LA, USA, May 6-9, 2019}}}\ (\bibinfo  {publisher} {OpenReview.net},\
  \bibinfo {year} {2019})\BibitemShut {NoStop}%
\bibitem [{\citenamefont {Hjelm}\ \emph {et~al.}(2019)\citenamefont {Hjelm},
  \citenamefont {Fedorov}, \citenamefont {Lavoie{-}Marchildon}, \citenamefont
  {Grewal}, \citenamefont {Bachman}, \citenamefont {Trischler},\ and\
  \citenamefont {Bengio}}]{Hjelm19}%
  \BibitemOpen
  \bibfield  {author} {\bibinfo {author} {\bibfnamefont {R.~D.}\ \bibnamefont
  {Hjelm}}, \bibinfo {author} {\bibfnamefont {A.}~\bibnamefont {Fedorov}},
  \bibinfo {author} {\bibfnamefont {S.}~\bibnamefont {Lavoie{-}Marchildon}},
  \bibinfo {author} {\bibfnamefont {K.}~\bibnamefont {Grewal}}, \bibinfo
  {author} {\bibfnamefont {P.}~\bibnamefont {Bachman}}, \bibinfo {author}
  {\bibfnamefont {A.}~\bibnamefont {Trischler}}, \ and\ \bibinfo {author}
  {\bibfnamefont {Y.}~\bibnamefont {Bengio}},\ }in\ \href
  {https://openreview.net/forum?id=Bklr3j0cKX} {\emph {\bibinfo {booktitle}
  {7th International Conference on Learning Representations, {ICLR} 2019, New
  Orleans, LA, USA, May 6-9, 2019}}}\ (\bibinfo  {publisher} {OpenReview.net},\
  \bibinfo {year} {2019})\BibitemShut {NoStop}%
\bibitem [{\citenamefont {Song}\ and\ \citenamefont {Ermon}(2020)}]{Song20}%
  \BibitemOpen
  \bibfield  {author} {\bibinfo {author} {\bibfnamefont {J.}~\bibnamefont
  {Song}}\ and\ \bibinfo {author} {\bibfnamefont {S.}~\bibnamefont {Ermon}},\
  }in\ \href@noop {} {\emph {\bibinfo {booktitle} {International Conference on
  Learning Representations}}}\ (\bibinfo {year} {2020})\BibitemShut {NoStop}%
\bibitem [{\citenamefont {G\"okmen}\ \emph {et~al.}(2021)\citenamefont
  {G\"okmen}, \citenamefont {Ringel}, \citenamefont {Huber},\ and\
  \citenamefont {Koch-Janusz}}]{Gokmen21}%
  \BibitemOpen
  \bibfield  {author} {\bibinfo {author} {\bibfnamefont {D.~E.}\ \bibnamefont
  {G\"okmen}}, \bibinfo {author} {\bibfnamefont {Z.}~\bibnamefont {Ringel}},
  \bibinfo {author} {\bibfnamefont {S.~D.}\ \bibnamefont {Huber}}, \ and\
  \bibinfo {author} {\bibfnamefont {M.}~\bibnamefont {Koch-Janusz}},\ }\href
  {\doibase 10.1103/PhysRevE.104.064106} {\bibfield  {journal} {\bibinfo
  {journal} {Phys. Rev. E}\ }\textbf {\bibinfo {volume} {104}},\ \bibinfo
  {pages} {064106} (\bibinfo {year} {2021})}\BibitemShut {NoStop}%
\bibitem [{\citenamefont {Kullback}\ and\ \citenamefont
  {Leibler}(1951)}]{Kullback51}%
  \BibitemOpen
  \bibfield  {author} {\bibinfo {author} {\bibfnamefont {S.}~\bibnamefont
  {Kullback}}\ and\ \bibinfo {author} {\bibfnamefont {R.~A.}\ \bibnamefont
  {Leibler}},\ }\href@noop {} {\bibfield  {journal} {\bibinfo  {journal} {Ann.
  Math. Statist.}\ }\textbf {\bibinfo {volume} {22}},\ \bibinfo {pages} {79}
  (\bibinfo {year} {1951})}\BibitemShut {NoStop}%
\bibitem [{\citenamefont {Donsker}\ and\ \citenamefont
  {Varadhan}(1983)}]{Donsker83}%
  \BibitemOpen
  \bibfield  {author} {\bibinfo {author} {\bibfnamefont {M.}~\bibnamefont
  {Donsker}}\ and\ \bibinfo {author} {\bibfnamefont {S.}~\bibnamefont
  {Varadhan}},\ }\href {\doibase 10.1002/cpa.3160360204} {\bibfield  {journal}
  {\bibinfo  {journal} {Communications on Pure and Applied Mathematics}\
  }\textbf {\bibinfo {volume} {36}},\ \bibinfo {pages} {183} (\bibinfo {year}
  {1983})}\BibitemShut {NoStop}%
\bibitem [{\citenamefont {Chen}\ \emph {et~al.}(2018)\citenamefont {Chen},
  \citenamefont {Li}, \citenamefont {Grosse},\ and\ \citenamefont
  {Duvenaud}}]{Chen18}%
  \BibitemOpen
  \bibfield  {author} {\bibinfo {author} {\bibfnamefont {R.~T.~Q.}\
  \bibnamefont {Chen}}, \bibinfo {author} {\bibfnamefont {X.}~\bibnamefont
  {Li}}, \bibinfo {author} {\bibfnamefont {R.~B.}\ \bibnamefont {Grosse}}, \
  and\ \bibinfo {author} {\bibfnamefont {D.~K.}\ \bibnamefont {Duvenaud}},\
  }in\ \href
  {https://proceedings.neurips.cc/paper/2018/file/1ee3dfcd8a0645a25a35977997223d22-Paper.pdf}
  {\emph {\bibinfo {booktitle} {Advances in Neural Information Processing
  Systems}}},\ Vol.~\bibinfo {volume} {31},\ \bibinfo {editor} {edited by\
  \bibinfo {editor} {\bibfnamefont {S.}~\bibnamefont {Bengio}}, \bibinfo
  {editor} {\bibfnamefont {H.}~\bibnamefont {Wallach}}, \bibinfo {editor}
  {\bibfnamefont {H.}~\bibnamefont {Larochelle}}, \bibinfo {editor}
  {\bibfnamefont {K.}~\bibnamefont {Grauman}}, \bibinfo {editor} {\bibfnamefont
  {N.}~\bibnamefont {Cesa-Bianchi}}, \ and\ \bibinfo {editor} {\bibfnamefont
  {R.}~\bibnamefont {Garnett}}}\ (\bibinfo  {publisher} {Curran Associates,
  Inc.},\ \bibinfo {year} {2018})\BibitemShut {NoStop}%
\bibitem [{\citenamefont {{Sedaghat}}\ \emph {et~al.}(2021)\citenamefont
  {{Sedaghat}}, \citenamefont {{Romaniello}}, \citenamefont {{Carrick}},\ and\
  \citenamefont {{Pineau}}}]{Sedaghat21}%
  \BibitemOpen
  \bibfield  {author} {\bibinfo {author} {\bibfnamefont {N.}~\bibnamefont
  {{Sedaghat}}}, \bibinfo {author} {\bibfnamefont {M.}~\bibnamefont
  {{Romaniello}}}, \bibinfo {author} {\bibfnamefont {J.~E.}\ \bibnamefont
  {{Carrick}}}, \ and\ \bibinfo {author} {\bibfnamefont {F.-X.}\ \bibnamefont
  {{Pineau}}},\ }\href {\doibase 10.1093/mnras/staa3540} {\bibfield  {journal}
  {\bibinfo  {journal} {Monthly Notices of the Royal Astronomical Society}\
  }\textbf {\bibinfo {volume} {501}},\ \bibinfo {pages} {6026} (\bibinfo {year}
  {2021})},\ \Eprint {http://arxiv.org/abs/2009.12872} {arXiv:2009.12872}
  \BibitemShut {NoStop}%
\bibitem [{\citenamefont {{Ait Kerroum}}\ \emph {et~al.}(2010)\citenamefont
  {{Ait Kerroum}}, \citenamefont {Hammouch},\ and\ \citenamefont
  {Aboutajdine}}]{Kerroum10}%
  \BibitemOpen
  \bibfield  {author} {\bibinfo {author} {\bibfnamefont {M.}~\bibnamefont {{Ait
  Kerroum}}}, \bibinfo {author} {\bibfnamefont {A.}~\bibnamefont {Hammouch}}, \
  and\ \bibinfo {author} {\bibfnamefont {D.}~\bibnamefont {Aboutajdine}},\
  }\href {\doibase https://doi.org/10.1016/j.patrec.2009.11.010} {\bibfield
  {journal} {\bibinfo  {journal} {Pattern Recognition Letters}\ }\textbf
  {\bibinfo {volume} {31}},\ \bibinfo {pages} {1168} (\bibinfo {year}
  {2010})},\ \bibinfo {note} {pattern Recognition in Remote
  Sensing}\BibitemShut {NoStop}%
\bibitem [{\citenamefont {Eirola}\ \emph {et~al.}(2014)\citenamefont {Eirola},
  \citenamefont {Lendasse},\ and\ \citenamefont {Karhunen}}]{Eirola14}%
  \BibitemOpen
  \bibfield  {author} {\bibinfo {author} {\bibfnamefont {E.}~\bibnamefont
  {Eirola}}, \bibinfo {author} {\bibfnamefont {A.}~\bibnamefont {Lendasse}}, \
  and\ \bibinfo {author} {\bibfnamefont {J.}~\bibnamefont {Karhunen}},\ }in\
  \href {\doibase 10.1109/IJCNN.2014.6889561} {\emph {\bibinfo {booktitle}
  {2014 International Joint Conference on Neural Networks (IJCNN)}}}\ (\bibinfo
  {year} {2014})\ pp.\ \bibinfo {pages} {1606--1613}\BibitemShut {NoStop}%
\bibitem [{\citenamefont {Lan}\ \emph {et~al.}(2006)\citenamefont {Lan},
  \citenamefont {Erdogmus}, \citenamefont {Ozertem},\ and\ \citenamefont
  {Huang}}]{Lan06}%
  \BibitemOpen
  \bibfield  {author} {\bibinfo {author} {\bibfnamefont {T.}~\bibnamefont
  {Lan}}, \bibinfo {author} {\bibfnamefont {D.}~\bibnamefont {Erdogmus}},
  \bibinfo {author} {\bibfnamefont {U.}~\bibnamefont {Ozertem}}, \ and\
  \bibinfo {author} {\bibfnamefont {Y.}~\bibnamefont {Huang}},\ }in\ \href
  {\doibase 10.1109/IJCNN.2006.247209} {\emph {\bibinfo {booktitle} {The 2006
  IEEE International Joint Conference on Neural Network Proceedings}}}\
  (\bibinfo {year} {2006})\ pp.\ \bibinfo {pages} {5034--5039}\BibitemShut
  {NoStop}%
\bibitem [{\citenamefont {Leiva-Murillo}\ and\ \citenamefont
  {Art{\'e}s-Rodr{\'i}guez}(2004)}]{Leiva04}%
  \BibitemOpen
  \bibfield  {author} {\bibinfo {author} {\bibfnamefont {J.~M.}\ \bibnamefont
  {Leiva-Murillo}}\ and\ \bibinfo {author} {\bibfnamefont {A.}~\bibnamefont
  {Art{\'e}s-Rodr{\'i}guez}},\ }in\ \href@noop {} {\emph {\bibinfo {booktitle}
  {Independent Component Analysis and Blind Signal Separation}}},\ \bibinfo
  {editor} {edited by\ \bibinfo {editor} {\bibfnamefont {C.~G.}\ \bibnamefont
  {Puntonet}}\ and\ \bibinfo {editor} {\bibfnamefont {A.}~\bibnamefont
  {Prieto}}}\ (\bibinfo  {publisher} {Springer Berlin Heidelberg},\ \bibinfo
  {address} {Berlin, Heidelberg},\ \bibinfo {year} {2004})\ pp.\ \bibinfo
  {pages} {271--278}\BibitemShut {NoStop}%
\bibitem [{\citenamefont {Nilsson}\ \emph {et~al.}(2002)\citenamefont
  {Nilsson}, \citenamefont {Gustaftson}, \citenamefont {Vang~Andersen},\ and\
  \citenamefont {Kleijn}}]{Nilsson02}%
  \BibitemOpen
  \bibfield  {author} {\bibinfo {author} {\bibfnamefont {M.}~\bibnamefont
  {Nilsson}}, \bibinfo {author} {\bibfnamefont {H.}~\bibnamefont {Gustaftson}},
  \bibinfo {author} {\bibfnamefont {S.}~\bibnamefont {Vang~Andersen}}, \ and\
  \bibinfo {author} {\bibfnamefont {W.~B.}\ \bibnamefont {Kleijn}},\ }in\ \href
  {\doibase 10.1109/ICASSP.2002.5743770} {\emph {\bibinfo {booktitle} {2002
  IEEE International Conference on Acoustics, Speech, and Signal
  Processing}}},\ Vol.~\bibinfo {volume} {1}\ (\bibinfo {year} {2002})\ pp.\
  \bibinfo {pages} {I--525--I--528}\BibitemShut {NoStop}%
\bibitem [{\citenamefont {Polo}\ and\ \citenamefont {Vicente}(2022)}]{Polo22}%
  \BibitemOpen
  \bibfield  {author} {\bibinfo {author} {\bibfnamefont {F.~M.}\ \bibnamefont
  {Polo}}\ and\ \bibinfo {author} {\bibfnamefont {R.}~\bibnamefont {Vicente}},\
  }\href@noop {} {\bibfield  {journal} {\bibinfo  {journal} {Neural Computing
  and Applications}\ ,\ \bibinfo {pages} {1}} (\bibinfo {year}
  {2022})}\BibitemShut {NoStop}%
\bibitem [{\citenamefont {Ueda}\ \emph {et~al.}(1998)\citenamefont {Ueda},
  \citenamefont {Nakano}, \citenamefont {Ghahramani},\ and\ \citenamefont
  {Hinton}}]{Ueda98}%
  \BibitemOpen
  \bibfield  {author} {\bibinfo {author} {\bibfnamefont {N.}~\bibnamefont
  {Ueda}}, \bibinfo {author} {\bibfnamefont {R.}~\bibnamefont {Nakano}},
  \bibinfo {author} {\bibfnamefont {Z.}~\bibnamefont {Ghahramani}}, \ and\
  \bibinfo {author} {\bibfnamefont {G.}~\bibnamefont {Hinton}},\ }in\ \href
  {\doibase 10.1109/NNSP.1998.710657} {\emph {\bibinfo {booktitle} {Neural
  Networks for Signal Processing VIII. Proceedings of the 1998 IEEE Signal
  Processing Society Workshop (Cat. No.98TH8378)}}}\ (\bibinfo {year} {1998})\
  pp.\ \bibinfo {pages} {274--283}\BibitemShut {NoStop}%
\bibitem [{\citenamefont {{Bovy}}\ \emph {et~al.}(2011)\citenamefont {{Bovy}},
  \citenamefont {{Hogg}},\ and\ \citenamefont {{Roweis}}}]{Bovy11}%
  \BibitemOpen
  \bibfield  {author} {\bibinfo {author} {\bibfnamefont {J.}~\bibnamefont
  {{Bovy}}}, \bibinfo {author} {\bibfnamefont {D.~W.}\ \bibnamefont {{Hogg}}},
  \ and\ \bibinfo {author} {\bibfnamefont {S.~T.}\ \bibnamefont {{Roweis}}},\
  }\href {\doibase 10.1214/10-AOAS439} {\bibfield  {journal} {\bibinfo
  {journal} {Annals of Applied Statistics}\ }\textbf {\bibinfo {volume} {5}},\
  \bibinfo {pages} {1657} (\bibinfo {year} {2011})},\ \Eprint
  {http://arxiv.org/abs/0905.2979} {arXiv:0905.2979 [stat.ME]} \BibitemShut
  {NoStop}%
\bibitem [{\citenamefont {Shireman}\ \emph {et~al.}(2016)\citenamefont
  {Shireman}, \citenamefont {Steinley},\ and\ \citenamefont
  {Brusco}}]{Emilie16}%
  \BibitemOpen
  \bibfield  {author} {\bibinfo {author} {\bibfnamefont {E.~M.}\ \bibnamefont
  {Shireman}}, \bibinfo {author} {\bibfnamefont {D.}~\bibnamefont {Steinley}},
  \ and\ \bibinfo {author} {\bibfnamefont {M.~J.}\ \bibnamefont {Brusco}},\
  }\href {\doibase 10.1080/00273171.2016.1160359} {\bibfield  {journal}
  {\bibinfo  {journal} {Multivariate Behavioral Research}\ }\textbf {\bibinfo
  {volume} {51}},\ \bibinfo {pages} {466} (\bibinfo {year} {2016})},\ \bibinfo
  {note} {pMID: 27494191},\ \Eprint
  {http://arxiv.org/abs/https://doi.org/10.1080/00273171.2016.1160359}
  {https://doi.org/10.1080/00273171.2016.1160359} \BibitemShut {NoStop}%
\bibitem [{\citenamefont {Baudry}\ and\ \citenamefont
  {Celeux}(2015)}]{Baudry15}%
  \BibitemOpen
  \bibfield  {author} {\bibinfo {author} {\bibfnamefont {J.-P.}\ \bibnamefont
  {Baudry}}\ and\ \bibinfo {author} {\bibfnamefont {G.}~\bibnamefont
  {Celeux}},\ }\href {\doibase 10.1007/s11222-015-9561-x} {\bibfield  {journal}
  {\bibinfo  {journal} {Statistics and Computing}\ }\textbf {\bibinfo {volume}
  {25}},\ \bibinfo {pages} {713} (\bibinfo {year} {2015})}\BibitemShut
  {NoStop}%
\bibitem [{\citenamefont {{Melchior}}\ and\ \citenamefont
  {{Goulding}}(2018)}]{Melchior18}%
  \BibitemOpen
  \bibfield  {author} {\bibinfo {author} {\bibfnamefont {P.}~\bibnamefont
  {{Melchior}}}\ and\ \bibinfo {author} {\bibfnamefont {A.~D.}\ \bibnamefont
  {{Goulding}}},\ }\href {\doibase 10.1016/j.ascom.2018.09.013} {\bibfield
  {journal} {\bibinfo  {journal} {Astronomy and Computing}\ }\textbf {\bibinfo
  {volume} {25}},\ \bibinfo {pages} {183} (\bibinfo {year} {2018})},\ \Eprint
  {http://arxiv.org/abs/1611.05806} {arXiv:1611.05806 [astro-ph.IM]}
  \BibitemShut {NoStop}%
\bibitem [{\citenamefont {Dempster}\ \emph {et~al.}(1977)\citenamefont
  {Dempster}, \citenamefont {Laird},\ and\ \citenamefont {Rubin}}]{Dempster77}%
  \BibitemOpen
  \bibfield  {author} {\bibinfo {author} {\bibfnamefont {A.~P.}\ \bibnamefont
  {Dempster}}, \bibinfo {author} {\bibfnamefont {N.~M.}\ \bibnamefont {Laird}},
  \ and\ \bibinfo {author} {\bibfnamefont {D.~B.}\ \bibnamefont {Rubin}},\
  }\href {http://www.jstor.org/stable/2984875} {\bibfield  {journal} {\bibinfo
  {journal} {Journal of the Royal Statistical Society. Series B
  (Methodological)}\ }\textbf {\bibinfo {volume} {39}},\ \bibinfo {pages} {1}
  (\bibinfo {year} {1977})}\BibitemShut {NoStop}%
\bibitem [{\citenamefont {Lloyd}(1982)}]{Lloyd82}%
  \BibitemOpen
  \bibfield  {author} {\bibinfo {author} {\bibfnamefont {S.}~\bibnamefont
  {Lloyd}},\ }\href {\doibase 10.1109/TIT.1982.1056489} {\bibfield  {journal}
  {\bibinfo  {journal} {IEEE Transactions on Information Theory}\ }\textbf
  {\bibinfo {volume} {28}},\ \bibinfo {pages} {129} (\bibinfo {year}
  {1982})}\BibitemShut {NoStop}%
\bibitem [{\citenamefont {Arthur}\ and\ \citenamefont
  {Vassilvitskii}(2007)}]{Arthur07}%
  \BibitemOpen
  \bibfield  {author} {\bibinfo {author} {\bibfnamefont {D.}~\bibnamefont
  {Arthur}}\ and\ \bibinfo {author} {\bibfnamefont {S.}~\bibnamefont
  {Vassilvitskii}},\ }in\ \href@noop {} {\emph {\bibinfo {booktitle}
  {Proceedings of the Eighteenth Annual ACM-SIAM Symposium on Discrete
  Algorithms}}},\ \bibinfo {series and number} {SODA '07}\ (\bibinfo
  {publisher} {Society for Industrial and Applied Mathematics},\ \bibinfo
  {address} {USA},\ \bibinfo {year} {2007})\ p.\ \bibinfo {pages}
  {1027–1035}\BibitemShut {NoStop}%
\bibitem [{\citenamefont {Akaike}(1974)}]{Akaike74}%
  \BibitemOpen
  \bibfield  {author} {\bibinfo {author} {\bibfnamefont {H.}~\bibnamefont
  {Akaike}},\ }\href {\doibase 10.1109/TAC.1974.1100705} {\bibfield  {journal}
  {\bibinfo  {journal} {IEEE Transactions on Automatic Control}\ }\textbf
  {\bibinfo {volume} {19}},\ \bibinfo {pages} {716} (\bibinfo {year}
  {1974})}\BibitemShut {NoStop}%
\bibitem [{\citenamefont {Schwarz}(1978)}]{Schwarz78}%
  \BibitemOpen
  \bibfield  {author} {\bibinfo {author} {\bibfnamefont {G.}~\bibnamefont
  {Schwarz}},\ }\href {\doibase 10.1214/aos/1176344136} {\bibfield  {journal}
  {\bibinfo  {journal} {The Annals of Statistics}\ }\textbf {\bibinfo {volume}
  {6}},\ \bibinfo {pages} {461 } (\bibinfo {year} {1978})}\BibitemShut
  {NoStop}%
\bibitem [{\citenamefont {Ross}(2014)}]{Ross14}%
  \BibitemOpen
  \bibfield  {author} {\bibinfo {author} {\bibfnamefont {B.~C.}\ \bibnamefont
  {Ross}},\ }\href {\doibase 10.1371/journal.pone.0087357} {\bibfield
  {journal} {\bibinfo  {journal} {PLOS ONE}\ }\textbf {\bibinfo {volume} {9}},\
  \bibinfo {pages} {1} (\bibinfo {year} {2014})}\BibitemShut {NoStop}%
\bibitem [{\citenamefont {Kingma}\ and\ \citenamefont
  {Welling}(2014)}]{Kingma14}%
  \BibitemOpen
  \bibfield  {author} {\bibinfo {author} {\bibfnamefont {D.~P.}\ \bibnamefont
  {Kingma}}\ and\ \bibinfo {author} {\bibfnamefont {M.}~\bibnamefont
  {Welling}},\ }in\ \href@noop {} {\emph {\bibinfo {booktitle} {2nd
  International Conference on Learning Representations, {ICLR} 2014, Banff, AB,
  Canada, April 14-16, 2014, Conference Track Proceedings}}}\ (\bibinfo {year}
  {2014})\ \Eprint {http://arxiv.org/abs/http://arxiv.org/abs/1312.6114v10}
  {http://arxiv.org/abs/1312.6114v10} \BibitemShut {NoStop}%
\bibitem [{\citenamefont {Darbellay}\ and\ \citenamefont
  {Vajda}(2000)}]{Darbellay00}%
  \BibitemOpen
  \bibfield  {author} {\bibinfo {author} {\bibfnamefont {G.}~\bibnamefont
  {Darbellay}}\ and\ \bibinfo {author} {\bibfnamefont {I.}~\bibnamefont
  {Vajda}},\ }\href {\doibase 10.1109/18.825848} {\bibfield  {journal}
  {\bibinfo  {journal} {IEEE Transactions on Information Theory}\ }\textbf
  {\bibinfo {volume} {46}},\ \bibinfo {pages} {709} (\bibinfo {year}
  {2000})}\BibitemShut {NoStop}%
\bibitem [{\citenamefont {Haeri}\ and\ \citenamefont
  {Ebadzadeh}(2014)}]{Haeri14}%
  \BibitemOpen
  \bibfield  {author} {\bibinfo {author} {\bibfnamefont {M.~A.}\ \bibnamefont
  {Haeri}}\ and\ \bibinfo {author} {\bibfnamefont {M.~M.}\ \bibnamefont
  {Ebadzadeh}},\ }\href {\doibase 10.1007/s10700-014-9178-0} {\bibfield
  {journal} {\bibinfo  {journal} {Fuzzy Optim. Decis. Mak.}\ }\textbf {\bibinfo
  {volume} {13}},\ \bibinfo {pages} {287} (\bibinfo {year} {2014})}\BibitemShut
  {NoStop}%
\bibitem [{\citenamefont {Kim}\ and\ \citenamefont {Mnih}(2018)}]{Kim18}%
  \BibitemOpen
  \bibfield  {author} {\bibinfo {author} {\bibfnamefont {H.}~\bibnamefont
  {Kim}}\ and\ \bibinfo {author} {\bibfnamefont {A.}~\bibnamefont {Mnih}},\
  }in\ \href {https://proceedings.mlr.press/v80/kim18b.html} {\emph {\bibinfo
  {booktitle} {Proceedings of the 35th International Conference on Machine
  Learning}}},\ \bibinfo {series} {Proceedings of Machine Learning Research},
  Vol.~\bibinfo {volume} {80},\ \bibinfo {editor} {edited by\ \bibinfo {editor}
  {\bibfnamefont {J.}~\bibnamefont {Dy}}\ and\ \bibinfo {editor} {\bibfnamefont
  {A.}~\bibnamefont {Krause}}}\ (\bibinfo  {publisher} {PMLR},\ \bibinfo {year}
  {2018})\ pp.\ \bibinfo {pages} {2649--2658}\BibitemShut {NoStop}%
\bibitem [{\citenamefont {Burgess}\ and\ \citenamefont
  {Kim}(2018)}]{Burgess18}%
  \BibitemOpen
  \bibfield  {author} {\bibinfo {author} {\bibfnamefont {C.}~\bibnamefont
  {Burgess}}\ and\ \bibinfo {author} {\bibfnamefont {H.}~\bibnamefont {Kim}},\
  }\href@noop {} {\enquote {\bibinfo {title} {3{D} {S}hapes {D}ataset},}\
  }\bibinfo {howpublished} {\url{https://github.com/deepmind/3d-shapes}}
  (\bibinfo {year} {2018})\BibitemShut {NoStop}%
\bibitem [{\citenamefont {{Dodelson}}(2003)}]{Dodelson03}%
  \BibitemOpen
  \bibfield  {author} {\bibinfo {author} {\bibfnamefont {S.}~\bibnamefont
  {{Dodelson}}},\ }\href@noop {} {\emph {\bibinfo {title} {{Modern
  Cosmology}}}}\ (\bibinfo  {publisher} {Academic Press},\ \bibinfo {year}
  {2003})\BibitemShut {NoStop}%
\bibitem [{\citenamefont {{Navarro}}\ \emph {et~al.}(1996)\citenamefont
  {{Navarro}}, \citenamefont {{Frenk}},\ and\ \citenamefont
  {{White}}}]{Navarro96}%
  \BibitemOpen
  \bibfield  {author} {\bibinfo {author} {\bibfnamefont {J.~F.}\ \bibnamefont
  {{Navarro}}}, \bibinfo {author} {\bibfnamefont {C.~S.}\ \bibnamefont
  {{Frenk}}}, \ and\ \bibinfo {author} {\bibfnamefont {S.~D.~M.}\ \bibnamefont
  {{White}}},\ }\href {\doibase 10.1086/177173} {\bibfield  {journal} {\bibinfo
   {journal} {\apj}\ }\textbf {\bibinfo {volume} {462}},\ \bibinfo {pages}
  {563} (\bibinfo {year} {1996})},\ \Eprint
  {http://arxiv.org/abs/astro-ph/9508025} {arXiv:astro-ph/9508025 [astro-ph]}
  \BibitemShut {NoStop}%
\bibitem [{\citenamefont {Tormen}(1997)}]{Tormen97}%
  \BibitemOpen
  \bibfield  {author} {\bibinfo {author} {\bibfnamefont {G.}~\bibnamefont
  {Tormen}},\ }\href {\doibase 10.1093/mnras/290.3.411} {\bibfield  {journal}
  {\bibinfo  {journal} {Monthly Notices of the Royal Astronomical Society}\
  }\textbf {\bibinfo {volume} {290}},\ \bibinfo {pages} {411} (\bibinfo {year}
  {1997})},\ \Eprint
  {http://arxiv.org/abs/https://academic.oup.com/mnras/article-pdf/290/3/411/18540204/290-3-411.pdf}
  {https://academic.oup.com/mnras/article-pdf/290/3/411/18540204/290-3-411.pdf}
  \BibitemShut {NoStop}%
\bibitem [{\citenamefont {Jenkins}\ \emph {et~al.}(1998)\citenamefont
  {Jenkins}, \citenamefont {Frenk}, \citenamefont {Pearce} \emph
  {et~al.}}]{Jenkins98}%
  \BibitemOpen
  \bibfield  {author} {\bibinfo {author} {\bibfnamefont {A.}~\bibnamefont
  {Jenkins}}, \bibinfo {author} {\bibfnamefont {C.~S.}\ \bibnamefont {Frenk}},
  \bibinfo {author} {\bibfnamefont {F.~R.}\ \bibnamefont {Pearce}},  \emph
  {et~al.},\ }\href {\doibase 10.1086/305615} {\bibfield  {journal} {\bibinfo
  {journal} {The Astrophysical Journal}\ }\textbf {\bibinfo {volume} {499}},\
  \bibinfo {pages} {20} (\bibinfo {year} {1998})}\BibitemShut {NoStop}%
\bibitem [{\citenamefont {{Navarro}}\ \emph {et~al.}(1997)\citenamefont
  {{Navarro}}, \citenamefont {{Frenk}},\ and\ \citenamefont
  {{White}}}]{Navarro97}%
  \BibitemOpen
  \bibfield  {author} {\bibinfo {author} {\bibfnamefont {J.~F.}\ \bibnamefont
  {{Navarro}}}, \bibinfo {author} {\bibfnamefont {C.~S.}\ \bibnamefont
  {{Frenk}}}, \ and\ \bibinfo {author} {\bibfnamefont {S.~D.~M.}\ \bibnamefont
  {{White}}},\ }\href {\doibase 10.1086/304888} {\bibfield  {journal} {\bibinfo
   {journal} {\apj}\ }\textbf {\bibinfo {volume} {490}},\ \bibinfo {pages}
  {493} (\bibinfo {year} {1997})},\ \Eprint
  {http://arxiv.org/abs/astro-ph/9611107} {arXiv:astro-ph/9611107 [astro-ph]}
  \BibitemShut {NoStop}%
\bibitem [{\citenamefont {Huss}\ \emph {et~al.}(1999)\citenamefont {Huss},
  \citenamefont {Jain},\ and\ \citenamefont {Steinmetz}}]{Huss99}%
  \BibitemOpen
  \bibfield  {author} {\bibinfo {author} {\bibfnamefont {A.}~\bibnamefont
  {Huss}}, \bibinfo {author} {\bibfnamefont {B.}~\bibnamefont {Jain}}, \ and\
  \bibinfo {author} {\bibfnamefont {M.}~\bibnamefont {Steinmetz}},\ }\href
  {\doibase 10.1086/307161} {\bibfield  {journal} {\bibinfo  {journal} {The
  Astrophysical Journal}\ }\textbf {\bibinfo {volume} {517}},\ \bibinfo {pages}
  {64} (\bibinfo {year} {1999})}\BibitemShut {NoStop}%
\bibitem [{\citenamefont {Wang}\ and\ \citenamefont {White}(2009)}]{Wang09}%
  \BibitemOpen
  \bibfield  {author} {\bibinfo {author} {\bibfnamefont {J.}~\bibnamefont
  {Wang}}\ and\ \bibinfo {author} {\bibfnamefont {S.~D.~M.}\ \bibnamefont
  {White}},\ }\href {\doibase 10.1111/j.1365-2966.2009.14755.x} {\bibfield
  {journal} {\bibinfo  {journal} {Monthly Notices of the Royal Astronomical
  Society}\ }\textbf {\bibinfo {volume} {396}},\ \bibinfo {pages} {709}
  (\bibinfo {year} {2009})},\ \Eprint
  {http://arxiv.org/abs/https://academic.oup.com/mnras/article-pdf/396/2/709/3386539/mnras0396-0709.pdf}
  {https://academic.oup.com/mnras/article-pdf/396/2/709/3386539/mnras0396-0709.pdf}
  \BibitemShut {NoStop}%
\bibitem [{\citenamefont {{Pepe}}\ \emph {et~al.}(2002)\citenamefont {{Pepe}},
  \citenamefont {{Mayor}}, \citenamefont {{Rupprecht}} \emph
  {et~al.}}]{Pepe02}%
  \BibitemOpen
  \bibfield  {author} {\bibinfo {author} {\bibfnamefont {F.}~\bibnamefont
  {{Pepe}}}, \bibinfo {author} {\bibfnamefont {M.}~\bibnamefont {{Mayor}}},
  \bibinfo {author} {\bibfnamefont {G.}~\bibnamefont {{Rupprecht}}},  \emph
  {et~al.},\ }\href@noop {} {\bibfield  {journal} {\bibinfo  {journal} {The
  Messenger}\ }\textbf {\bibinfo {volume} {110}},\ \bibinfo {pages} {9}
  (\bibinfo {year} {2002})}\BibitemShut {NoStop}%
\bibitem [{\citenamefont {{Mayor}}\ \emph {et~al.}(2003)\citenamefont
  {{Mayor}}, \citenamefont {{Pepe}}, \citenamefont {{Queloz}}, \citenamefont
  {{Bouchy}}, \citenamefont {{Rupprecht}}, \citenamefont {{Lo Curto}} \emph
  {et~al.}}]{Mayor03}%
  \BibitemOpen
  \bibfield  {author} {\bibinfo {author} {\bibfnamefont {M.}~\bibnamefont
  {{Mayor}}}, \bibinfo {author} {\bibfnamefont {F.}~\bibnamefont {{Pepe}}},
  \bibinfo {author} {\bibfnamefont {D.}~\bibnamefont {{Queloz}}}, \bibinfo
  {author} {\bibfnamefont {F.}~\bibnamefont {{Bouchy}}}, \bibinfo {author}
  {\bibfnamefont {G.}~\bibnamefont {{Rupprecht}}}, \bibinfo {author}
  {\bibfnamefont {G.}~\bibnamefont {{Lo Curto}}},  \emph {et~al.},\ }\href@noop
  {} {\bibfield  {journal} {\bibinfo  {journal} {The Messenger}\ }\textbf
  {\bibinfo {volume} {114}},\ \bibinfo {pages} {20} (\bibinfo {year}
  {2003})}\BibitemShut {NoStop}%
\bibitem [{\citenamefont {{Dinh}}\ \emph {et~al.}(2014)\citenamefont {{Dinh}},
  \citenamefont {{Krueger}},\ and\ \citenamefont {{Bengio}}}]{Dinh14}%
  \BibitemOpen
  \bibfield  {author} {\bibinfo {author} {\bibfnamefont {L.}~\bibnamefont
  {{Dinh}}}, \bibinfo {author} {\bibfnamefont {D.}~\bibnamefont {{Krueger}}}, \
  and\ \bibinfo {author} {\bibfnamefont {Y.}~\bibnamefont {{Bengio}}},\
  }\href@noop {} {\bibfield  {journal} {\bibinfo  {journal} {arXiv e-prints}\
  ,\ \bibinfo {eid} {arXiv:1410.8516}} (\bibinfo {year} {2014})},\ \Eprint
  {http://arxiv.org/abs/1410.8516} {arXiv:1410.8516 [cs.LG]} \BibitemShut
  {NoStop}%
\bibitem [{\citenamefont {Rezende}\ and\ \citenamefont
  {Mohamed}(2015)}]{Rezende15}%
  \BibitemOpen
  \bibfield  {author} {\bibinfo {author} {\bibfnamefont {D.~J.}\ \bibnamefont
  {Rezende}}\ and\ \bibinfo {author} {\bibfnamefont {S.}~\bibnamefont
  {Mohamed}},\ }in\ \href@noop {} {\emph {\bibinfo {booktitle} {Proceedings of
  the 32nd International Conference on International Conference on Machine
  Learning - Volume 37}}},\ \bibinfo {series and number} {ICML'15}\ (\bibinfo
  {publisher} {JMLR.org},\ \bibinfo {year} {2015})\ p.\ \bibinfo {pages}
  {1530–1538}\BibitemShut {NoStop}%
\bibitem [{\citenamefont {Burnham}\ and\ \citenamefont
  {Anderson}(2004)}]{Burnham04}%
  \BibitemOpen
  \bibfield  {author} {\bibinfo {author} {\bibfnamefont {K.~P.}\ \bibnamefont
  {Burnham}}\ and\ \bibinfo {author} {\bibfnamefont {D.~R.}\ \bibnamefont
  {Anderson}},\ }\href {\doibase 10.1177/0049124104268644} {\bibfield
  {journal} {\bibinfo  {journal} {Sociological Methods \& Research}\ }\textbf
  {\bibinfo {volume} {33}},\ \bibinfo {pages} {261} (\bibinfo {year} {2004})},\
  \Eprint {http://arxiv.org/abs/https://doi.org/10.1177/0049124104268644}
  {https://doi.org/10.1177/0049124104268644} \BibitemShut {NoStop}%
\bibitem [{\citenamefont {{Holoien}}\ \emph {et~al.}(2017)\citenamefont
  {{Holoien}}, \citenamefont {{Marshall}},\ and\ \citenamefont
  {{Wechsler}}}]{Holoien17}%
  \BibitemOpen
  \bibfield  {author} {\bibinfo {author} {\bibfnamefont {T.~W.~S.}\
  \bibnamefont {{Holoien}}}, \bibinfo {author} {\bibfnamefont {P.~J.}\
  \bibnamefont {{Marshall}}}, \ and\ \bibinfo {author} {\bibfnamefont {R.~H.}\
  \bibnamefont {{Wechsler}}},\ }\href {\doibase 10.3847/1538-3881/aa68a1}
  {\bibfield  {journal} {\bibinfo  {journal} {The Astronomical Journal}\
  }\textbf {\bibinfo {volume} {153}},\ \bibinfo {eid} {249} (\bibinfo {year}
  {2017})},\ \Eprint {http://arxiv.org/abs/1611.00363} {arXiv:1611.00363
  [astro-ph.IM]} \BibitemShut {NoStop}%
\end{thebibliography}%

\newpage

\appendix

\section{Comparison of convergence criteria for Gaussian mixture models}
\label{app:AIC_BIC}
By default, our proposed procedure considers the validation log-likelihood to select the best number of components of the GMM model. Alternatively, one can use the Akaike or the Bayesian information criteria (AIC or BIC, respectively), which are defined as:
\begin{gather}
\mbox{AIC}=2p - 2\ell \ , \label{eq:IC1}\\
\mbox{BIC}=p\ln{N} - 2\ell \ ,
\label{eq:IC2}
\end{gather}
where $\ell$ is the log-likelihood on the training data (the entire dataset in this case), and $p=6c-1$ is the total number of GMM parameters, with $c$ the number of GMM components. These criteria include a term for the goodness of fit ($\ell$), plus a penalization term to avoid overfitted models. The model with the lowest AIC or BIC should be chosen, and ample discussions are available as to which criterion works best \cite{Burnham04, Bovy11, Holoien17}.

\begin{figure}
        \centering
        \includegraphics[width=0.95\columnwidth]{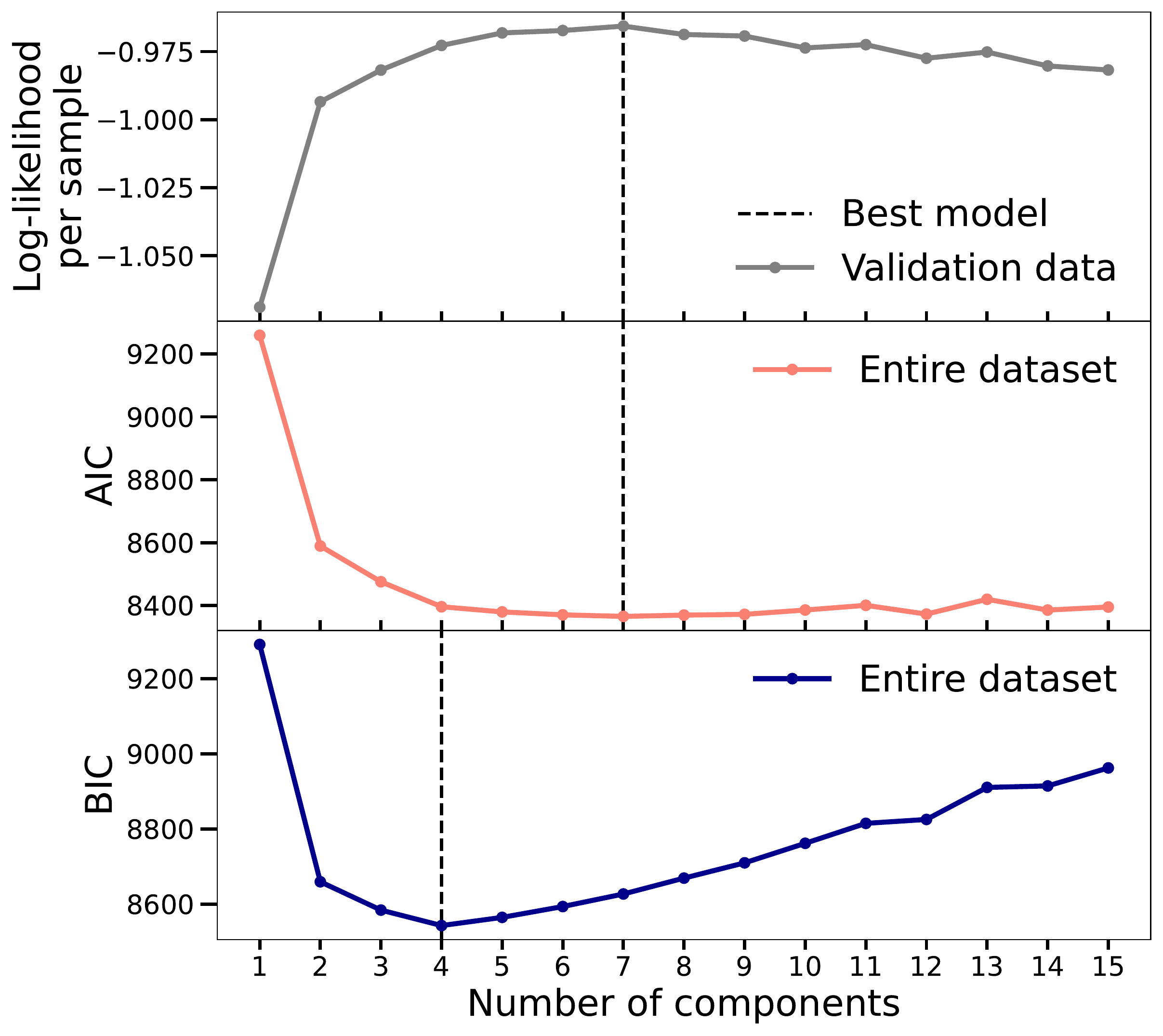}
        \caption{\textit{Top panel:} Validation log-likelihood per sample as a function of the number of Gaussian mixture model (GMM) components, when fitting joint samples from the ground-truth density for the first radial bin and the second latent variable of the $\mbox{IVE}_{\rm{infall}}$ model \cite{LucieSmith22}. The GMM reaches the highest log-likelihood at 7 components (dashed black line), which is the value used to then estimate mutual information according to the procedure described in Sect.~\ref{sec:proposed_procedure}, since the change in the log-likelihood is always bigger than the specified threshold of $10^{-5}$. \textit{Middle and bottom panels:} Trend of the Akaike information criterion (AIC) and Bayesian information criterion (BIC), defined in Eq.~(\ref{eq:IC1}) and Eq.~(\ref{eq:IC2}), as a function of the number of components. The lowest values of these quantities (marked by dashed black lines) indicate the models that best fit the data. The mutual information (MI) values returned by GMM-MI are $0.026 \pm 0.003$ nat and $0.025 \pm 0.003$ nat with 7 and 4 components respectively, showing that the different number of GMM components found does not significantly impact the final estimates of MI.}
        \label{fig:ll_components}
\end{figure}

As an example, we compare the trend of the validation log-likelihood, AIC and BIC in the context of the dark matter halo density profiles (Sect.~\ref{sec:dm_haloes}) when considering the second latent variable (latent `B') and the density in the first radial bin. We increase the number of GMM components from 1 to 15, and report the results in Fig.~\ref{fig:ll_components}. The validation log-likelihood reaches its maximum at 7 components, and then starts to slowly decrease. The AIC also prefers 7 components, while the BIC is in favor of fewer components (4). This is not surprising, since the penalization term is stronger in the BIC case, given the high number of samples. All three metrics considered are efficient to compute, and since the MI estimates returned by GMM-MI with 7 and 4 components are $0.026 \pm 0.003$ nat and $0.025 \pm 0.003$, respectively, we conclude that our approach is robust to the choice of the metric used to select the number of GMM components.

\section{Derivation of the mutual information between a continuous and a categorical variable}
\label{app:MI_der}
While Eq.~(\ref{eq:MI_cat}) is not novel, in this appendix we detail the assumptions made in its derivation. We first rewrite Eq.~(\ref{eq:MI}) as:
\begin{align}
    I(X, Y) = \int_{\mathcal{X}\times\mathcal{Y}}  p_{(X|Y)} (x | y) p_Y(y) \ln{\frac{p_{(X|Y)}(x|y)}{ p_X(x) }} \dif x \dif y  \ .
    \label{eq:MI_cond}
\end{align}
Then, we assume a generalized probability density function for the categorical variable $F$ over $\mathcal{F}$:
\begin{align}
    p_F(f) = \sum_{i=1}^{v} p_F(f=f_i) \delta(f-f_i) = \frac{1}{v} \sum_{i=1}^{v}  \delta(f-f_i) \ ,
    \label{eq:f_pdf}
\end{align}
where $\delta$ is the Dirac delta function, and in the last step we assumed that $F$ can take the values $f_{1:v}$ with equal probability. Combining the last two equations, we obtain:
\begin{align}
    \nonumber I (X, F)
    &= \int_{\mathcal{X}\times\mathcal{F}}  \dif x \dif f \ p_{(X|F)}(x | f) p_F(f) \ln{\frac{p_{(X|F)}(x|f)}{ p_X(x) }}  \\ 
    &= \frac{1}{v} \sum_{i=1}^v  \int_{\mathcal{X}} \dif x \ p_{(X|F)}(x | f_i) \left[ \ln{ p_{(X|F)}(x|f_i)} -\ln{\frac{1}{v} \sum_{j=1}^v p_{(X|F)}(x|f_j) } \right] \ ,
    \label{eq:MI_final}
\end{align}
as reported in Eq.~(\ref{eq:MI_cat}).

\section{Ground truth values of mutual information}
\label{app:true_mi}
We report the true values of MI for the bivariate distributions considered in Sect.~\ref{sec:validation}. These values can be obtained via direct integration of Eq.~(\ref{eq:MI}), and depend on a real-valued parameter $\alpha>0$. For the gamma-exponential distribution \cite{Darbellay99, Darbellay00, Kraskov04, Haeri14} as defined in Eq.~(\ref{eq:gammaexp}):
\begin{align}
\label{eq:mi_true_1}
    I(X, Y) = \psi(\alpha+1) - \ln{\alpha} \ , \end{align}
where $\psi$ is the digamma function, defined as:
\begin{align}
    \psi(x) = \frac{\Gamma'(x)}{\Gamma(x)} \ .
\end{align}
For the ordered Weinman exponential distribution \cite{Darbellay99, Darbellay00, Kraskov04, Haeri14} as defined in Eq.~(\ref{eq:weinman}):
\begin{align}
\label{eq:mi_true_2}
I(X, Y) = \begin{cases} 
     \ln{\left(\frac{1-2\alpha}{2\alpha}\right)} + \psi\left( \frac{1}{1-2\alpha}\right) -\psi(1) & \alpha<\frac{1}{2} \\
    -\psi(1) & \alpha=\frac{1}{2}\\
    \ln{\left(\frac{2\alpha-1}{2\alpha}\right)} + \psi\left( \frac{2\alpha}{2\alpha-1}\right) -\psi(1) & \alpha>\frac{1}{2}
    \end{cases} \ .
\end{align}

\end{document}